\documentstyle[12pt,amstex,righttag]{article}  
\numberwithin{equation}{section}
\newcommand{\sd}{\times}
\newcommand{\sk}{\phantom{-}}    
\newcommand{\iti}{\phantom{1}}   %
\newcommand{\pa}{\partial}       
\newcommand{\vT}{\Theta}        
\newcommand{\vat}{\vartheta}        
\newcommand{\vp}{\varpi}           
\newcommand{\gv}{\text{\tiny GV}}  %
\newcommand{\jacobi}{T}            
\newcommand{\gop}{Gopakumar--Vafa\ }
\newcommand{\cy}{Calabi--Yau\ }     
\newcommand{\gw}{Gromov--Witten\ }
\newcommand{\pf}{Picard--Fuchs\ }
\newcommand{\inst}{\text{inst}}    
\newcommand{\E}{\bo{E}}             
\newcommand{\Zseven}{\chi}
\newcommand{\one}{\tilde{1}}
\newcommand{\TE}{\varTheta_{\E_8}}   
\newcommand{\TEn}{\varTheta_{\E_8}^{(n)}}%
\newcommand{\ed}{{\cal E}}          
\newcommand{\fiber}{[\delta]}       
\newcommand{\h}{{\frak h}^{*}}      
\newcommand{\td}{\text{d}}          
\newcommand{\ti}{\text{i}}
\newcommand{\ox}{\overline{x}}
\newcommand{\oy}{\overline{y}}
\newcommand{\Einf}{E_{\infty}}
\newcommand{\lr}{\longrightarrow}
\newcommand{\SLtwo}{{SL}(2;\text{\bf Z})}
\newcommand{\Amb}{\bo{A}}
\newcommand{\K}{K\"ahler\ }
\newcommand{\dP}{del Pezzo\ }
\newcommand{\nn}{\nonumber}
\newcommand{\bo}[1]{\boldsymbol{#1}}    
\newcommand{\soeji}[1]{(#1)}
\newcommand{\IJMPA}[4]{Int. J. Mod. Phys. \text{\bf A{#1}} ({#2})
{#3}--{#4}} 
\newcommand{\NPB}[4]{Nucl. Phys. \text{\bf B{#1}} ({#2})
{#3}--{#4}} 
\newcommand{\ATMP}[4]{Adv. Theor. Math. Phys. \text{\bf {#1}} ({#2})
{#3}--{#4}} 
\newcommand{\CMP}[4]{Commun. Math. Phys. \text{\bf {#1}} ({#2})
{#3}--{#4}} 
\newcommand{\RMP}[5]{Rev. Math. Phys. \text{\bf {#1}} No. {#2} ({#3})
{#4}--{#5}} 
\begin{document}
\begin{titlepage}
\begin{flushright}
{\tt hep-th/0110121} \\
October, 2001
\end{flushright}
\vspace{0.5cm}
\begin{center}
{\LARGE Exceptional String: Instanton Expansions 
\vspace{0.1cm}

and Seiberg--Witten Curve
\par}
\lineskip .75em
\vskip2.5cm
{\large Kenji Mohri}
\footnote{{\em E-mail address:} mohri@@het.ph.tsukuba.ac.jp}
\vskip 1.5em
{\large\it Institute of Physics, University of Tsukuba \\
\vskip 0.4em
Ibaraki 305-8571, Japan}
\end{center}
\vskip3cm
\begin{abstract}
We investigate instanton expansions of
partition functions of several toric E-string models 
using local mirror symmetry and
elliptic modular forms.
We also develop a method to determine 
the Seiberg--Witten curve 
of E-string with the help of elliptic functions.
\vspace{3cm}

\end{abstract}
\end{titlepage}
\baselineskip=0.7cm




\tableofcontents
\section{Introduction}
Exceptional string (E-string in short)
has originally been discovered as 
the effective six dimensional (6D) theory
associated with a small $\E_8$ instanton in heterotic string
\cite{GH}. 
It has become clear since then that physical content of 
the toroidal compactification of E-string down to 4D
is extremely rich \cite{Ga1,GMS,KMV,LMW,MNW,MNVW,HST}.

4D E-string theory can be realized as type IIA string 
``compactified'' on the canonical line bundle 
of a rational elliptic surface $B_9$.
However we must rely on mirror symmetry for 
any quantitative analysis.
In fact, we have two versions of mirror symmetry.
One is local mirror symmetry \cite{CKYZ}
applied to $B_9$, which must be realized torically \cite{HKTY}.
At the expense of the restriction on \K moduli,
this method enables us to investigate systematically
the BPS spectrum even at higher genera \cite{HST}.
The other describes E-string by the Seiberg--Witten curve
\cite{Ga1,GMS} based on the fact that 
$B_9$ is self-mirror; 
the E-string is mapped to
type IIB string on a non-compact \cy manifold
containing a rational elliptic surface $S_9$ \cite{MNVW},
the complex moduli of which replaces the \K moduli of $B_9$.
The purpose of this paper is then to explore further 
these two descriptions of E-string.

This paper is organized as follows.
In section~\ref{E-string}, we collect general results  on 
the \K moduli parameters and the partition functions of 
E-string.
The remaining sections are divided into two parts.
The first part consists of 
sections~\ref{torus-model}--\ref{higher-genus-partition},
where we investigate the six toric E-string models 
by means of local mirror symmetry;
in section~\ref{torus-model}, we study 
the four torus models associated with
the E-string models, 
where the relation between the periods and 
elliptic modular forms is the central problem,
in section~\ref{model-building} we analyze the Picard--Fuchs system
of the E-string models;
we then investigate the partition functions of the E-string models
of genus zero, one in  sections~\ref{genus-zero-partition}, 
\ref{genus-one-partition} respectively;
finally in section~\ref{higher-genus-partition},  
partition functions
of higher genera are considered
in connection with elliptic modular forms,
\gop invariants and $\E_8$ Jacobi forms.

The second part deals with the Seiberg--Witten curve
of E-string; after a review of the period map of 
rational elliptic surface 
in section~\ref{seiberg-witten-curve},
we obtain a procedure to determine   
the Seiberg--Witten curve for given Wilson lines 
using elliptic functions in section~\ref{inverse-problem}.

\section{E-String}
\label{E-string}
\subsection{Homology lattice and affine root lattice}
Let us first consider type IIA string compactified on 
a \cy threefold
which contains a rational elliptic surface $B_9$ 
with a section \cite{MP} as a divisor,
where we use the symbol $B_N$ for $\E_N$ \dP surface;
a rational elliptic surface is alternatively
called an almost $\E_9$ \dP surface.

The resulting 4D theory is a $N\!=\!2$ supergravity.
The large radius limit of  
the normal direction to $B_9$ then kills almost all
the degrees of freedom of the original \cy threefold;
effectively we are left with 
$K_{B_9}$, the canonical line bundle of $B_9$, as a compactification
manifold, and the 4D theory reduces to a E-string theory
which does not contain gravity.
We are interested in the physical quantities of the E-string
that depend only on the complexified \K class
$J\in H_2(B_9;\text{\bf C})$ of $B_9$
inherited from the \cy threefold.
It should be noted that we can take  
the complex structure of $B_9$ to be generic,
that is, we can assume that
the elliptic fibration $\pi\colon B_9\to \text{\bf P}^1$
has the twelve singular fibers of the $\text{I}_1$ type.

We recall here some properties of the 
second homology classes  $H_2(B_9)$.
The lattice structure in $H_2(B_9)$ induced by  
intersection parings is given by
\begin{align}
H_2(B_9)&=
\text{\bf Z}l \oplus
\text{\bf Z}\ed_1
\oplus \cdots \oplus\text{\bf Z}\ed_9,
\\
l\cdot l&=1, \quad 
l\cdot \ed_i=0, \quad
\ed_i\cdot \ed_j=-\delta_{ij}.
\label{generators}
\end{align} 
$B_9$ is realized as a blow-up of $\text{\bf P}^2$ at 
the nine points which are the base points of a cubic pencil.
The class $l$ is given by the total transform of a line in $\text{\bf P}^2$,
while $\ed_i$ the exceptional divisor associated with the $i$th
base point.
The first Chern class $c_1(B_9)$ is represented by 
the fiber class $\fiber$ of the elliptic fibration 
$\pi\colon B_9\to \text{\bf P}^1$, 
which can be written 
in terms of the generators (\ref{generators}) as
$\fiber=3l-\ed_1-\cdots -\ed_9$.
It is readily verified that
$\fiber\cdot l=3$,
$\fiber\cdot \ed_i=1$, and
$\fiber\cdot \fiber=0$.

The sublattice $\fiber^{\perp}$ of $H_2(B_9)$,
which is the orthogonal complement of $\fiber$ in $H_2(B_9)$,
is naturally identified with
the root lattice of the affine Lie algebra $\E_8^{(1)}$:
$
L(\E_8^{(1)})=\bigoplus_{i=0}^{8}\text{\bf Z}\alpha_i,
$
where $\{\alpha_i\}_{i=0}^{8}$ is the simple roots of $\E_8^{(1)}$. 
This can be seen as follows.
First we find the generators of $\fiber^{\perp}$
as a free $\text{\bf Z}$-module 
{\allowdisplaybreaks 
\begin{align}
\fiber^{\perp}&=\bigoplus_{i=0}^{8}\text{\bf Z}[\alpha_i],
\qquad
H_2(B_9)\cong \fiber^{\perp}\oplus \text{\bf Z}\ed_9,
\label{bunkai}
\\
[\alpha_0]=\ed_8-\ed_9, 
\quad
[\alpha_i]&=\ed_i-\ed_{i+1},
\ \  i=1,\dots,7,
\quad
[\alpha_8]=l-\ed_1-\ed_2-\ed_3.
\nn
\end{align}}
Then we see that the intersection pairings 
$([\alpha_i]\cdot [\alpha_j])$ coincides with
the minus of the Cartan matrix of $\E_8^{(1)}$. 
Note that the element of $L(\E_8^{(1)})$  
corresponding to $\fiber\in \fiber^{\perp}$ is
\begin{equation}
\fiber=
[\alpha_0]+
2[\alpha_1]+
4[\alpha_2]+
6[\alpha_3]+
5[\alpha_4]+
4[\alpha_5]+
3[\alpha_6]+
2[\alpha_7]+
3[\alpha_8],
\nn
\end{equation}
which is in accord with the standard notation 
of the affine Lie algebra;
$\alpha_0=\delta-\theta$, 
with $\theta$ the highest root of $\E_8$.

Let $\{\omega_i\}_{i=1}^{8}$ be the fundamental weights of $\E_8$.
Then $(\omega_i|\alpha_j)=\delta_{i,j}$ and $\theta=\omega_7$.
We denote the corresponding elements
of $\fiber^{\perp}$ by
$\{[\omega_i]\}_{i=1}^{8}$, which satisfy 
$[\omega_i]\cdot [\alpha_j]=-\delta_{i,j}$. 

The dual of the Cartan subalgebra of $\E_8^{(1)}$,
which we denote by $\h$, contains  
the zeroth fundamental weight $\Lambda_0$
as a generator in addition to the simple roots
$\{\alpha_i\}$, that is,
\[
\h=\text{\bf C}\Lambda_0
\oplus \text{\bf C}\delta
\oplus \text{\bf C}\alpha_1
\oplus \cdots
\oplus \text{\bf C}\alpha_8.
\] 
$\Lambda_0$ satisfies 
$(\Lambda_0|\alpha_i)=0,\  i\ne 0$
and $(\Lambda_0|\delta)=1$,
which leads to the final identification
$\ed_9=-[\Lambda_0]-1/2\fiber$.
Thus we have another basis of the vector space 
\begin{equation}
H_2(B_9;\text{\bf Q})=\bigoplus_{i=1}^{8}\text{\bf Q}[\alpha_i]
\oplus \text{\bf Q}[\delta]\oplus \text{\bf Q}[\Lambda_0],
\end{equation}
from which it follows that  
$H_2(B_9;\text{\bf C})$ can be identified with $\h$, 
the CSA of $\E_8^{(1)}$.

To summarize, we find the isomorphism:
$\h\ni x \mapsto [x]\in H_2(B_9;\text{\bf C})$
of the $\text{\bf C}$-vector spaces
such that
$(x|y)=-[x]\cdot[y]$ for any $x,y\in \h$.

The important consequence on the above observation 
is that the Weyl group $W(\E_8^{(1)})$  
of $\E_8^{(1)}$ acts on $H_2(B_9;\text{\bf C})$.
To see the effect of the Weyl action on the \K moduli
parameters of the E-string, which should be a physical symmetry,
we put the \K class $J\in H_2(B_9;\text{\bf C})\cong \h$
in the canonical form
\begin{equation}
J=(\tfrac12\tau+\sigma)[\delta]-\tau[\Lambda_0]
-\sum_{i=1}^{8}\mu_i[\omega_i],
\label{canonical}
\end{equation}   
where $\tau$ is the complex modulus of the torus $\text{\bf T}^2$,
on which the 6D E-string theory is compactified to 4D,
$\tau+\sigma$  the \K modulus of $\text{\bf T}^2$,
or equivalently, the E-string tension and the self-dual $B$-flux
on $\text{\bf T}^2$, and $(\mu_i)$ the $\E_8$ Wilson lines, that is,
the moduli of the flat $\E_8$ bundles on $\text{\bf T}^2$ 
\cite{Ga1,GMS}.

%
There exists a semi-direct product structure:
$W(\E_8^{(1)})=W(\E_8)\sd T$,
where $W(\E_8)$ is the finite Weyl group and
$T:=\{t_{\beta}|\beta\in L(\E_8)\}$, 
the translation by the root lattice $L(\E_8)$.

$W(\E_8)$ affects only the Wilson lines 
$\mu:=\sum_{i=1}^{8}\mu_i\omega_i$;
it is clear that $\mu_i$ transforms in the same way 
as the simple root $\alpha_i$ of $\E_8$.
The embedding $\imath$ of the finite root lattice 
$L(\E_8)$ in the Euclidean space $\text{\bf R}^8$ defined by
\begin{equation}
\imath:
\begin{pmatrix}
\alpha_1\\
\alpha_2\\
\alpha_3\\
\alpha_4\\
\alpha_5\\
\alpha_6\\
\alpha_7\\
\alpha_8
\end{pmatrix}\rightarrow
\begin{pmatrix}
\sk \frac12  & -\frac12   &  -\frac12  &  -\frac12  &  -\frac12  
&  -\frac12   &  -\frac12  &  \sk \frac12  \\
 -1 & \sk 1 & \sk 0 & \sk 0 & \sk 0 &\sk 0 & \sk 0 &\sk  0 \\
 \sk 0 & -1 &  \sk 1 & \sk 0 & \sk  0 & \sk 0 & \sk  0 &\sk  0 \\
 \sk 0 &\sk  0 &  -1 & \sk 1 & \sk  0 &\sk 0 & \sk  0 & \sk 0 \\
 \sk 0 &\sk  0 &\sk  0 & -1 &\sk  1 &\sk 0 &\sk  0 & \sk 0 \\
\sk 0 &\sk 0 &\sk  0 &\sk 0 &  -1 &\sk 1 &\sk  0 & \sk 0 \\
\sk 0 &\sk 0 &\sk  0 &\sk 0 &\sk  0 & -1 &\sk  1 & \sk 0 \\
\sk 1 &\sk 1 &\sk  0 &\sk 0 &\sk  0 &\sk 0 &\sk  0 & \sk 0 
\end{pmatrix}
\begin{pmatrix}
\text{\bf e}_1\\
\text{\bf e}_2\\
\text{\bf e}_3\\
\text{\bf e}_4\\
\text{\bf e}_5\\
\text{\bf e}_6\\
\text{\bf e}_7\\
\text{\bf e}_8
\end{pmatrix},
\label{euclid}
\end{equation}
where $\{\text{\bf e}_i\}$ is an orthonormal basis of $\text{\bf R}^8$,
greatly simplifies the description of 
$W(\E_8)$; it is generated by 
(i) the permutations of $\{\text{\bf e}_i\}$,
(ii) the  sign flips of even number of $\text{\bf e}_i$s and 
(iii) the involution,
which is the Weyl reflection with respect to the root
$\theta\!-\!\sum_{i=1}^{7}\alpha_i$,
$\text{\bf e}_i\mapsto \text{\bf e}_i-1/4\sum_{j=1}^{8}\text{\bf e}_j$
\cite{Ga1,GMS}.
If $\imath(\mu)=\sum_{i=1}^{8}m_i(\mu)\text{\bf e}_i$,
we call $(m_i(\mu))$ the Euclidean coordinates of $\mu$.

On the other hand, the translation $t_{\beta}$
by $\beta\in L(\E_8)$ is given by
\begin{equation}
t_{\beta}(x)=x+(x|\delta)\beta
-\left\{\frac{(\beta|\beta)}{2}(x|\delta)+(x|\beta)\right\}\delta.
\end{equation}
Substituting $x=J$ (\ref{canonical}),
we see the $t_{\beta}$ action on the \K moduli parameters:
\begin{equation}
\sigma\mapsto 
\sigma+\frac12 (\beta|\beta)\tau+(\mu|\beta),
\quad \tau \mapsto \tau,
\quad
\mu\mapsto \mu+\tau\beta,
\label{Eichler--Siegel}
\end{equation}
which is familiar as a symmetry in classical theta functions.
Note that to realize  another symmetry translation
$\sigma\to \sigma$,
$\tau\to\tau$,
$\mu\to\mu+\alpha$,
as a Weyl group action,
we need to consider the doubly-affinized $\E_8$ algebra,
$\E_9^{(1)}$ \cite{DHIZ,Saito,Satake},
which might be possible only if we extend $H_2(B_9)$
to the full homology lattice 
$H_{0}(B_9)\oplus H_2(B_9)\oplus H_4(B_9)$.

\subsection{Partition functions}
The \gw partition function $Z_{g;n}(\tau|\mu)$
of genus $g$ and winding number $n$
of the local $B_9$ model with the
\K moduli $J$ given in (\ref{canonical}) 
is defined by the expansion coefficient of 
the genus $g$ potential $F_g$ \cite{BCOV2}:
\begin{equation}
F_g(\sigma,\tau,\mu)=
\sum_{n=1}^{\infty}p^n
Z_{g;n}(\tau|\mu),
\quad p:=\text{e}^{2\pi \ti \sigma}.
\label{genus-g-F}
\end{equation}
We remark here that the genus $g$ refers to that of type IIA string,
while the winding number $n$ refers to that of E-string.

Using $\varphi(\tau):=\prod_{n=1}^{\infty}(1-q^n)$,
with $q=\text{e}^{2\pi \ti\tau}$, we can write
$Z_{g;n}$ as 
\begin{equation}
Z_{g;n}(\tau|\mu)=
\frac{\jacobi_{g;n}(\tau|\mu)}{\varphi(\tau)^{12n}}.
\label{bunshi-bunbo}
\end{equation}
The numerator  $\jacobi_{g;n}$ 
is so-called a Weyl-invariant $\E_8$ {\em quasi}-Jacobi form
\cite{KY} of index $n$ and weight $2g-2+6n$, 
which means that
$\jacobi_{g;n}(\tau|\mu)$ is invariant 
under the  $W(\E_8)$ action on $\mu$,
and it has the following transformation properties:
\begin{align}
\jacobi_{g;n}
(\tau|\mu\!+\!\alpha\!+\!\beta\tau)&=
\text{e}^{-\pi \ti n[(\beta|\beta)\tau+2(\beta|\mu)]}\,
\jacobi_{g;n}(\tau|\mu), 
 \
\alpha,\beta \in L(\E_8),
\label{Jac-1}
\\
\hat{\jacobi}_{g;n}\left.\left(\frac{a\tau\!+\!b}{c\tau\!+\!d}
\right|\frac{\mu}{c\tau\!+\!d}\right)
&=(c\tau+d)^{2g-2+6n}\,
\text{e}^{\frac{n\pi\ti c}{c\tau+d}(\mu|\mu)}\,
\hat{\jacobi}_{g;n}(\tau|\mu),
\nn\\
\begin{pmatrix}
a & b\\
c & d
\end{pmatrix}
&\in \SLtwo,
\label{Jac-2}
\end{align}
where $\hat{\jacobi}_{g;n}$ is obtained from $\jacobi_{g;n}$
by replacement of each $E_2(\tau)$ in it with
$\hat{E}_2(\tau):=E_2(\tau)-3/(\pi\text{Im}\tau)$,
which gets rid of the modular anomaly of $Z_{g;n}$
which comes from 
\begin{equation}
E_2\left(\frac{a\tau\!+\!b}{c\tau\!+\!d}\right)
=(c\tau\!+\!d)^2\,
\left(E_2(\tau)+\frac{12}{2\pi \ti}\,
\frac{c}{c\tau\!+\!d}\right),
\label{E2}
\end{equation}
at the sacrifice of holomorphy \cite{SW}.
(\ref{Jac-1}) shows
that $\text{e}^{2\pi\ti n\sigma}\jacobi_{g;n}(\tau|\mu)$
is invariant under the translation $t_{\beta}$ 
(\ref{Eichler--Siegel}) as expected.
In particular, the genus zero, singly winding partition function
can be given by the classical level one $\E_8$ theta function 
\cite{KMV,CL,MNVW}: 
\begin{equation}
\jacobi_{0;1}(\tau|\mu)=\TE(\tau|\mu)
:=\sum_{\alpha\in L(\E_8)}
\text{e}^{\pi \ti
(\alpha|\alpha) \tau
+2\pi \ti (\alpha|\mu)}.
\label{kihon-shiki}
\end{equation}  
$\TE$ can be written in terms of the Euclidean coordinates
for the Wilson lines as 
\begin{equation}
\TE(\tau|\mu)=\frac12 \sum_{a=1}^{4}
\prod_{i=1}^{8}\vat_{a}(\tau|m_i),
\qquad
\iota(\mu)=\sum_{i=1}^8 m_i\, \text{\bf e}_i.
\end{equation}
The relation to curve counting problem is as follows \cite{HST}:
in $Z_{0;1}(\tau|\mu)$,
the $\E_8$ theta function part is regarded as the contribution 
to $Z_{0;1}$ from the Mordell--Weil lattice \cite{FYY}, while
the denominator part $\varphi^{12}$ \cite{KMV,CL} 
from the twelve degenerate elliptic fibres 
of the fibration $\pi:B_9\to \text{\bf P}^1$.

Furthermore it is clear from the analysis of the BPS states
in \cite{HST,HST2}
that higher genus partition functions 
of the singly winding sector can be obtained by 
\begin{equation}
\Phi_{1}:=
\sum_{g=0}^{\infty}Z_{g;1}(\tau|\mu) x^{2g-2}
=\left(\frac{\eta(\tau)^3}{\vat_1(\tau|\tfrac{x}{2\pi})}\right)^2
Z_{0;1}(\tau|\mu).
\label{ansatz-for-genus-g}
\end{equation}
The genus expansion of the right hand side reads \cite{SW}, 
\begin{equation*}
\left(\frac{\eta(\tau)^3}{\vat_1(\tau|\tfrac{x}{2\pi})}\right)^2
=\frac{1}{x^2}
\exp\left[\sum_{k=1}^{\infty}\frac{(-1)^{k+1}b_{2k}}{k(2k)!}
E_{2k}(\tau)x^{2k}\right]
=\sum_{g=0}^{\infty}x^{2g-2}A_g(\tau),
\end{equation*}
where $b_n$ and 
the $(2k)$th Eisenstein series $E_{2k}(\tau)$ are defined by
\begin{equation*}
\frac{x}{\text{e}^x-1}=\sum_{n=0}^{\infty}b_n\frac{x^n}{n!},
\quad
E_{2k}(\tau)=1-\frac{4k}{b_{2k}}\sum_{n=1}^{\infty}
\sigma_{2k-1}(n)q^n,
\quad \sigma_{k}(n):=\sum_{m|n}m^k.
\end{equation*}
As $E_{2k}\!\in\!\text{\bf Q}[E_4,E_6]$, that is, 
$E_8\!=\!E_4^2$,
$E_{10}\!=\!E_4E_6$,
$E_{12}\!=\!1/691(250E_6^2+441E_4^3)$,
$E_{14}\!=\!E_4^2E_6$, for example,
the coefficients $A_g$ above can be reduced to 
{\allowdisplaybreaks
\begin{align*}
A_1&=\frac{1}{12}E_2,\
A_2=\frac{1}{1440}(E_4+5E_2^2),\
A_3=
\frac{1}{362880}(4E_6+21E_2E_4+35E_2^3),
\\
A_4&=
\frac{1}{87091200}(39E_4^2+80E_2E_6+210E_2^2E_4+175E_2^4),
\\
A_5&=
\frac{1}{11496038400}
(136E_4E_6+429E_2E_4^2+440E_2^2E_6+770E_2^3E_4+385E_2^5),
\dots.
\end{align*}}
%
%

The partition functions $Z_{g;n}$ obeys 
the modular anomaly equation 
\begin{align}
\frac{\pa Z_{g;n}(\tau|\mu)}{\pa E_2(\tau)}
&=\frac{1}{24}
\sum_{h+h'=g}
\sum_{k+k'=n}
kk'Z_{h;k}(\tau|\mu)\,
Z_{h';k'}(\tau|\mu)
\nn\\
&+\frac{1}{24}
n(n+1)Z_{g-1;n}(\tau|\mu),
\label{modular-anomaly}
\end{align}
which has first been found in genus zero partition functions
\cite{MNW}, and then generalized to higher genus 
partition functions in \cite{HST}.
This can be rewritten using the 
genus $g$ potential (\ref{genus-g-F}) 
with each $E_2(\tau)$ in it replaced by
$\hat{E}_2(\tau)$, 
$\hat{F}_g=\sum_{n=1}^{\infty}p^n\hat{Z}_{g;n}$, as 
\begin{equation*}
-16\pi \ti \left(\text{Im}\tau\right)^2
\Big(\frac{\pa \hat{F}_g}{\pa \bar{\tau}}\Big)=
\sum_{h=0}^{g}\vT_p\hat{F}_{h}
\vT_p\hat{F}_{g-h}
+(\vT_p\!+\!1)\vT_p\hat{F}_{g-1},
\end{equation*}
where $\vT_p=p\partial/\partial p$ is the Euler derivative.
Compare this with the holomorphic anomaly equation 
for the topological string amplitude $F_g$ 
for $g\!>\!1$ on a \cy threefold $X$ 
expressed in a general coordinate system \cite{BCOV2} 
\begin{equation*}
\pa_{\bar{i}}\hat{F}_g=
\frac{1}{2}C_{\bar{i}\bar{j}\bar{k}}\,
\text{e}^{2K}
G^{j{\bar j}}G^{k{\bar k}}
\Big(\sum_{h=1}^{g-1}D_j\hat{F}_hD_k\hat{F}_{g-h}
+D_jD_k\hat{F}_{g-1}\Big).
\end{equation*}
where $G_{i{\bar j}}=\pa_i\pa_{{\bar j}}K$ 
is the Weil--Petersson--Zamolodchikov metric on the complex moduli space
${\cal M}(X^{*})$ of the mirror $X^{*}$, 
$\text{e}^{-K}$ a natural Hermite metric on a positive 
line bundle $L\to {\cal M}(X^*)$, with 
the holomorphic three-form of $X^*$ being a local section,
and $C_{ijk}$ the $\text{\bf 27}^3$ Yukawa coupling.
The covariant derivatives above have contributions 
not only from the Levi-Civita connection 
$\varGamma_{ij}^{k}=G^{k\bar{l}}\pa_jG_{i{\bar l}}$
but also from the Hermite connection $-\pa_iK$ on $L$
because $\hat{F}_g$ is a section of $L^{2g-2}$,
\begin{equation*}
D_i\hat{F}_g=\left[\pa_i+(2\!-\!2g)\pa_iK\right]\hat{F}_g,
\quad
D_iD_j\hat{F}_g=
\left[\pa_i+(2\!-\!2g)\pa_iK\right]D_j\hat{F}_g
-\varGamma_{ij}^kD_k\hat{F}_g.
\end{equation*}

Even more important is the  potential $\Phi_n$ 
of fixed winding number $n$:
\begin{equation}
\Phi_n(\lambda;\tau,\mu)=
\sum_{g=0}^{\infty}x^{2g-2}Z_{g;n}(\tau|\mu),
\qquad 
\lambda:=\text{e}^{\frac{1}{12}E_2(\tau)x^2}.
\label{fixed-winding}
\end{equation}
Introduction of $\lambda$ and the full potential  
${\cal A}:=\sum_{g=0}^{\infty}x^{2g-2}F_g
=\sum_{n=1}^{\infty}p^n\Phi_n$
greatly simplifies  the modular anomaly equation 
(\ref{modular-anomaly}):
\begin{equation}
\vT_{\lambda}\exp({\cal A})
=\frac{1}{2}\vT_p(\vT_p+1)\exp({\cal A}).
\label{kantan}
\end{equation}
If we substitute the $p$-expansion
$\exp({\cal A})=\sum_{n=0}^{\infty}p^n\psi_n(\Phi_1,\dots,\Phi_n)$,
where $\psi_n$ is the $n$th Schur polynomial,
then we obtain 
$\vT_{\lambda}\psi_n=n(n\!+\!1)/2\,\psi_n$, hence
\begin{equation}
\psi_n(\Phi_1,\dots,\Phi_n)=\lambda^{\frac{1}{2}n(n+1)}
\psi_n(\Phi_1^0,\dots,\Phi_n^0),
\label{simplify}
\end{equation}
where $\Phi_n^0=\Phi_n|_{E_2=0}$ is the anomaly-free part;
in particular, $\Phi_1=\lambda\Phi_1^0$ from 
(\ref{ansatz-for-genus-g}).

Finally (\ref{simplify}) enables us to  
express the solution for the modular anomaly equation
(\ref{modular-anomaly}) concisely by $\Phi_n$ with 
the anomaly-free part $\Phi_n^0$ as its integration constant:
{\allowdisplaybreaks
\begin{align}
\Phi_2&=\frac{1}{2}(\lambda-1)(\Phi_1)^2+\lambda^3\Phi_2^0,
\label{winding-2}\\
\Phi_3&=-\frac{1}{6}(\lambda-1)^2(2\lambda+1)
(\Phi_1)^3+(\lambda^2-1)\Phi_1\Phi_2+\lambda^6\Phi_3^0,
\nn\\
\Phi_4&=\frac{1}{24}(\lambda-1)^3
(6\lambda^3+6\lambda^2+3\lambda+1)(\Phi_1)^4
+\frac{1}{2}(\lambda^4-1)(\Phi_2)^2
+(\lambda^3-1)\Phi_1\Phi_3
\nn\\
&-\frac{1}{2}(\lambda-1)^2(\lambda+1)(2\lambda^2+\lambda+1)
(\Phi_1)^2\Phi_2
+\lambda^{10}\Phi_4^0.
\nn
\end{align}}
We are interested in $\Phi_n$  also because 
it encapsulates the interaction of $n$ E-strings,
which can be quite different from 
that of fundamental strings \cite{MNVW}.

We also give the prediction of the leading term of
the partition function:
\begin{equation}
Z_{g;n}(\tau|\mu)=\beta_{g,0}\,n^{2g-3}+O(q^n),
\qquad
\beta_{g,0}:=\frac{|b_{2g}(2g\!-\!1)|}{(2g)!},
\label{leading}
\end{equation} 
which generalizes the genus zero result in \cite{MNW}.
This can be used to partially fix the integration constants 
of the partition function.

\section{Four Torus Models}
\label{torus-model}
In this section, we investigate  the instanton expansion
by means of elliptic modular forms \cite{Scho}
of the periods of four one-parameter families 
of elliptic curves, $\E_{N}$, $N\!=\!5,6,7,8$,
each of which is defined as a complete intersection 
in a toric variety as shown in the Table \ref{teeburu}.
The results of this section will play an essential role
in the instanton expansion of E-string model 
in the next section.
\begin{table}[h]
\caption{Toric models}
\label{teeburu}
\begin{center}
\begin{tabular}{|l|c|}
\hline
elliptic curve  & $\E_9$ \dP \\
\hline
$\E_5:\ \text{\bf P}^3[2,2]$ & $\E_5$  \phantom{\LARGE A}${\E}_{\one}$\\
\hline
$\E_6:\ \text{\bf P}^2[3]$ & $\E_6$  \phantom{\LARGE A}${\E}_0$ \\
\hline
$\E_7:\ \text{\bf P}_{1,1,2}[4]$ & $\E_7$  \phantom{\LARGE A}\phantom{$E_0$} \\
\hline
$\E_8:\ \text{\bf P}_{1,2,3}[6]$ & $\E_8$ \phantom{\LARGE A}\phantom{$E_0$}\\
\hline
\end{tabular}
\end{center}
\end{table}

The $\E_N$ torus model have been studied  
in connection with the one-parameter
families of the local $\E_N$ del Pezzo models
\cite{KMV,LMW,CKYZ,MOhY,MOnY}, and  
\cy threefolds with these elliptic fibers have been used to 
describe 4D string models the $\E_8$ gauge symmetry of which 
is broken to $\E_N$ by Wilson lines \cite{AFIU,KMV}.

\subsection{Periods of elliptic curves}
The \pf operator for the periods
of the $\E_N$ family of elliptic curves is given by
\begin{equation}
{\cal D}^{\soeji{N}}_{\text{ell}}=
\vT^2-z(\vT+\alpha^{\soeji{N}})(\vT+\beta^{\soeji{N}}), 
\label{PF-torus}
\end{equation}
where $\vT=z\td/\td z$ is the Euler derivative
with respect to the {\em bare} modulus $z$, and
$\alpha$, $\beta$ is defined by
$\alpha+\beta=1$, and 
$(\alpha^{\soeji{5}},\alpha^{\soeji{6}},
\alpha^{\soeji{7}},\alpha^{\soeji{8}})
=(1/2,1/3,1/4,1/6)$.

The Gauss hypergeometric equation defined by
(\ref{PF-torus}) has the regular singular points
at $z=0,1,\infty$.
The solutions around $z\!=\!0$,
which corresponds to  the large radius limit point 
of the sigma model with the torus as target, are given by
{\allowdisplaybreaks \begin{align}
\vp(z)&=
\sum_{n=0}^{\infty}a(n)z^{n}
={}_2F_1(\alpha,\beta;1;z),
\label{sol-00}
\\
\vp_{D}(z)&=
\vp(z)\log\left(\frac{z}{\kappa}\right)+
\sum_{n=1}^{\infty}a(n)b(n)z^{n},
\label{sol-01}
\end{align}}
where $(\kappa^{\soeji{5}}, \kappa^{\soeji{6}},
\kappa^{\soeji{7}}, \kappa^{\soeji{8}})=(16,27,64,432)$,
and
\begin{equation*}
a(n)=\frac{(\alpha)_n(\beta)_n}{(n!)^2},\quad
b(n)=\sum_{k=0}^{n-1}
\left(\frac{1}{k\!+\!\alpha}+\frac{1}{k\!+\!\beta}
-\frac{2}{k\!+\!1}\right),
\end{equation*}
with $(\alpha)_n:=\Gamma(n\!+\!\alpha)/\Gamma(\alpha)$ 
the Pochhammer symbol.

The mirror map of the torus is defined by
\begin{equation}
2\pi \ti \tau=\frac{\vp_D(z)}{\vp(z)},
\label{torus-mirror}
\end{equation}
where $\tau$ is the \K parameter of the torus.

In terms of the local coordinate $u=1\!-\!z$
at another regular singular point $z\!=\!1$, 
the \pf operator takes the same form
as one at $z\!=\!0$ (\ref{PF-torus}), and
the continuation of the solutions above becomes \cite{MOnY}
\begin{equation}
\begin{pmatrix}
\vp(z)\\
\vp_D(z)
\end{pmatrix}
\to 
-\begin{pmatrix}
0 & x\\
x^{-1} & 0
\end{pmatrix}
\begin{pmatrix}
 \vp(u)\\
 \vp_D(u)
\end{pmatrix},
\qquad x=\frac{\sin(\pi\alpha)}{\pi}.
\label{sol-1}
\end{equation}
Let $M_{0,1,\infty}$ be the monodromy matrices 
around the regular singular points $z\!=\!0,1,\infty$
with respect to the basis
$\{\vp_D(z)/(2\pi \ti), \vp(z)\}$.
We can compute $M_0$ and $M_1$ from (\ref{sol-00}),
(\ref{sol-01}) and (\ref{sol-1})
respectively:
\begin{equation}
M_0=T=
\begin{pmatrix}
1 & 1 \\
0  & 1 
\end{pmatrix},
\quad
M_1=-ST^{(9-N)}S=
\begin{pmatrix}
1 & 0 \\
-(9\!-\!N)  & 1 
\end{pmatrix},
\end{equation}
where $S$ and $T$ are the standard generators of
$\SLtwo$. 
The remaining one $M_{\infty}$ can be obtained from  
$M_{\infty}=M_1M_0$.
The monodromy group $\varGamma_{\text{ell}}$ is generated by
$M_0$ and $M_1$;
in particular, 
for $N\ne 8$,
$\varGamma_{\text{ell}}^{\soeji{N}}\cong \varGamma_0(9\!-\!N)$, 
which is the Hecke subgroup of $\SLtwo$ \cite{Scho}
\[
\varGamma_0(h):=\left\{
\left.
\begin{pmatrix}
a & b\\
c & d
\end{pmatrix}
\in \SLtwo
\ \right|\ 
c\equiv 0 \mod h  \right\}.
\]
The structure of $M_{*}(\varGamma_0(9\!-\!N))$,
the graded ring of 
of the modular forms of even degree 
of $\varGamma_0(9\!-\!N)$, should be clear from
\begin{align}
\E_5:\ \
M_{*}(\varGamma_0(4))
&=\text{\bf C}[\vartheta_3(2\tau)^4,\vartheta_4(2\tau)^4], 
\\
\E_6:\ \
M_{*}(\varGamma_0(3))&=
\left.\text{\bf C}[\vp^{(6)},H]\right|_{\text{even}},
\ \
H:=\frac{\eta(\tau)^9}{\eta(3\tau)^3},
\\
\E_7:\ \
M_{*}(\varGamma_0(2))&=\text{\bf C}[A,B],
\ \ 
A:=\frac{1}{2}(\vat_3^4+\vat_4^4)(\tau),\
B:=\vat_3^4\vat_4^4(\tau).
\end{align}

The fundamental solution $\vp(z)$  
admits the following expressions:
\begin{alignat}{2}
\vp^{\soeji{5}}(z)\ 
&=\varTheta_{\bo{A}_1\oplus \bo{A}_1}(\tau) &  \
&=\vat_3(2\tau)^2,
\\
\vp^{\soeji{6}}(z)\ 
&=\varTheta_{\bo{A}_2}(\tau) & \
&=\vat_3(2\tau)\vat_3(6\tau)
+\vat_2(2\tau)\vat_2(6\tau), 
\\
\vp^{\soeji{7}}(z)^2\ &=\varTheta_{\bo{D}_4}(\tau) & \
&=A(\tau),
\\
\vp^{\soeji{8}}(z)^4
&=\TE(\tau) & \
&=E_4(\tau),
\end{alignat}
where $\varTheta_{\bo{K}}(\tau)$ is the
theta function associated with the root lattice of $\bo{K}$.

Despite of the modular anomaly (\ref{E2}),
$hE_2(h\tau)-E_2(\tau)$ is a sound modular form of 
$\varGamma_0(h)$ for each $h\in \text{\bf N}$.
The following identity then shows that 
$\vp(z)^2$ is an element of $M_2(\varGamma_{\text{ell}}^{(N)})$
for $N=5,6,7$:
\begin{equation}
(8\!-\!N)\omega^{(N)}(z)^2=(9\!-\!N)E_2((9\!-\!N)\tau)-E_2(\tau).
\end{equation}



Let us define for later use the modular function $e_{2k}$ by
$e_{2k}(\tau)=E_{2k}(\tau)\vp^{-2k}$ for each model,
which turn out to be written in terms of $y:=1/(1-z)$:
\begin{alignat}{3}
\E_5&: \quad & 
e_4(\tau)&=16-\frac{16}{y}+\frac{1}{y^2},
\quad &
e_6(\tau)&
=-64+\frac{96}{y}-\frac{30}{y^2}-\frac{1}{y^3},
\nn\\
\E_6&: \quad & 
e_4(\tau)&=9-\frac{8}{y},
\quad &
e_6(\tau)&=-27+\frac{36}{y}-\frac{8}{y^2},
\nn\\
\E_7&: \quad & 
e_4(\tau)&=4-\frac{3}{y},
\quad &
e_6(\tau)&=-8+\frac{9}{y},
\nn\\
\E_8&: \quad & 
e_4(\tau)&=1,
\quad &
e_6(\tau)&=-1+\frac{2}{y}.
\nn
\end{alignat}
We see that 
$e_4$ and $e_6$ are subject to the algebraic relation:
{\allowdisplaybreaks 
\begin{alignat*}{2}
\E_5&: \quad &
0&=e_4^3-e_6^2-108\,e_4e_6-8640\,e_6-1620\,e_4^2-17280\,e_4+27648,
\\
\E_6&: \quad &
0&=8\,e_6+18\,e_4+e_4^2-27,
\\
\E_7&: \quad  &
0&=3\,e_4+e_6-4.
\end{alignat*}}
%
Let $T^{\soeji{N}}(\tau)$ be the modular function of 
$\varGamma_{\text{ell}}^{\soeji{N}}$ defined by
\begin{align}
T^{\soeji{5}}(\tau)&=\left(\frac{\eta(\tau)}{\eta(4\tau)}\right)^8,\ \
T^{\soeji{6}}(\tau)=\left(\frac{\eta(\tau)}{\eta(3\tau)}\right)^{12},
\ \
T^{\soeji{7}}(\tau)=\left(\frac{\eta(\tau)}{\eta(2\tau)}\right)^{24},
\nn \\
T^{\soeji{8}}(\tau)&=
\bigg(\frac{E_4(\tau)^{\frac32}\!+\!E_6(\tau)}{2\eta(\tau)^{12}}\bigg)^2,
\end{align}
where 
$\eta(\tau):=q^{1/24}\prod_{n=1}^{\infty}(1\!-\!q^n)$ 
is the Dedekind eta function with $q\!=\!\text{e}^{2\pi \ti \tau}$.
Then the inversion of the mirror map (\ref{torus-mirror}) 
can be given by
\begin{equation}
\frac{1}{z(\tau)}=1+\frac{1}{\kappa}\, T(\tau).
\end{equation}
We note here that when $N\!=\!5,6,7$,
$\varGamma_{\text{ell}}^{(N)}$ is a genus zero modular group,
that is, the upper half plane divided by the action 
of $\varGamma_{\text{ell}}^{(N)}$ is completed to $\text{\bf P}^1$, 
and the $q$-expansion of $T^{(N)}(\tau)$ reproduces 
the  Thompson series associated with $\varGamma_{\text{ell}}^{(N)}$
\cite{LY}.

Another useful expressions of $1-z$ by modular forms are
\begin{alignat}{2}
\E_5:& \quad
1-z(\tau)=
\left(\frac{\vat_4(2\tau)}{\vat_3(2\tau)}\right)^4, & \qquad
\E_6:& \quad
1-z(\tau)=\frac{H}{\vp^3},
\\
\E_7:& \quad
1-z(\tau)=
\frac{B}{A^2},
& \qquad
\E_8:& \quad
1-z(\tau)=\frac12(1+E_6 E_4^{-\frac32}(\tau)).
\end{alignat}

For each model, $\vp(z)$ and $z$ satisfy the following equations:
\begin{equation}
\frac{1}{2\pi \ti}\frac{\td z}{\td \tau}
=z(1-z)\,\vp(z)^2,
\quad
{\kappa}\eta(\tau)^{24}=z(1-z)^{9-N}\vp(z)^{12}.
\label{rec-ni-hitsuyou}
\end{equation}

\subsection{Modular identities}
The following power series will play an important role 
in the instanton expansion of the E-string models:
\begin{align}
\xi^{\soeji{0}}(z)&=\sum_{n=1}^{\infty}a^{\soeji{6}}(n)h(3n)z^n,
\label{modular-0}
\\
\xi^{\soeji{\one}}(z)&=\sum_{n=1}^{\infty}a^{\soeji{5}}(n)h(2n)z^n,
\label{modular-1}
\\
\xi^{\soeji{N}}(z)&=\sum_{n=1}^{\infty}a^{\soeji{N}}(n) h(n)z^n,
\quad \text{for\ } N=5,6,7,8.
\label{modular-N}
\end{align}
where $h(n)=\sum_{k=1}^{n}k^{-1}$ is the harmonic function.
A computer experiment gives the identities: 
\begin{alignat}{2}
&\E_6: \quad & 
\psi^{\soeji{0}}(\tau)
&=\exp\left(-\frac{\xi^{\soeji{0}}(z)}{\vp(z)}\right)
=\left(\frac{27 q}{z}\right)^{\frac16}(1\!-\!z)^{\frac12}
\label{identity-0}
\\
&\E_5: \quad & 
\psi^{\soeji{\one}}(\tau)
&=\exp\left(-\frac{\xi^{\soeji{\one}}(z)}{\vp(z)}\right)
=\left(\frac{16 q}{z}\right)^{\frac14}(1\!-\!z)^{\frac12},
\label{identity-1}
\\
&\E_N: \quad & 
\psi^{\soeji{N}}(\tau)&=\exp\left(-\frac{\xi^{\soeji{N}}(z)}{\vp(z)}\right)
=\left(\frac{\kappa^{\soeji{N}} q}{z}\right)^{\frac12}(1\!-\!z)^{\frac12}
=\left(qT^{\soeji{N}}(\tau)\right)^{\frac12}.
\label{identity-N}
\end{alignat}

\section{Model Building of E-Strings}
\label{model-building}
In this section, we give six models:
 $\E_N$, $N\!=\!0,\one,5,6,7,8$,
of $\E_9$ almost del Pezzo surface listed in Table \ref{teeburu},
which is realized as a complete intersection in a toric variety,
and analyze the Picard--Fuchs systems defined by them
at the large radius point.
Among the six models, we call the four 
$\E_{5,6,7,8}$ the principal series,
the reason of which will become clear 
in the investigation of the Picard--Fuchs system of them.
\subsection{Principal series}
The $\E_{6,7,8}$ models are obtained as hypersurfaces 
in ambient toric threefolds with their \K classes
inherited  from those of the ambient spaces.

To describe these, let us first define the action of 
$(\text{\bf C}^{*})^2$ on $\text{\bf C}^5$ for each model by
\begin{alignat}{2}
&\E_6:\quad &
(x_1,x_2,x_3,x_4,x_5)
&\to (\lambda x_1,\lambda x_2, \lambda^{-1}\mu x_3,\mu x_4,\mu x_5),
\\
&\E_7:\quad &
(x_1,x_2,x_3,x_4,x_5)
&\to (\lambda x_1,\lambda x_2, \lambda^{-1}\mu x_3,\mu x_4,\mu^2 x_5),
\\
&\E_8:\quad &
(x_1,x_2,x_3,x_4,x_5)
&\to (\lambda x_1,\lambda x_2, \lambda^{-1}\mu x_3,\mu^2 x_4,\mu^3 x_5),
\end{alignat}
where $\lambda,\mu\in \text{\bf C}^{*}$.
Then the ambient toric variety $\bo{A}$ can be realized by the quotient
$\bo{A}:=(\text{\bf C}^5\!-\!F)/(\text{\bf C}^{*})^2,$
where $F=\{x_1\!=\!x_2\!=\!0\}\cup \{x_3\!=\!x_4\!=\!x_5\!=\!0\}$ 
is the bad point set of the $(\text{\bf C}^*)^2$ action.

It is easy to see that the ambient space $\Amb$ 
for the $\E_{6,7,8}$ model
has a structure of weighted projective surface bundle over 
$\text{\bf P}^1$
with fiber 
$\text{\bf P}^2$,
$\text{\bf P}_{1,1,2}$,
$\text{\bf P}_{1,2,3}$
respectively. 

We can now define the $\E_9$ almost \dP surface $B_9$ 
for the $\E_{6,7,8}$ model as a hypersurface in $\Amb$ of bidegree
$(0,3)$, $(0,4)$ and $(0,6)$ respectively,
where the bidegree refers to the $(\lambda,\mu)$ charge of
the defining polynomial.
The \K classes for $B_9$ induced from those of
the ambient space are given by
\begin{align}
J^{\soeji{6}}&=\sigma\fiber+\tau(\fiber+\ed_7+\ed_8+\ed_9),
\\
J^{\soeji{7}}&=\sigma\fiber+\tau(\fiber+\ed_8+\ed_9),
\\
J^{\soeji{8}}&=\sigma\fiber+\tau(\fiber+\ed_9),
\end{align}
where we recall that $\fiber$ is the first Chern class
of $B_9$.
The local model of $B_9$ embedded in a \cy threefold
is described by the total space of 
the canonical line bundle $K_{B_9}$,
which is again realized as a hypersurface in the non-compact
toric variety ${\cal O}_{\Amb}(-1,0)$.  
{}From this fact, 
we can identify the Mori vectors of the local \cy
models, which plays an essential role 
in the formulation of mirror symmetry \cite{HKTY}:
\begin{alignat}{3}
&\E_6:\quad & 
l_{1}&=(0; 1,1,-1,0,0,-1), \quad & 
l_{2}&=(-3;0,0,1,1,1,0),
\label{Mori-E6}
\\
&\E_7:\quad & 
l_{1}&=(0; 1,1,-1,0,0,-1), \quad & 
l_{2}&=(-4;0,0,1,1,2,0),
\label{Mori-E7}
\\
&\E_8:\quad & 
l_{1}&=(0; 1,1,-1,0,0,-1), \quad & 
l_{2}&=(-6;0,0,1,2,3,0).
\label{Mori-E8}
\end{alignat}
It will suffice here to point out that 
the first components of the vectors $l_1, l_2$
are (minus) the degrees of the hypersurface, while the rest 
the $\text{\bf C}^{*}$ charges of the homogeneous coordinates
of $K_{\bo{A}}$, 
where the last one corresponds to the non-compact direction.

%
Next we consider the $\E_5$ model.
The ambient toric variety in this case is 
a $\text{\bf P}^3$ bundle over $\text{\bf P}^1$.
This space admits the quotient realization 
$(\text{\bf C}^6\!-\!F)/(\text{\bf C}^*)^2$, where 
the $(\text{\bf C}^{*})^2$ action on $\text{\bf C}^6$ is defined by  
$
(x_1,x_2,x_3,x_4,x_5,x_6)\to
(\lambda x_1,\lambda x_2,
\lambda^{-1}\mu x_3,\mu x_4,\mu x_5,\mu x_6),
$
and the bad point set 
$F=\{x_1\!=\!x_2\!=\!0\}\cup \{x_3\!=\!x_4\!=\!x_5\!=\!x_6\!=\!0\}$.

The surface $B_9$ is defined by a complete intersection 
of two hypersurfaces of bidegree $(0,2)$. 
The \K class of the $\E_5$ model induced from the ambient space 
turns out to be
\begin{equation}
J^{\soeji{5}}
=\sigma\fiber+\tau(\fiber+\ed_6+\ed_7+\ed_8+\ed_9).
\end{equation}
The Mori vectors in this case can be seen  
as in the case of the $\E_{6,7,8}$ models above: 
\begin{equation}
\E_5:\quad  
l_{1}=(0,0; 1,1,-1,0,0,0,-1), \quad
l_{2}=(-2,-2;0,0,1,1,1,1,0).
\label{Mori-E5}
\end{equation}

\subsection{$\E_0$ and $\E_{\one}$ models}
We present here the remaining two models,
$\E_0$ and $\E_{\one}$. Because the ambient spaces for these
models are products of projective spaces, rather than a twisted fiber
bundle as in the case of principal series,
the toric construction of them may be omitted.

The $\E_0$ model is realized as a hypersurface of
bidegree $(1,3)$ in $\text{\bf P}^1\!\times\!\text{\bf P}^2$,
while the $\E_{\one}$ model as one of tridegree $(1,2,2)$ 
in $\text{\bf P}^1\!\times\!\text{\bf P}^1\!\times\!\text{\bf P}^1$. 

The \K classes induced from the ambient spaces are given by
\begin{align}
J^{\soeji{0}}&=\sigma\fiber+\tau l,
\\
J^{\soeji{\one}}&=\sigma\fiber
+\tau_1(l-\ed_1)+\tau_2(l-\ed_2).
\end{align} 
Note that for the $\E_{\one}$ model,
we must put the restriction $\tau_1=\tau_2=\tau$ 
to obtain a two-parameter model.

The Mori vectors of these two models are 
{\allowdisplaybreaks
\begin{alignat}{3}
&\E_0:\quad &
l_1&=(-1;1,1,0,0,0,-1),\quad &
l_2&=(-3;0,0,1,1,1,0),
\label{Mori-E0}
\\
&\E_{\one}:\quad &
l_1&=(-1;1,1,0,0,0,0,-1),\quad &
l_2&=(-2;0,0,1,1,0,0,0),
\nn\\
&{} \quad &
l_3&=(-2;0,0,0,0,1,1,0).\quad & {}&
\label{Mori-E1}
\end{alignat}}

\subsection{\K classes}
Each of the \K classes of the six models 
obtained above admits two important representations,
the one of which reveals the blow-up/down scheme 
of the two-parameter family of $B_9$s,
and the other the canonical form 
(\ref{canonical}) suggested by the investigation of
6D E-string compactified on $\text{\bf T}^2$ \cite{Ga1,GMS}: 
{\allowdisplaybreaks
\begin{alignat}{2}
J^{\soeji{0}}&=\sigma c_1(B_9)+\frac{\tau}{3} c_1(B_0)
&
&=(\sigma+\tfrac32 \tau)\fiber
-3\tau [\Lambda_0]
-\tau [\omega_8],
\\
\label{K-0}
J^{\soeji{\one}}&=\sigma c_1(B_9)
+\frac{\tau}{2} c_1(B_{\one})
& 
&=(\sigma+2\tau) \fiber
-4\tau [\Lambda_0]
-\tau [\omega_2],
\label{K-1}
\\
J^{\soeji{5}}&=\sigma c_1(B_9)+\tau c_1(B_5)
&
&=(\sigma+2\tau) \fiber
-4\tau [\Lambda_0]
-\tau [\omega_5],
\label{K-5}
\\
J^{\soeji{6}}&=\sigma c_1(B_9)+\tau c_1(B_6)
&
&=(\sigma+\tfrac32\tau) \fiber
-3\tau [\Lambda_0]
-\tau [\omega_6],
\label{K-6}
\\
J^{\soeji{7}}&=\sigma c_1(B_9)+\tau c_1(B_7)
&
&=(\sigma +1\tau)\fiber
-2\tau [\Lambda_0]
-\tau[\omega_7],
\\
\label{K-7}
J^{\soeji{8}}&=\sigma c_1(B_9)+\tau c_1(B_8)
&
&=(\sigma+\tfrac12\tau) \fiber
-1 \tau [\Lambda_0],
\label{K-8}
\end{alignat}}
where we recall that
$\fiber=c_1(B_9)=3l-\sum_{i=1}^9\ed_i$,
$\ed_9\!=\!-[\Lambda_0]\!-\!1/2\fiber$,
$B_N$ the $\E_N$ \dP surface with 
$c_1(B_N)=3l-\sum_{i=1}^N\ed_i$,
$B_0=\text{\bf P}^2$ with $c_1(B_0)=3l$,
and 
$B_{\one}=\text{\bf P}^1\times\text{\bf P}^1$ 
with
$c_1(B_{\one})=2(2l\!-\!\ed_1\!-\!\ed_2)$.

The existence of the blow-down of each model to 
the one parameter family of the corresponding \dP surface,
where the name of the model comes, 
leads to the relation between \gw invariants of
these models as we shall see below.
The appearance of a single fundamental weight in the Wilson line
term in each model is also suggestive.

There are a few words to be said on 
what should be the $E_N$ models for $N=1,2,3,4$.
Their \K moduli are expected to be written as
{\allowdisplaybreaks
\begin{align}
J^{(1)}&=
\sigma c_1(B_9)+\tau c_1(B_1)
=(\sigma+4\tau)\fiber
-8\tau[\Lambda_0]
-\tau([\omega_1]\!+\!2[\omega_8]),
\\
J^{(2)}&=
\sigma c_1(B_9)+\tau c_1(B_2)
=(\sigma+\tfrac72\tau)\fiber
-7\tau[\Lambda_0]
-\tau([\omega_2]+[\omega_8]),
\\
J^{(3)}&=
\sigma c_1(B_9)+\tau c_1(B_3)
=(\sigma+3\tau)\fiber
-6\tau[\Lambda_0]
-\tau[\omega_3],
\\
J^{(4)}&=
\sigma c_1(B_9)+\tau c_1(B_4)
=(\sigma+\tfrac52\tau)\fiber
-5\tau[\Lambda_0]
-\tau [\omega_4],
\end{align}}
In view of the fact that $B_{1,2,3}$ are themselves
toric surfaces, we might expect 
that at least for $N=1,2,3$,
$\E_{N}$ models can be realized as a 
hypersurface in the toric threefold 
$B_{N}\!\times\!\text{\bf P}^1$,
which we will not pursue further.

\subsection{Periods}
The solutions to the \pf differential equations 
around the large radius limit point 
$(z_1,z_2)=(0,0)$ can be obtained by the Frobenius method
\cite{HKTY}.
We first define the formal power series 
$\Omega^{\soeji{N}}_{\rho}(z_1,z_2)$
from the Mori vectors $l_1,l_2$:
\begin{equation}
\Omega_{\rho}(z_1,z_2)=
\sum_{n_1=0}^{\infty}\sum_{n_2=0}^{\infty}
A(n_1\!+\!\rho_1,n_2\!+\!\rho_2)
z_1^{n_1+\rho_1}z_2^{n_2+\rho_2},
\end{equation}
%
where $A(n_1,n_2)$ for the $\E_0$ and $\E_{\one}$ models are
{\allowdisplaybreaks
\begin{align}
A^{\soeji{0}}(n_1,n_2)&=
\left(\frac{1}{27}\right)^{n_2}
\frac{\Gamma(1+n_1+3n_2)}
{\Gamma(1-n_1)\Gamma(1+n_1)^2\Gamma(1+n_2)^3},
\\
A^{\soeji{\one}}(n_1,n_2)&=
\left(\frac{1}{16}\right)^{n_2}
\frac{\Gamma(1+n_1+2n_2)\Gamma(1+2n_2)}
{\Gamma(1-n_1)\Gamma(1+n_1)^2\Gamma(1+n_2)^4},
\label{E1-A}
\end{align}}
while for the principal series $\E_{N}$, $N=5,6,7,8$, 
\begin{equation}
A^{\soeji{N}}(n_1,n_2):=
\frac{\Gamma(\alpha+n_2)\Gamma(\beta+n_2)}
{\Gamma(\alpha)\Gamma(\beta)
\Gamma(1+n_2)\Gamma(1+n_1)^2
\Gamma(1-n_1)\Gamma(1+n_2-n_1)}.
\end{equation}

We remark that the $\E_{\one}$ model has been defined as the 
three parameter model. 
The three Mori vectors (\ref{Mori-E1}) produce the formal power series
{\allowdisplaybreaks
\begin{align*}
\Omega_{\rho}(z_1,w_{1},w_{2})
&=\sum_{n_1=0}^{\infty}
\sum_{k_{1}=0}^{\infty}
\sum_{{k}_{2}=0}^{\infty}
A(n_1\!+\!\rho_1,k_{1}\!+\!\rho_{21},{k}_{2}\!+\!{\rho}_{22})
z_1^{n_1+\rho_1}
w_{1}^{k_{1}+\rho_{21}}
{w}_{2}^{{k}_{2}+{\rho}_{22}},
\\
A(n_1,k_1,k_2)&=\frac{\Gamma(1+n_1+2k_1+2k_2)}
{\Gamma(1-n_1)\Gamma(1+n_1)^2\Gamma(1+k_1)^2\Gamma(1+k_2)^2}.
\end{align*}}
Then the reduction to the two parameter model
is achieved by setting $w_1\!=\!w_2\!=\!z_2/16$;
correspondingly, the coefficient of the power series 
reduces to the one described in (\ref{E1-A}): 
{\allowdisplaybreaks
\begin{align*}
16^{n_2}A(n_1,n_2)&=\sum_{k_1+k_2=n_2}A(n_1,k_1,k_2)
\\
&=\frac{\Gamma(1+n_1+2n_2)}{\Gamma(1-n_1)\Gamma(1+n_1)^2\Gamma(1+n_2)^2}
\sum_{k_1+k_2=n_2}\left(\frac{(n_2)!}{(k_1)! (k_2)!}\right)^2
\\
&=\frac{\Gamma(1+n_1+2n_2)\Gamma(1+2n_2)}
{\Gamma(1-n_1)\Gamma(1+n_1)^2\Gamma(1+n_2)^4}.
\end{align*}}

%
The computation of the \pf operators
can be done in a standard manner
once we know the form of $A(n_1,n_2)$ \cite{HKTY}.
For $\E_0$ and $\E_{\one}$ model we have
{\allowdisplaybreaks
\begin{align}
{\cal D}_1^{\soeji{0}}&=\vT_1^2
+z_1\vT_1(\vT_1+3\vT_2+1),
\nn \\
{\cal D}_2^{\soeji{0}}&=9\vT_2^2-3\vT_1\vT_2
-z_2(3\vT_2+\vT_1+1)(3\vT_2+\vT_1+2)
-3z_1\vT_1\vT_2,
\label{PF-0}
\\
{\cal D}_1^{({1})}&=\vT_1^2
+z_1\vT_1(\vT_1+2\vT_2+1),
\nn\\
{\cal D}_2^{({1})}
&=4\vT_2^2-2\vT_1\vT_2-z_2(2\vT_2+1)(2\vT_2+\vT_1+1)
-2z_1\vT_1\vT_2,
\label{PF-1}
\end{align}}
and for the principal series $\E_N$ with $N=5,6,7,8$,
\begin{align}
{\cal D}_1^{\soeji{N}}&=
\vT_1^2
-z_1\vT_1(\vT_1-\vT_2),
\nn \\
{\cal D}_2^{\soeji{N}}&=\vT_2(\vT_2-\vT_1)
-z_2(\vT_2+\alpha^{\soeji{N}})(\vT_2+\beta^{\soeji{N}}),
\label{PF-N}
\end{align}
where 
$\vT_1=z_1\pa_{z_1}$ and $\vT_2=z_2\pa_{z_2}$.
In passing, we remark here that the \pf system for the principal
series gives an Appell--Horn hypergeometric system
in four ways according to the choice of the 
base point $(z_{1}^{\pm 1}\!=\!0,z_{2}^{\pm 1}\!=\!0)$.
It would be quite interesting to analyze their monodromies.

The Frobenius method then gives 
the four solutions of the \pf system 
(\ref{PF-0}), (\ref{PF-1}), (\ref{PF-N}):
{\allowdisplaybreaks
\begin{align}
\vp(z_2)&=\left. \Omega_{\rho}(z_1,z_2)\right|_{\rho=0},
\\
\vp_D(z_2)&=\left. \pa_{\rho_2}\Omega_{\rho}(z_1,z_2)
\right|_{\rho=0},
\\
\phi(z_1,z_2)&=\left. \pa_{\rho_1}\Omega_{\rho}(z_1,z_2)
\right|_{\rho=0},
\\
\phi_D^{\soeji{0}}(z_1,z_2)&=\left.
\left(\pa_{\rho_1}\pa_{\rho_2}+\frac16\pa_{\rho_2}^2\right) 
 \Omega_{\rho}^{\soeji{0}}(z_1,z_2)\right|_{\rho=0},
\\
\phi^{\soeji{\one}}_D(z_1,z_2)&=\left.
\left(\pa_{\rho_1}\pa_{\rho_2}+\frac14\pa_{\rho_2}^2\right) 
 \Omega_{\rho}^{\soeji{\one}}(z_1,z_2)\right|_{\rho=0},
\\
\phi^{\soeji{N}}_D(z_1,z_2)&=\left.
\left(\pa_{\rho_1}\pa_{\rho_2}+\frac12\pa_{\rho_2}^2\right) 
 \Omega_{\rho}^{\soeji{N}}(z_1,z_2)\right|_{\rho=0},
\ \  N=5,6,7,8.
\end{align}}
The first two solutions $\vp$ and $\vp_D$ are the same as 
those of the corresponding torus 
(\ref{sol-00}), (\ref{sol-01}), which is evident from the structure of 
the \pf operators above.

The third one is given by
\begin{equation}
\phi(z_1,z_2)
=\vp(z_2)\log(z_1)
+\xi^{\soeji{N}}(z_2)
+\sum_{n_1=1}^{\infty}{\cal L}^{\soeji{N}}_{n_1}\vp(z_2)z_1^{n_1},
\end{equation}
where $\xi^{\soeji{N}}(z_2)$ has been defined in 
(\ref{modular-0}), (\ref{modular-1}) and (\ref{modular-N})
and ${\cal L}^{\soeji{N}}_n$ is the differential operator defined by 
\begin{alignat}{2}
{\cal L}^{\soeji{0}}_{n}&=\frac{(-1)^n}{n\cdot n!}
\prod_{k=1}^{n}(3\Theta_2+k),\qquad & &{}
\\
{\cal L}_{n}^{\soeji{\one}}&=\frac{(-1)^n}{n\cdot n!}
\prod_{k=1}^{n}(2\Theta_2+k),\qquad & &{}
\\
{\cal L}^{\soeji{N}}_{n}&=\frac{(-1)^n}{n\cdot n!}
\prod_{k=0}^{n-1}(\Theta_2-k), \qquad & N&=5,6,7,8.
\end{alignat}

The flat coordinates $\tau$, $\sigma$ are obtained by
the mirror map
\begin{equation}
2\pi \ti \tau=\frac{\vp_D(z_2)}{\vp(z_2)},
\quad
2\pi \ti \sigma=\frac{\phi(z_1,z_2)}{\vp(z_2)}.
\label{mirror-E9}
\end{equation}

Let $c_n:={\cal L}_n\vp/\vp$, then (\ref{mirror-E9})
with (\ref{identity-0}), (\ref{identity-1}), (\ref{identity-N}) yields
\begin{equation}
\text{e}^{2\pi \ti \sigma}\psi^{\soeji{N}}(\tau)
=z_1\exp\Big(\sum_{n_1=1}^{\infty}c_{n_1}(\tau)z_1^{n_1}\Big),
\end{equation}
from which we know that 
the inversion of the mirror for $z_1$ takes the form
\begin{equation}
z_1=\tilde{p}\Big(1+\sum_{n=1}^{\infty}d_n(\tau)\tilde{p}^n\Big),
\label{z1-tenkai}
\end{equation}
where $\tilde{p}:=\text{e}^{2\pi \ti \sigma}\psi(\tau)$,
and 
$d_n\in \text{\bf Q}[c_1,\dots,c_n]$; explicitly, 
\begin{align}
d_1&=-c_1,\quad
d_2=\frac32\, c_1^2-c_2,\quad
d_3=-\frac83\,c_1^3+4\,c_1c_2-c_3,
\nn\\
d_4&=\frac{125}{24}\,c_1^4-\frac{25}{2}\,c_1^2c_2
+5\,c_1c_3+\frac52\,c_2^2-c_4,
\dots.
\label{dn-tenkai}
\end{align}

After some tedious calculations, we find that
the last solution $\phi_D$ is given by  
\begin{align}
\phi^{\soeji{0}}_D(z_1,z_2)&=
(2\pi \ti)^2\vp(z_2)
\Big(\sigma\tau+\frac16\tau^2
-\frac{1}{6}\Big)
-\sum_{n=1}^{\infty}\frac{f_{n}(z_2)}{\vp(z_2)}\,z_1^{n},
\label{4-cycle-0}
\\
\phi^{\soeji{\one}}_D(z_1,z_2)&=
(2\pi \ti)^2\vp(z_2)
\Big(\sigma\tau+\frac14\tau^2
-\frac{1}{8}\Big)
-\sum_{n=1}^{\infty}\frac{f_{n}(z_2)}{\vp(z_2)}\,z_1^{n},
\label{4-cycle-1}
\\
\phi^{(\scriptstyle{N})}_D(z_1,z_2)&=
(2\pi \ti)^2\vp(z_2)
\Big(\sigma\tau+\frac12\tau^2
-\frac{1}{2(9\!-\!N)}\Big)
-\sum_{n=1}^{\infty}\frac{f_{n}(z_2)}{\vp(z_2)}\,z_1^{n},
\label{4-cycle-N}
\end{align}
where $f_{n}(z_2)$ is the ``higher Wronskian'' defined by 
\begin{equation}
f_{n}(z_2)=-\left(\vp{\cal L}_n\vp_D-\vp_D{\cal L}_n\vp\right)(z_2).
\end{equation}
%
It must be noted here that 
a crucial ingredient in obtaining the concise expression for
$\phi_D$ above is the  following formula of combinatoric nature:
\begin{equation}
\boxed{\Big(\sum_{n=1}^{\infty}a(n)k(n)z_2^n\Big)
\Big(\sum_{n=0}^{\infty}a(n)z_2^n\Big)
=
\Big(\sum_{n=1}^{\infty}a(n)b(n)z_2^n\Big)
\Big(\sum_{n=1}^{\infty}a(n)s(n)z_2^n\Big)},
\end{equation}
where $s(n)$ and $k(n)$ are defined by
\[
s(n)=\sum_{l=0}^{n-1}
\Big(\frac{1}{l\!+\!\alpha}+\frac{1}{l\!+\!\beta}\Big),\
k(n)=s(n)
\Big(s(n)-2\sum_{l=0}^{n-1}\frac{1}{l\!+\!1}\Big)
-\sum_{l=0}^{n-1}
\Big(\frac{1}{(l\!+\!\alpha)^2}+\frac{1}{(l\!+\!\beta)^2}
\Big).
\]

The instanton part of the genus zero prepotential of the model 
$F_0$ is defined by
\begin{equation}
\frac{1}{2\pi \ti}\left(\frac{\pa F_0}{\pa \sigma}\right)
=h\sum_{n=1}^{\infty}
\frac{f_{n}(z_2)}{\vp(z_2)^2}\,z_1^{n},
\label{genus-zero-inst}
\end{equation}
where $(h^{\soeji{0}},h^{\soeji{\one}},
h^{\soeji{5}},h^{\soeji{6}},h^{\soeji{7}},
h^{\soeji{8}})=(3,4,4,3,2,1)$
is the normalization factor,
which may be found, for example,  
by computation of the classical central charge 
of a D-branes system corresponding to a coherent sheaf ${\cal F}$
with the \K class $J^{\soeji{N}}$ \cite{MOhY}:
\[
Z^{\text{class}}({\cal F})=
-\Big[\exp(-J^{\soeji{N}})\cdot \text{ch}({\cal F})
\cdot \big([B_9]+\tfrac12\fiber+\tfrac12[\text{pt}]\big)\Big].
\] 

\subsection{Relation to Seiberg--Witten periods}
In this subsection
we show how to realize the two periods 
$\phi$ and  $\phi_D$,
which have been obtained as the solutions of the \pf
differential equations,
as the Seiberg--Witten integrals
of the 6D non-critical string theory \cite{GH} 
compactified on $\text{\bf T}^2$ to 4D \cite{Ga1,GMS,LMW,MNW}.

Let $\phi_0(z_1,z_2)$ be a solution of  the \pf system 
(\ref{PF-0}), (\ref{PF-1}), or (\ref{PF-N}).
It is easy to see that if $\vT_1\phi_0=0$, then ${\cal D}_1\phi_0=0$ 
is automatic 
and ${\cal D}_2\phi_0=0$ reduces to the \pf equation for the 
corresponding elliptic curve: ${\cal D}_{\text{ell}}\phi_0=0$;
hence $\phi_0$ is a linear combination of $\vp$ and $\vp_D$.

Consider the case $\vT_1\phi_0\ne 0$.
The first equation ${\cal D}_1\phi_0=0$ gives a constraint on the 
function form of $\vT_1\phi_0$:
{\allowdisplaybreaks
\begin{alignat}{2}
\E_0:\quad 
\vT_1\phi_0(z_1,z_2)&=\frac{1}{1\!+\!z_1}\,\omega(\tilde{z}_2),
&\quad \tilde{z}_2:&=\frac{z_2}{(1\!+\!z_1)^3},
\label{om-0}
\\
\E_{\one}:\quad 
\vT_1\phi_0(z_1,z_2)&=\frac{1}{1\!+\!z_1}\,\omega(\tilde{z}_2),
&\quad \tilde{z}_2:&=\frac{z_2}{(1\!+\!z_1)^2},
\label{om-1}
\\
\E_{5,6,7,8}:\quad
\vT_1\phi_0(z_1,z_2)&=
\omega(\tilde{z}_2),
&\quad \tilde{z}_2:&=z_2(1-z_1).
\label{om-N}
\end{alignat}}
Upon the action of $\vT_1$,
the second equation ${\cal D}_2 \phi_0=0$ 
reduces to the \pf equation of the corresponding torus
in terms of the variable $\tilde{z}_2$:
{\allowdisplaybreaks 
\begin{alignat}{2}
\E_0&: \qquad & 
\vT_1{\cal D}_2\phi_0(z_1,z_2)&=
3\,{\cal D}^{\soeji{6}}_{\text{ell}}\omega(\tilde{z}_2),\\
\E_{\one}&: \qquad &
\vT_1{\cal D}_2\phi_0(z_1,z_2)&=
4\,{\cal D}^{\soeji{5}}_{\text{ell}}\omega(\tilde{z}_2),\\
\E_{5,6,7,8}&: \qquad & 
\vT_1{\cal D}_2\phi_0(z_1,z_2)&=\frac{1}{1\!-\!z_1}\,
{\cal D}^{\soeji{N}}_{\text{ell}}\omega(\tilde{z}_2),
\end{alignat}}
that is, $\omega(\tilde{z}_2)$ is a period of the elliptic curve.

We have thus arrived at the representation of the general solution 
$\phi_0$ for the \pf system in terms of a period  
$\omega$ of the fiber torus, which is closely related to the
Seiberg--Witten periods:
{\allowdisplaybreaks
\begin{align}
\E_0:\qquad
\phi_0(z_1,z_2)&=\int^{z_1}\frac{\td z_1}{z_1}\,
\frac{1}{1\!+\!z_1}\,
\omega\left(\frac{z_2}{(1\!+\!z_1)^3}\right)+c(z_2),
\\
\E_{\one}:\qquad
\phi_0(z_1,z_2)&=\int^{z_1}\frac{\td z_1}{z_1}\,
\frac{1}{1\!+\!z_1}\,
\omega\left(\frac{z_2}{(1\!+\!z_1)^2}\right)+c(z_2),
\\
{\E}_{5,6,7,8}:\qquad
\phi_0(z_1,z_2)&=\int^{z_1}\frac{\td z_1}{z_1}\,
\omega(z_2-z_1z_2)+c(z_2),
\end{align}}
where  $c(z_2)$ is a counterterm to ensure
${\cal D}_2\phi_0=0$.

Curiously, the deformation of the single-variable function
$\omega(z_2)$ to the one appeared in  the right hand side of
(\ref{om-0}), (\ref{om-1}) and (\ref{om-N}) admits the following
description in terms of the differential operator 
${\cal L}_n^{\soeji{N}}$
respectively:
\begin{align}
\frac{1}{1\!+\!z_1}\,\omega\left(\frac{z_2}{(1\!+\!z_1)^3}\right)&=
\left(1+\sum_{n=1}^{\infty}nz_1^{n}{\cal L}_{n}^{\soeji{0}}\right)
\omega(z_2), 
\label{henkei-0}
\\
\frac{1}{1\!+\!z_1}\,\omega\left(\frac{z_2}{(1\!+\!z_1)^2}\right)&=
\left(1+\sum_{n=1}^{\infty}nz_1^{n}{\cal L}_{n}^{\soeji{\one}}\right)
\omega(z_2),
\label{henkei-1}
\\ 
\omega(z_2- z_1z_2)&=
\left(1+\sum_{n=1}^{\infty}nz_1^{n}{\cal L}_{n}^{\soeji{N}}\right)
\omega(z_2),
\quad N=5,6,7,8.
\label{henkei-N}
\end{align}

For the period $\phi$, we obtain the following 
integral formula:
for $\E_0$ and $\E_{\one}$ models,
\begin{align}
\E_0:\
\phi^{\soeji{0}}(z_1,z_2)&=
\int_{\epsilon}^{z_1}\frac{\td z_1}{z_1}\frac{1}{1\!+\!z_1}
\vp^{\soeji{6}}\left(\frac{z_2}{(1\!+\!z_1)^3}\right)
+\xi^{\soeji{0}}(z_2),
\label{phi0}
\\
\E_{\one}:\
\phi^{\soeji{\one}}(z_1,z_2)&=
\int_{\epsilon}^{z_1}\frac{\td z_1}{z_1}\frac{1}{1\!+\!z_1}
\vp^{\soeji{5}}\left(\frac{z_2}{(1\!+\!z_1)^2}\right)
+\xi^{\soeji{\one}}(z_2),
\label{phi1}
\end{align}
and for the principal series
\begin{align}
\E_{5,6,7,8}:\
\phi^{\soeji{N}}(z_1,z_2)&=
\int_{\epsilon}^{z_1}\frac{\td z_1}{z_1}
\vp^{\soeji{N}}(z_2-z_1z_2)+\xi^{\soeji{N}}(z_2)
\label{phiN}
\\
&=\int_1^{z_1}\frac{\td z_1}{z_1}
\vp^{\soeji{N}}(z_2-z_1z_2),
\nn
\end{align}
where we must discard a $\log(\epsilon)$ term
before taking the limit $\epsilon \to 0$.
The last equation  implies
that for the principal series, 
$\phi$ is a vanishing period at $z_1=1$. 

Let $\tilde{\tau}$ be the coupling constant of the Seiberg--Witten 
theory with the bare parameters $(z_1,z_2)$. It is given by
the deformed mirror map
\begin{equation}
2\pi \ti \tilde{\tau}=\frac{\vp_D(\tilde{z}_2)}{\vp(\tilde{z}_2)},
\end{equation}
where $\tilde{z}_2=\tilde{z}_2(z_1,z_2)$ is given 
in (\ref{om-0}), (\ref{om-1}), and (\ref{om-N}) respectively.

We can also show using 
(\ref{henkei-0}), 
(\ref{henkei-1}), 
(\ref{henkei-N}), 
that the instanton part of the period $\phi_D$ can be 
written as a Seiberg--Witten period,
where we have no need to introduce the cut-off parameter $\epsilon$
in contrast to the case of $\phi$:
\begin{alignat}{2}
\E_0&:\ & 
-\frac{1}{2\pi \ti}
\sum_{n=1}^{\infty}
\left(\frac{f^{\soeji{0}}_n}{\vp^{\soeji{6}}}\right)(z_2)\,
z_1^n&=
\int_0^{z_1}\frac{\td z_1}{z_1}\frac{1}{1\!+\!z_1}\,
(\tilde{\tau}-\tau)\,
\vp^{\soeji{6}}\left(\frac{z_2}{(1\!+\!z_1)^3}\right),
\label{tokuni0}
\\
\E_{\one}&:\ & 
-\frac{1}{2\pi \ti}
\sum_{n=1}^{\infty}
\left(\frac{f^{\soeji{\one}}_n}{\vp^{\soeji{5}}}\right)(z_2)\,
z_1^n&=
\int_0^{z_1}\frac{\td z_1}{z_1}\frac{1}{1\!+\!z_1}\,
(\tilde{\tau}-\tau)\,
\vp^{\soeji{5}}\left(\frac{z_2}{(1\!+\!z_1)^2} \right),
\label{tokuni1}
\\
\E_{5,6,7,8}&:\ & 
-\frac{1}{2\pi \ti}
\sum_{n=1}^{\infty}
\left(\frac{f^{\soeji{N}}_n}{\vp^{\soeji{N}}}\right)(z_2)\,
z_1^n&=
\int_0^{z_1}\frac{\td z_1}{z_1}
(\tilde{\tau}-\tau)\,
\vp^{\soeji{N}}(z_2-z_1z_2).
\label{tokuni}
\end{alignat}
Note that in particular (\ref{tokuni}) for the $\E_8$ model
corresponds to the formula \cite[(3.5)]{MNW} obtained 
by direct evaluation of the Seiberg--Witten periods
from the curve (\ref{SW-e-hachi}).
The integration variable $v$ there satisfies 
$E_4(\tilde{\tau})^3/E_6(\tilde{\tau})^2
=E_4(\tau)^3/(E_6(\tau)\!-\!v)^2$,
from which we see the correspondence of the bare variables
\begin{equation*}
v:=\left(\frac{27}{\pi^6}\right)\frac{1}{u}
=-2E_4(\tau)^{3/2}z_2z_1
=[E_6(\tau)\!-\!E_4(\tau)^{3/2}]z_1.
\end{equation*}

\section{Genus Zero Partition Functions}
\label{genus-zero-partition}
\subsection{Recursion relations}
In order to obtain the instanton expansion of the 
genus zero potential $F_0$, 
we have to convert the two sequences of 
functions of $z_2$:
\begin{equation*}
c_n=\frac{{\cal L}_n\vp}{\vp},
\quad
f_n=-\left(\vp{\cal L}_n \vp_D- \vp_D{\cal L}_n \vp\right)
\end{equation*}
into modular functions of $\varGamma_{\text{ell}}$,
which is achieved by finding
the recursion relations for $\{c_n\}$ and $\{f_n\}$.
To this end, let us make the ansatz 
$c_n(\tau)=B_n e_2(\tau)+D_n$, 
with $e_{2k}:=E_{2k}\vp^{-2k}$.
It will become clear from the recursion relations
that $B_n, D_n, f_n$ are degree $n$ polynomials in $y$,
where $y=(1\!-\!z_2)^{-1}$;
in particular they are all {\em anomaly-free} modular functions of 
$\varGamma_{\text{ell}}$.

The equations (\ref{rec-ni-hitsuyou}) enable us to evaluate 
the following logarithmic derivatives that are indispensable to
establish the recursion relations:
\begin{align}
\vT_2\vp&=\frac{\vp}{12}
\left\{ye_2-[(10\!-\!N)-(9\!-\!N)y]\right\},
\label{relation-1}
\\
\vT_2(e_2\vp)&=\frac{\vp}{12}
\left\{-ye_4+e_2[(10\!-\!N)-(9\!-\!N)y]\right\},
\label{relation-2}
\end{align} 
where $\vp$ is the fundamental period of the $\E_N$ elliptic curves with
$N=5,6,7,8$.
\paragraph{Principal series}
For them, from the relation
$(n\!+\!1)^2{\cal L}_{n+1}=-n(\vT_2-n){\cal L}_n$,
\[
\vp c_{n+1}=-\frac{n}{(n\!+\!1)^2}\,(\vT_2-n)(\vp c_n).
\]
Then the equations (\ref{relation-1}) and (\ref{relation-2})
yields the recursion relation for $\{c_n\}$:
\begin{equation*}
\begin{pmatrix}
B_{n+1}\\
D_{n+1}
\end{pmatrix}
=-\frac{n}{(n\!+\!1)^2}
\left[y(y\!-\!1)\frac{\td}{\td y}+
\begin{pmatrix}
D_1-n & -B_1\\
e_4 B_1 & -D_1-n
\end{pmatrix}
\right]
\begin{pmatrix}
B_{n}\\
D_{n}
\end{pmatrix},
\end{equation*}
where $e_4=(16-16y^{-1}+y^{-2})$, $(9-8y^{-1})$,
$(4-3y^{-1})$, $1$, for $N=5,6,7,8$ respectively.
Note that  we have replaced $\vT_2$ by $y(y\!-\!1)\td/\td y$ 
in the recursion relation above 
because $B_n$ and $D_n$ depend on $z_2$ only through $y$.

Since $c_1=-\vT_2\vp/\vp$,
the first term $(B_1,D_1)$ can be seen immediately
from (\ref{relation-1}):
\[
B_1=-\frac{y}{12},\quad
D_1=\frac{1}{12}[(10\!-\!N)-(9\!-\!N)y].
\]

The recursion relation for $\{f_n\}$ can be obtained 
in a similar manner:
\begin{equation*}
f_{n+1}=-\frac{n}{(n\!+\!1)^2}
\Big\{\big[y(y\!-\!1)\frac{\td}{\td y}-n+c_1\big]f_n-f_1c_n \Big\},
\quad f_1=y.
\end{equation*}
Furthermore $\{f_n\}$ and $\{c_n=B_ne_2+D_n\}$ 
are related each other by 
\begin{equation*}
B_n=-\frac{1}{12}f_n,
\quad
D_n=\frac{1}{y}
\Big\{\frac{(n\!+\!1)^2}{n}f_{n+1}
+\big[y(y\!-\!1)\frac{\td}{\td y}-n+D_1\big]f_n\Big\}.
\end{equation*}

Here we list the first few elements of $\{(B_n,D_n)\}$ only 
for the $\E_8$ model:
{\allowdisplaybreaks
\begin{align*}
B_1&=-\frac{1}{12}\,y, \
D_1=\frac{1}{12}\,(2-y),\
B_2=\frac{1}{48}\,y(y-2),\
D_2=\frac{1}{144}\,(7-7\,y+3\,y^2),\\
B_3&=-\frac{1}{7776}\,y(211-211\,y+72\,y^2),\
D_3=-\frac{1}{7776}\,(y-2)(72\,y^2-91\,y+91),\\
B_4&=\frac{1}{10368}\,y(y-2)(54\,y^2-103\,y+103),\\
D_4&=\frac{1}{124416}\,(1729-3458\,y+4477\,y^2-2748\,y^3+648\,y^4).
\end{align*}}
\paragraph{$\E_{\one}$ model}
The procedure to get the recursion relations for 
$\{c_n\}$ and $\{f_n\}$ are similar to the case of principal series.
In this case the $\E_5$ torus is relevant.
 
Using
$(n\!+1\!)^2{\cal L}_{n+1}=-n(2\vT_2\!+\!n\!+\!1){\cal L}_n$
we get the recursion relation for $\{c_n\}$
\begin{equation*}
\begin{pmatrix}
B_{n+1}\\
D_{n+1}
\end{pmatrix}
=-\frac{n}{(n\!+\!1)^2}
\left[2y(y\!-\!1)\frac{\td}{\td y}+
\begin{pmatrix}
D_1+n+2 & -B_1\\
e_4 B_1 & -D_1+n
\end{pmatrix}
\right]
\begin{pmatrix}
B_{n}\\
D_{n}
\end{pmatrix},
\end{equation*} 
with $(B_1,D_1)=(-y/6,-(4y\!+\!1)/6)$ and
$e_4=(16-16y^{-1}+y^{-2})$.

The recursion relation for $\{f_n\}$ reads
\begin{equation*}
f_{n+1}=-\frac{n}{(n\!+\!1)^2}
\Big\{\big[2y(y\!-\!1)\frac{\td}{\td y}+2+n+c_1\big]f_n
-f_1c_n\Big\},\quad f_1=2y.
\end{equation*}
The relation between $\{f_n\}$ and $\{c_n\}$ is
\begin{equation*}
B_n=-\frac{1}{12}f_n,
\quad
D_n=\frac{1}{2y}\,
\Big\{\frac{(n\!+\!1)^2}{n}f_{n+1}
+\big[2y(y-1)\frac{\td}{\td y}+2+n+D_1\big]f_n \Big\}.
\end{equation*}
We list the first few members:
{\allowdisplaybreaks
\begin{align*}
B_1&=-\frac{1}{6}\,y,\
D_1=-\frac{1}{6}\,(1+4\,y),\
B_2=\frac{1}{24}\,y(2\,y+1),\
D_2=\frac{1}{24}\,(8\,y^2+1),
\\
B_3&=-\frac{1}{108}\,y(8\,y^2+y+2),\
D_3=-\frac{1}{108}\,(1+4\,y)(8\,y^2-5\,y+2),
\\
B_4&=\frac{1}{288}\,y(24\,y^3-4\,y^2+2\,y+3),\
D_4=\frac{1}{288}\,(96\,y^4-64\,y^3+7\,y^2+5\,y+3).
\end{align*}}
\paragraph{$\bo{\E_0}$ model}
This case has been analyzed in \cite{HST}, 
which we briefly repeat here for convenience. 
Recall that underlying torus model is $\E_6$.

First, the relation among the operators
$(n+1)^2{\cal L}_{n+1}=-n(3\vT_2+n+1){\cal L}_n$
yields 
\begin{equation*}
\begin{pmatrix}
B_{n+1}\\
D_{n+1}
\end{pmatrix}
=-\frac{n}{(n\!+\!1)^2}
\left[3y(y\!-\!1)\frac{\td}{\td y}+
\begin{pmatrix}
D_1+n+2 & -B_1\\
e_4 B_1 & -D_1+n
\end{pmatrix}
\right]
\begin{pmatrix}
B_{n}\\
D_{n}
\end{pmatrix},
\end{equation*}
with $(B_1,D_1)=(-y/4,-3y/4)$ and 
$e_4=(9-8y^{-1})$.

The recursion relation for $\{f_n\}$ becomes
\begin{equation*}
f_{n+1}=-\frac{n}{(n\!+\!1)^2}
\Big\{\big[3y(y\!-\!1)\frac{\td}{\td y}+n+2+c_1\big]f_n-c_nf_1\Big\},
\quad f_1=3y,
\end{equation*}
with the relation between  $\{f_n\}$ and $\{c_n\}$
\begin{equation*}
B_n=-\frac{1}{12}\, f_n,\
D_n=\frac{1}{3y}\Big\{\frac{(n\!+\!1)^2}{n}f_{n+1}+
\big[3y(y\!-\!1)\frac{\td}{\td y}+n+2+D_1\big]f_n \Big\}.
\end{equation*} 
The first few members of $\{(B_n,D_n)\}$ are
\begin{align*}
B_1&=-\frac{1}{4}\, y,\
D_1=-\frac{3}{4}\, y,\
B_2=\frac{3}{16}\,y^2,\
D_2=\frac{1}{16}\,y(9\,y-4),
\\
B_3&=-\frac{1}{72}\,y^2(18\,y-7),\
D_3=-\frac{1}{72}\,y(54\,y^2-45\,y+4),
\\
B_4&=\frac{1}{192}\,y^2(27\,y-2)(3\,y-2),\
D_4=\frac{1}{192}\,y^2(243\,y^2-288\,y+68).
\end{align*}

Finally, we have observed for any of the six models  that 
\begin{equation}
\frac{f_n(y)}{\vp(z_2)^2}=-12\,\frac{\pa c_n(\tau)}{\pa E_2(\tau)},
\label{E2-bibun}
\end{equation}
which turns out to be of fundamental importance both for
the instanton expansion and for the investigation 
of the modular anomaly equations of the genus zero 
and one \gw potentials below.

\subsection{Instanton expansion}
The instanton expansion of the genus zero
potential $F_0$ is obtained by
conversion of $z_1$ and $z_2$ in  $F_0$ 
to the function of $q:=\text{e}^{2\pi \ti \tau}$ and  
$p:=\text{e}^{2\pi \ti \sigma}$.
We define the genus zero \gw invariant
$N_{0;n,m}\in \text{\bf Q}$ of bidegree $(n,m)$ 
by the $q$-expansion of $Z_{0; n}(\tau)$
\begin{equation}
Z_{0; n}(\tau)=\sum_{m=0}^{\infty}N_{0;n,m}q^m.
\end{equation}

We find a useful expression of the genus zero potential:  
\begin{equation}
\frac{1}{2\pi \ti}\,
\left(\frac{\pa F_0}{\pa \sigma}\right)
=\vT_p F_0
=12h\left(\frac{\pa z_1}{\pa E_2}\right)
\left(\frac{1}{\tilde{p}}\frac{\pa \tilde{p}}{\pa z_1}\right),
\label{F0-kakikae}
\end{equation}
which follows from the fact that
$\tilde{p}$ does not depend on $E_2$:
\[
\left(\frac{d\tilde{p}}{d E_2}\right)
=\left(\frac{\pa z_1}{\pa E_2}\right)
\left(\frac{\pa \tilde{p}}{\pa z_1}\right)+
\sum_{n=1}^{\infty}
\left(\frac{\pa c_n}{\pa E_2}\right)
\left(\frac{\pa \tilde{p}}{\pa c_n}\right)=0,
\]
because $\tilde{p}=p\psi(\tau)$ with $\psi(\tau)$ 
anomaly-free, and the substitution of the identity (\ref{E2-bibun}).

Then we see from the expansion of the right hand side of
(\ref{F0-kakikae}) that the general form of the partition function 
$Z_{0;n}(\tau)$ reads 
\begin{align}
Z_{0;1}&=(9-N)\, \left(\frac{y\psi}{\vp^2}\right),
\quad N=0,\one,5,6,7,8,
\\
Z_{0; n}
&=(Z_{0;1})^n
\vp^{2(n-1)}\,
{\cal P}_{0; n-1}(e_2,y^{-1}),
\label{Z0n}
\end{align} 
where ${\cal P}_{0; n-1}$ is a degree $n\!-\!1$ polynomial 
in $e_2=E_2(\tau)/\vp(z_2)^2$ and $y^{-1}=1\!-\!z_2$.
The relation between $Z_{0;1}$ and the special value of 
the $\E_8$ theta function prescribed by the \K class 
$J^{\soeji{N}}$ defined in (\ref{K-0})--(\ref{K-8})
will be discussed later.

As space is limited, we give only 
the first four terms of the $p$-expansions of (\ref{F0-kakikae}):
{\allowdisplaybreaks
\begin{align}
\vT_pF_0^{\soeji{0}} &=
9\,\frac{y\tilde{p}}{\vp^2}
+\frac{(y\tilde{p})^2}{\vp^2}\frac{9}{4}\,e_2
+\frac{(y\tilde{p})^3}{\vp^2}\frac{1}{32}\,
\big(27e_2^2+45-\frac{40}{y}\big)
\nn\\
&+\frac{(y\tilde{p})^4}{\vp^2}
\frac{1}{32}\,
\big[12e_2^3
+e_2\big(45-\frac{40}{y}\big)
-27+\frac{36}{y}-\frac{8}{y^{2}}\big],
\nn\\
\vT_pF_0^{\soeji{\one}}&=8\,\frac{y\tilde{p}}{\vp^2}
+\frac{(y\tilde{p})^2}{\vp^2}
\frac{2}{3}\big(2e_2+2-\frac{1}{y}\big)
+\frac{(y\tilde{p})^3}{\vp^2}
\frac{1}{9}\big[3e_2^2
+e_2\big(6-\frac{3}{y}\big)
+8-\frac{8}{y}+\frac{2}{y^{2}}\big]
\nn\\
&+\frac{(y\tilde{p})^4}{\vp^2}
\frac{1}{162}\big[
16e_2^3
+e_2^2\big(48-\frac{24}{y}\big)
+e_2\big(108-\frac{108}{y}+\frac{27}{y^{2}}\big)
+40-\frac{60}{y}+\frac{48}{y^{2}}-\frac{14}{y^{3}}
\big],
\nn\\
\vT_pF_0^{\soeji{5}}&=4\,\frac{y\tilde{p}}{\vp^2}
+\frac{(y\tilde{p})^2}{\vp^2}
\frac{1}{3}\big(e_2+1+\frac{1}{y}\big)
+\frac{(y\tilde{p})^3}{\vp^2}
\frac{1}{72}\big[3e_2^2
+e_2\big(6+\frac{6}{y}\big)
+8+\frac{4}{y}+\frac{5}{y^2}\big]
\nn\\
&+\frac{(y\tilde{p})^4}{\vp^2}
\frac{1}{648}\big[
4e_2^3
+e_2^2\big(12+\frac{12}{y}\big)
+e_2\big(27+\frac{18}{y}+\frac{18}{y^2}\big)
+10+\frac{39}{y}+\frac{12}{y^2}+\frac{10}{y^3}
   \big],
\nn\\
\vT_pF_0^{\soeji{6}}&=3\,\frac{y\tilde{p}}{\vp^2}
+\frac{(y\tilde{p})^2}{\vp^2}
\frac{1}{4}\big(e_2+\frac{2}{y}\big)
+\frac{(y\tilde{p})^3}{\vp^2}
\frac{1}{864}\big(27e_2^2+108\frac{e_2}{y}
+45-\frac{4}{y}+\frac{112}{y^2}\big)
\nn\\
&+\frac{(y\tilde{p})^4}{\vp^2}
\frac{1}{2592}
\big[12e_2^3+72\frac{e_2^2}{y}
+e_2\big(45-\frac{4}{y}+\frac{148}{y^2}\big)
-27+\frac{144}{y}-\frac{8}{y^2}+\frac{104}{y^3}
\big],
\nn\\
\vT_pF_0^{\soeji{7}}&=
2\,\frac{y\tilde{p}}{\vp^2}
+\frac{(y\tilde{p})^2}{\vp^2}
\frac{1}{6}\big(e_2-1+\frac{3}{y}\big)
+\frac{(y\tilde{p})^3}{\vp^2}
\frac{1}{288}\big[
6e_2^2
-e_2\big(12-\frac{36}{y}\big)
+16-\frac{33}{y}+\frac{51}{y^2}\big]
\nn\\
&+\frac{(y\tilde{p})^4}{\vp^2}
\frac{1}{2592}
\big[8e_2^3
-e_2^2\big(24\!-\!\frac{72}{y}\big)
+e_2\big(54-\frac{135}{y}+\frac{207}{y^2}\big)
-56+\frac{189}{y}-\frac{180}{y^2}+\frac{189}{y^3}
\big],
\nn\\
\vT_pF_0^{\soeji{8}}&=
\frac{y\tilde{p}}{\vp^2}
+\frac{(y\tilde{p})^2}{\vp^2}
\frac{1}{12}\big(e_2-2+\frac{4}{y}\big)
+\frac{(y\tilde{p})^3}{\vp^2}
\frac{1}{2592}
\big[27e_2^2-e_2\big(108-\frac{216}{y}\big)
+153-\frac{394}{y}
\nn\\
&+\frac{394}{y^2}\big]
+\frac{(y\tilde{p})^4}{\vp^2}
\frac{1}{7776}
\big[12e_2^3
-e_2^2\big(72-\frac{144}{y}\big)
+e_2\big(189-\frac{538}{y}+\frac{538}{y^2}\big)
-213+\frac{734}{y}
\nn\\
&-\frac{924}{y^2}+\frac{616}{y^3}
\big].
\nn
\end{align}}

At this point it would be helpful to mention in advance
the general form of higher genus \gw partition functions 
$Z_{g; n}(\tau)$, which is indeed predicted from 
the modular anomaly equation (\ref{modular-anomaly}),
\begin{equation}
Z_{g; n}=
(Z_{0; 1})^n
\vp^{2(g+n-1)}\,
{\cal P}_{g; g+n-1}(e_2,y^{-1}),
\label{Zgn}
\end{equation}
where ${\cal P}_{g; g+n-1}$ is a degree 
$g\!+\!n\!-\!1$ polynomial 
in $e_2$ and $y^{-1}$.


\subsection{Partition functions and modular forms}
It is of primary importance
to ensure the relation between 
the singly wound partition function 
$Z_{0; 1}(\tau)$ and the $\E_8$ theta function
(\ref{kihon-shiki})
as a consistency check of our formalism.

The specializations of the $\E_8$ theta function to
the \K moduli of the $\E_{\one}$ and $\E_5$ models,
for example, are calculated as follows:
\begin{align*}
\tilde{\E}_1:\ 
\TE(4\tau|\omega_2\tau)
&=
\frac{1}{2}q^{-\frac32}
\big(
\sum_{a=2,3}\vat_{a}(4\tau|\tfrac{\tau}{2})^8
-\displaystyle{\sum_{b=1,4}}\vat_{b}(4\tau|\tfrac{\tau}{2})^8  
\big)
\\
&=\frac12 q^{-1}
\big(
\sum_{a=2,3}\vat_{a}(4\tau)^2\vat_{a}(4\tau|\tau)^6
-\vat_{4}(4\tau)^2
\vat_{4}(4\tau|\tau)^6  
\big),
\\
\E_5:\ 
\TE(4\tau|\omega_5\tau)
&=
\frac{1}{2}q^{-1}
\big(
\sum_{a=2,3} \vat_{a}(4\tau)^4\vat_{a}(4\tau|\tau)^4
-\vat_4(4\tau)^4\vat_4(4\tau|\tau)^4
\big).
\end{align*}

We find that $Z_{0;1}$  
can indeed be written by the $E_8$ theta function 
with the prescribed \K class $J^{\soeji{N}}$
as predicted in (\ref{kihon-shiki}),
in addition to the fact that
it admits a concise description by the Dedekind eta function: 
{\allowdisplaybreaks
\begin{alignat}{2}
Z^{\soeji{0}}_{0;1}(\tau)=
9\left(\frac{y\psi}{\vp^2}\right)^{\soeji{0}}(z_2)
&=9q^{\frac16}\,\frac{1}{\eta(\tau)^4} & \
&=Z_{0;1}(3\tau|\omega_8\tau),
\\
Z^{\soeji{\one}}_{0;1}(\tau)=
8\left(\frac{y\psi}{\vp^2}\right)^{\soeji{\one}}(z_2)
&=8q^{\frac14}\,\frac{1}{\eta(\tau)^2\eta(2\tau)^2} & \
&=Z_{0;1}(4\tau|\omega_2\tau),
\\
Z^{\soeji{5}}_{0;1}(\tau)=
4\left(\frac{y\psi}{\vp^2}\right)^{\soeji{5}}(z_2)
&=4q^{\frac12}\,\frac{\vp^{\soeji{5}}(z_2)}{\eta(2\tau)^6} & \
&=Z_{0;1}(4\tau|\omega_5\tau),
\\
Z^{\soeji{6}}_{0;1}(\tau)=
3\left(\frac{y\psi}{\vp^2}\right)^{\soeji{6}}(z_2)
&=3q^{\frac12}\,\frac{\vp^{\soeji{6}}(z_2)}{\eta(\tau)^3\eta(3\tau)^3} & \
&=Z_{0;1}(3\tau|\omega_6\tau),
\\
Z^{\soeji{7}}_{0;1}(\tau)=
2\left(\frac{y\psi}{\vp^2}\right)^{\soeji{7}}(z_2)
&=2q^{\frac12}\,\frac{\vp^{\soeji{7}}(z_2)^2}{\eta(\tau)^4\eta(2\tau)^4} & \
&=Z_{0;1}(2\tau|\omega_7\tau),
\\
Z^{\soeji{8}}_{0;1}(\tau)=
\iti \left(\frac{y\psi}{\vp_{0}^{2}}\right)^{\soeji{8}}(z_2)
&=\iti q^{\frac12}\,\frac{\vp^{\soeji{8}}(z_2)^4}{\eta(\tau)^{12}} & \
&=Z_{0;1}(\tau|0).
\end{alignat}}
$Z_{0;1}(\tau)$ of the $\E_{3}$ and the $\E_4$ model
calculated form  the $\E_8$ theta function using
(\ref{kihon-shiki}) can also be expressed by the eta functions:
\begin{alignat}{2}
\E_3:\ 
\frac{\TE(6\tau|\omega_3\tau)}{\varphi(6\tau)^{12}} &= \ & \
&\frac{6q^{\frac12}}{\eta(\tau)\eta(2\tau)\eta(3\tau)\eta(6\tau)},
\\
\E_4:\
\frac{\TE(5\tau|\omega_4\tau)}{\varphi(5\tau)^{12}} &=\  &\ 
&\frac{5q^{\frac12}}{\eta(\tau)^2\eta(5\tau)^2}.
\end{alignat}

As for partition functions of multiple winding number,
we here give the expressions of those in terms of 
modular forms only for the $\E_7$ and $\E_8$ models;
the latter has originally been obtained in \cite{MNW}.
\paragraph{$\E_7$ model}
Let $\chi:=Z_{0;1}^{(7)}/A
=2q^{1/2}/(\eta(\tau)\eta(2\tau))^4$, see (\ref{kai-no-teigi}). 
{\allowdisplaybreaks
\begin{align}
Z^{(7)}_{0;1} &= \Zseven A,
\nn\\
Z^{(7)}_{0;2} &= \frac{\Zseven^2}{48}A(AE_2-A^2+3B),
\nn\\ 
Z^{(7)}_{0;3} &= \frac{\Zseven^3}{6912}
A(6E_2^2A^2-12E_2A^3+36E_2AB+16A^4-33A^2B+51B^2),
\nn\\ 
Z^{(7)}_{0;4} &= \frac{\Zseven^4}{165888}
A(-56A^6+189A^4B-180B^2A^2+189B^3+8E_2^3A^3-24E_2^2A^4
+72E_2^2A^2B
\nn\\
&
+54E_2A^5-135E_2A^3B+207E_2AB^2), 
\nn\\
Z^{(7)}_{0;5} &= \frac{\Zseven^5}{1990656000}
A(123328A^8-514272A^6B+858987A^4B^2
+6250A^4E_2^4
-25000A^5E_2^3
\nn\\
&
+406215B^4
-585000A^3E_2B^2
+499500A^5E_2B
+607500AE_2B^3
-213750A^4E_2^2B
\nn\\
&
+326250A^2E_2^2B^2
-508680A^2B^3
-136000A^7E_2
+75000A^3E_2^3B+75000A^6E_2^2).
\nn
\end{align}}
\paragraph{$\E_8$ model}
{\allowdisplaybreaks
\begin{align}
Z_{0;1}^{(8)}&=
\frac{1}{\varphi^{12}}
E_4,\nn \\
Z_{0;2}^{(8)}&=
\frac{1}{24\varphi^{24}}
E_4(2E_6+E_2E_4),\nn \\
Z_{0;3}^{(8)}&=
\frac{1}{15552\varphi^{36}}
E_4(
109E_4^3+
197E_6^2+
216E_2E_4E_6+
54E_2^2E_4^2),\nn \\
Z_{0;4}^{(8)}&=
\frac{1}{62208\varphi^{48}}
E_4(
272E_4^3E_6+
154E_6^3+
109E_2E_4^4+
269E_2E_4E_6^2+
144E_2^2E_4^2E_6\nn \\
&+
24E_2^3E_4^3),\nn \\
Z_{0;5}^{(8)}&=
\frac{1}{37324800\varphi^{60}}
E_4(
426250E_2^2E_4^2E_6^2+
150000E_2^3E_4^3E_6+
207505E_6^4+
136250E_2^2E_4^5\nn \\
&+
772460E_4^3E_6^2+
116769E_4^6+
18750E_2^4E_4^4+
653000E_2E_4^4E_6+
505000E_2E_4E_6^3).\nn 
\end{align}}

\subsection{Modular anomaly equation}
We are now ready to give the modular anomaly equation, which 
determines the $E_2$ dependence of 
the genus zero partition function $Z_{0;n}(\tau)$
closely following \cite{MNW,HST}.

First by differentiation of (\ref{genus-zero-inst}) with respect to
$E_2(\tau)$, we obtain
\begin{equation*}
\frac{\pa }{\pa E_2}
\left(\vT_p F_0\right)=
h\sum_{n=1}^{\infty}\frac{f_n}{\vp^2}nz_1^{n-1}
\left(\frac{\pa z_1}{\pa E_2}\right).
\end{equation*}
Then the substitution of (\ref{F0-kakikae}) to the above equation 
brings about 
\begin{equation*}
\vT_p\left[24h\left(\frac{\pa F_0}{\pa E_2}\right)
-\left(\vT_pF_0\right)^2 \right]=0.
\end{equation*}
Eventually we arrive at 
the modular anomaly equation for genus zero:
\begin{equation*}
h\left(\frac{\pa F_0}{\pa E_2}\right)
=\frac{1}{24}\left(\vT_pF_0\right)^2.
\end{equation*}
Using the definition of the potential  
$F_0=\sum_{n=1}^{\infty}Z_{0;n}(\tau)p^n$,
the anomaly equation can be rewritten as
\begin{equation*}
h\frac{\pa Z_{0;n}(\tau)}{\pa E_2(\tau)}
=\frac{1}{24}\sum_{k=1}^{n-1}
k(n-k)Z_{0;k}(\tau)Z_{0;n-k}(\tau),
\end{equation*}
which is consistent with the general form 
of the anomaly equation (\ref{modular-anomaly}).
The appearance of the normalization factor 
$h$ (\ref{genus-zero-inst})
in the left hand side is explained by the fact that 
$hE_2(h\tau)-E_2(\tau)$ 
is an anomaly-free modular form of $\varGamma_0(h)$.

\subsection{Rational instanton numbers}
We denote the genus zero instanton number of bidegree
$(n,m)$  by
$N^{\inst}_{0;n,m}$,
which counts the `number' of the rational curves of a given degree
in the almost \dP surface $B_9$,
and we define its generating function with fixed $n$ by
%
\begin{equation}
{Z}^{\inst}_{0;n}(\tau)
=\sum_{m=0}^{\infty}N^{\inst}_{0; n,m}q^{m}.
\end{equation}
%
It is well-known that
the genus zero multiple covering formula found in 
\cite{CDGP} leads to the following decomposition
of the genus zero \gw partition function:
\begin{equation}
Z_{0;n}(\tau)=\sum_{k|n}
k^{-3}Z^{\inst}_{0;\frac{n}{k}}(k\tau).
\end{equation}
We can invert this equation using the M\"obius function 
$\mu:\text{\bf N}\to \{0,\pm 1\}$ as
\begin{equation}
Z^{\inst}_{0;n}(\tau)=
\sum_{k|n}\mu(k)k^{-3}Z_{0;\frac{n}{k}}(k\tau).
\label{Mobius-0}
\end{equation}  
Recall that the M\"obius function $\mu$ is defined as follows: 
$\mu(1)=1$, $\mu(n)=(-1)^l$ if $n$ is factorized into $l$ distinct
primes, and $\mu(n)=0$ if $n$ is not square-free.

We will give the first few terms of the expansions of 
$Z^{\inst}_{0;n}$, for each models below.
\paragraph{$\bo{\E_5}$ model}
{\allowdisplaybreaks \begin{align}
Z^{\inst}_{0;1}
&=4 
+ 16\, q 
+ 40\, q^2  
+ 96\, q^3  
+ 220\, q^{4}  
+ 464\, q^{5}  
+ 920\, q^{6}
+ 1760\, q^{7}
+ 3276\, q^{8}  
+ 5920\, q^{9}  
\nn\\
&  
+ 10408\, q^{10}   
+\cdots
\nn\\
Z^{\inst}_{0;2}
&=
-20\, q^2
-128\, q^3
-608\, q^4
-2304\, q^5
-7672\, q^6
-23040\, q^7
-64256\, q^8
-168448\, q^9
\nn\\
&
-419908\, q^{10}
-\cdots,
\nn\\
Z^{\inst}_{0;3}
&=
 48\, q^3
+588\, q^4
+4224\, q^5
+23112\, q^6
+105888\, q^7
+426624\, q^8
+1557216\, q^9
\nn\\
&
+5250816\, q^{10}
+\cdots,
\nn\\
Z^{\inst}_{0;4}
&=
-192\, q^4
-3328\, q^5
-32224\, q^6
-230400\, q^7
-1346944\, q^8
-6802432\, q^9
\nn\\
&
-30669248\, q^{10}
-\cdots,
\nn\\
Z^{\inst}_{0;5}
&=
 960\, q^5
+21320\, q^6
+260320\, q^7
+2298680\, q^8
+16354800\, q^9
+99283840\, q^{10}
+\cdots.
\nn
\end{align}}
%
%
%
\paragraph{$\bo{\E_6}$ model}
{\allowdisplaybreaks \begin{align}
Z^{\inst}_{0;1}
&=3
+27\, q
+81\, q^2
+255\, q^3
+702\, q^4
+1701\, q^5
+3930\, q^6
+8721\, q^7
+18225\, q^8
\nn
\\
&
+37056\, q^9
+73116\, q^{10}
+\cdots,
\nn 
\\
Z^{\inst}_{0;2}
&=
-54\, q^2
-492\, q^3
-3078\, q^4
-14904\, q^5
-61320\, q^6
-224532\, q^7
-751788\, q^8
\nn
\\
&
-2337264\, q^9
-6844338\, q^{10}
-\cdots,
\nn 
\\
Z^{\inst}_{0;3}
&=
 243\, q^3
+4131\, q^4
+40095\, q^5
+287307\, q^6
+1683018\, q^7
+8515449\, q^8
\nn
\\
&
+38457585\, q^9
+158463702\, q^{10}
+\cdots,
\nn 
\\
Z^{\inst}_{0;4}
&=
-1728\, q^4
-42120\, q^5
-559920\, q^6
-5344920\, q^7
-40835664\, q^8
\nn
\\
&
-264772872\, q^9
-1510286688\, q^{10}
-\cdots,
\nn 
\\
Z^{\inst}_{0;5}
&=
 15255\, q^5
+483585\, q^6
+8191530\, q^7
+97962210\, q^8
+925275420\, q^9
\nn\\
&
+7332946200\, q^{10}
+\cdots.
\nn 
\end{align}}
%
%
%
%
\paragraph{$\E_7$ model}
%
{\allowdisplaybreaks 
\begin{align}
Z^{\inst}_{0;1}
&=2
+56\, q
+276\, q^2
+1360\, q^3
+4718\, q^4
+15960\, q^5
+46284\, q^6
+130064\, q^7
\nn
\\
&
+334950\, q^8
+837872\, q^9
+1980756\, q^{10}
+\cdots,
\nn
\\
Z^{\inst}_{0;2}
&=
-272\, q^2
-4544\, q^3
-46416\, q^4
-335744\, q^5
-2008480\, q^6
-10255104\, q^7
\nn
\\
&
-46868416\, q^8
-194576128\, q^9
-749189328\, q^{10}
+\cdots,
\nn
\\
Z^{\inst}_{0;3}
&=
 3240\, q^3
+100134\, q^4
+1649088\, q^5
+18786852\, q^6
+168160176\, q^7
\nn
\\
&
+1255563072\, q^8
+8154689040\, q^9
+47265867648\, q^{10}
+\cdots,
\nn
\\
Z^{\inst}_{0;4}
&=
 -58432\, q^4
 -2633088\, q^5
 -60949696\, q^6
 -960253440\, q^7
\nn
\\
&
 -11638833216\, q^8
 -115871533568\, q^9
 -988372855168\, q^{10}
-\cdots,
\nn \\
Z^{\inst}_{0;5}
&=
 1303840\, q^5
 +77380260\, q^6
 +2323737360\, q^7
 +47046026140\, q^8
\nn
\\
&
 +724935311560\, q^9
 +9088122264000\, q^{10}
+\cdots. 
\nn
\end{align}}
%
%
%
%
%
\paragraph{$\E_8$ model}
{\allowdisplaybreaks 
\begin{align}
Z^{\inst}_{0;1}
&=1
+252\, q
+5130\, q^2
+54760\, q^3
+419895\, q^4
+2587788\, q^5
+13630694\, q^6
\nn
\\
&
+63618120\, q^7
+269531955\, q^8
+1054198840\, q^9
+3854102058\, q^{10}
+\cdots,
\nn
\\
Z^{\inst}_{0;2}
&=
-9252\, q^2
-673760\, q^3
-20534040\, q^4
-389320128\, q^5
-5398936120\, q^6
\nn 
\\
&
-59651033472\, q^7
-553157438400\, q^8
-4456706505600\, q^9
\nn\\
&
-31967377104276\, q^{10}
-\cdots,
\nn
\\
Z^{\inst}_{0;3}
&=
 848628\, q^3
+115243155\, q^4
+6499779552\, q^5
+219488049810\, q^6
\nn 
\\
&
+5218126709400\, q^7
+95602979109024\, q^8
+1428776049708360\, q^9
\nn\\
&
+18102884896663488\, q^{10}
+\cdots,
\nn
\\
Z^{\inst}_{0;4}
&=
-114265008\, q^4
-23064530112\, q^5
-1972983690880\, q^6
-100502238355200\, q^7
\nn
\\
&
-3554323792345440\, q^8
-95341997143018752\, q^9
\nn\\
&
-2053905830285978880\, q^{10}
-\cdots,
\nn
\\
Z^{\inst}_{0;5}
&=
  18958064400\, q^5
+5105167984850\, q^6
+594537323257800\, q^7
\nn
\\
&
+41416214037843150\, q^8
+1996136210493389700\, q^9
\nn\\
&
+72464241398191308000\, q^{10}
+\cdots.
\nn
\end{align}}
%
Note that for these principal series, 
$N^{\inst}_{0; n,n}$
coincides with the genus zero, degree $n$ instanton number  
of the local $\E_N$ \dP model computed in \cite{KMV,LMW}.

%
\paragraph{$\E_0$ model}
{\allowdisplaybreaks 
\begin{align}
Z^{\inst}_{0;1}
&=9 
+ 36\, q
+ 126\,  q^2
+ 360\,  q^3
+ 945\,  q^4
+ 2268\,  q^5
+ 5166\,  q^6
+ 11160\,  q^7
\nn\\
&
+ 23220\,  q^8
+ 46620\,  q^9  
+ 90972\, q^{10}   
+\cdots,
\nn\\
Z^{\inst}_{0;2}
&=
-18\, q 
- 252\, q^2 
- 1728\, q^3 
- 9000\,  q^4  
- 38808\,  q^5  
- 147384\, q^6
- 506880\,  q^7
\nn\\ 
&
- 1613088\,  q^8  
- 4813380\, q^9  
- 13609476\, q^{10}
-\cdots,
\nn\\
Z^{\inst}_{0;3}
&=3\, q 
+ 252\, q^2
+ 4158\, q^3  
+ 40173\, q^4  
+ 287415\,  q^5  
+ 1683450\,  q^6
+ 8516418\, q^7
\nn\\
&
+ 38458233\, q^8
+ 158467806\, q^9  
+ 605183100\, q^{10}
+\cdots,
\nn\\
Z^{\inst}_{0;4}
&=
-144\, q^2  
- 6048\, q^3  
- 107280\, q^4  
- 1235520\, q^5  
- 10796544\, q^6
- 77538240\, q^7  
\nn\\
&
- 479682720\, q^8  
- 2635776000\, q^9
- 13140695232\, q^{10}   
-\cdots,
\nn\\
Z^{\inst}_{0;5}
&=45\, q^2  
+ 5670\, q^3  
+ 189990\, q^4  
+ 3508920\, q^5  
+ 45151335\, q^6
+ 452510730\, q^7
\nn\\  
&
+ 3763732545\, q^8  
+ 27047637540\, q^9
+ 172619569800\, q^{10}   
+\cdots.
\nn
\end{align}}
We have checked  that 
$\{N^{\inst}_{0; 3n,n}\}=\{3,-6,27,-192,1695,\cdots\}$
coincides with the genus zero, degree $n$ instanton number of the local 
$\text{\bf P}^2$ model \cite{KMV,LMW}.
%
%
\paragraph{$\E_{\one}$ model} 
{\allowdisplaybreaks 
\begin{align}
Z^{\inst}_{0;1}
&=8
+16\, q
+56\, q^2
+112\, q^3
+280\, q^4
+528\, q^5
+1120\, q^6
+2016\, q^7
+3880\, q^8
\nn \\
&
+6720\, q^9
+12096\, q^{10}
+\cdots,
\nn
\\
Z^{\inst}_{0;2}
&=-4\, q
-56\, q^2
-280\, q^3
-1232\,  q^4
-4212\,  q^5
-13544\,  q^6
-38584\, q^7
-105200\,  q^8
\nn \\
&-266696\,  q^9
-653400\,  q^{10}
-\cdots,
\nn \\
Z^{\inst}_{0;3}
&=24\, q^2
+ 336\, q^3
+ 2688\,  q^4 
+ 15360\, q^5
+ 73584\, q^6
+303744\,  q^7
+ 1137192\, q^8  
\nn \\
&+3897648\, q^9  
+12515112\,  q^{10}
+\cdots,
\nn\\
Z^{\inst}_{0;4}
&=-4\, q^2
-224\, q^3
-3472\, q^4
-32704\,  q^5
-232280\,  q^6
-1351040\,  q^7
-6818336\,  q^8
\nn\\
&
-30695296\, q^9 
-126302196\,  q^{10}
-\cdots,
\nn\\
Z^{\inst}_{0;5}
&=80\,  q^3
+ 2800\,  q^4
+ 44800\,  q^5
+ 477160\,  q^6  
+ 3892240\,  q^7
+26296560\,  q^8
\nn\\
&+ 153653920\,  q^9
+ 800623600\,  q^{10}
+\cdots.
\nn
\end{align}}
We have checked that 
$\{N^{\inst}_{0; 2n,n}\}=\{-4,-4,-12,-48,-240,-1356,-8428,\cdots\}$
coincides with the genus zero, degree $n$ instanton number of the
local $\text{\bf P}^1\!\times\!\text{\bf P}^1$ \cite{CKYZ}.

\section{Genus One Partition Functions}
\label{genus-one-partition}
\subsection{Genus one potentials}
In general, the determination of the genus one potential
$F_1$ of a \cy threefold $X$ requires the knowledge of
the discriminant loci of the \pf system,
which represent the singularities of the mirror complex moduli space 
${\cal M}(X^{*})$, 
and the identification of the power indices associated with 
each of the irreducible components of the discriminant loci
\cite{BCOV1,HKTY}.

For our case of the local $\E_9$ almost \dP models,    
we find the following answers for the instanton parts: 
{\allowdisplaybreaks\begin{align}
\E_0: \
F_1
&=\frac12\log\left[
\left(\frac{[(1\!+\!z_1)^3\!-\!z_2]}{(1\!-\!z_2)}\right)^{-\frac16}
\left(\frac{\pa z_1}{\pa \tilde{p}}\right)
\right],
\\
\E_{\one}: \
F_1
&=\frac12\log\left[
\left(\frac{[(1\!+\!z_1)^2\!-\!z_2]}{(1\!-\!z_2)}\right)^{-\frac16}
(1\!+\!z_1)^{-\frac13}
\left(\frac{\pa z_1}{\pa \tilde{p}}\right)
\right],
\\
\E_{5,6,7,8}: \
F_1
&=\frac12\log\left[
\left(\frac{(1\!-\!z_2\!+\!z_1z_2)}{(1\!-\!z_2)}\right)^{-\frac16}
(1\!-\!z_1)^{-\frac{9-N}{6}}
\left(\frac{\pa z_1}{\pa \tilde{p}}\right)
\right].
\end{align}}
where $\tilde{p}:=p\psi(\tau)=\text{e}^{2\pi \ti \sigma}\psi(\tau)$.
The following identity will be used later:
\begin{equation}
\Big(1+\sum_{n=1}^{\infty}n c_{n}(\tau)z_1^{n}\Big)
\Big(\frac{\pa z_1}{\pa \tilde{p}}\Big)
=\exp\Big(-\sum_{n=1}^{\infty}c_{n}(\tau)z_1^{n}\Big).
\label{atode}
\end{equation}

It would suffice here to list
the first four terms of the $p$-expansions of
$F_1$:
{\allowdisplaybreaks
\begin{align*}
F_1^{\soeji{0}}&=\frac{(y\tilde{p})}{4}(e_2+2)
+\frac{(y\tilde{p})^2}{64}\big(5e_2^2+8e_2+3+\frac{8}{y}\big)
+\frac{(y\tilde{p})^3}{1152}
\big[39e_2^3+54e_2^2+e_2\big(117-\frac{8}{y}\big)
\\
&+54+\frac{32}{y^2}\big]
+\frac{(y\tilde{p})^4}{18432}
\big[309e_2^4+384e_2^3
+e_2^2\big(1746-\frac{784}{y}\big)
-e_2\big(72-\frac{1120}{y}-\frac{64}{y^2}\big)
\\
&+513-\frac{144}{y}+\frac{320}{y^2}\big],
\\
F_1^{\soeji{\one}}&=\frac{(y\tilde{p})}{6}(e_2+3)
+\frac{(y\tilde{p})^2}{144}
\big[5e_2^2+e_2\big(18-\frac{3}{y}\big)+16+\frac{8}{y}+\frac{4}{y^2}\big]
\\
&+\frac{(y\tilde{p})^3}{1296}
\big[13e_2^3
+e_2^2\big(57-\frac{15}{y}\big)
+e_2\big(114-\frac{33}{y}+\frac{24}{y^2}\big)
+88-\frac{20}{y^3}+\frac{84}{y^2}-\frac{24}{y}
\big]
\\
&+\frac{(y\tilde{p})^4}{31104}
\big[103e_2^4+e_2^3\big(540-\frac{174}{y}\big)
+e_2^2\big(1584-\frac{864}{y}+\frac{324}{y^2}\big)\\
&+e_2\big(1936+\frac{1560}{y^2}-\frac{404}{y^3}-\frac{1176}{y}\big)
+960-\frac{384}{y}
+\frac{1854}{y^2}
-\frac{966}{y^3}
+\frac{291}{y^4}
  \big],
\\
F_1^{\soeji{5}}&=\frac{y\tilde{p}}{12}(e_2+3)+
\frac{(y\tilde{p})^2}{576}
\big[5e_2^2
+e_2\big(18+\frac{6}{y}\big)
+16
+\frac{32}{y}
+\frac{19}{y^2}
\big]
\\
&+\frac{(y\tilde{p})^3}{10368}
\big[13e_2^3
+e_2^2\big(57+\frac{30}{y}\big)
+e_2\big(114+\frac{138}{y}+\frac{87}{y^2}\big)
+88+\frac{156}{y}+\frac{417}{y^2}+\frac{52}{y^3}\big]
\\
&+\frac{(y\tilde{p})^4}{497664}
\big[103e_2^4
+e_2^3\big(540+\frac{348}{y}\big)
+e_2^2\big(1584+\frac{1656}{y}+\frac{1062}{y^2}\big)
\\
&+e_2\big(1936+\frac{4152}{y}+\frac{6276}{y^2}+\frac{1252}{y^3}\big)
+960+\frac{3648}{y}+\frac{9198}{y^2}+\frac{7836}{y^3}+\frac{921}{y^4}
\big],
\\
F_1^{\soeji{6}}&=\frac{y\tilde{p}}{12}(e_2+2)
+\frac{(y\tilde{p})^2}{576}
\big[5e_2^2
+e_2\big(8+\frac{12}{y}\big)
+3+\frac{28}{y}+\frac{16}{y^2}\big]
\\
&+\frac{(y\tilde{p})^3}{31104}
\big[39e_2^3
+e_2^2\big(54+\frac{180}{y}\big)
+e_2\big(117+\frac{316}{y}+\frac{368}{y^2}\big)
+54+\frac{216}{y}+\frac{992}{y^2}+\frac{256}{y^3}\big]
\\
&+\frac{(y\tilde{p})^4}{1492992}
\big[309e_2^4
+e_2^3\big(384+\frac{2088}{y}\big)
+e_2^2\big(1746+\frac{3032}{y}+\frac{6112}{y^2}\big)
\\
&+e_2\big(-72+\frac{8536}{y}+\frac{14496}{y^2}+\frac{8384}{y^3}\big)
+513+\frac{2664}{y}+\frac{14720}{y^2}+\frac{25792}{y^3}
+\frac{4608}{y^4}\big],
\\
F_1^{\soeji{7}}&=\frac{y\tilde{p}}{12}(e_2+1)
+\frac{(y\tilde{p})^2}{1152}
\big[10e_2^2
-e_2\big(4-\frac{36}{y}\big)
+\frac{27}{y}+\frac{27}{y^2}\big]
\\
&+\frac{(y\tilde{p})^3}{20736}
\big[26e_2^3
-e_2^2\big(42-\frac{180}{y}\big)
+e_2\big(84-\frac{81}{y}+\frac{387}{y^2}\big)
-32+\frac{99}{y}+\frac{315}{y^2}+\frac{216}{y^3}\big],
\\
&+\frac{(y\tilde{p})^4}{3981312}
\big[824e_2^4
-e_2^3\big(2272-\frac{8352}{y}\big)
+e_2^2\big(6528-\frac{13176}{y}+\frac{30312}{y^2}\big)
\\
&-e_2\big(9728-\frac{32112}{y}+\frac{8640}{y^2}-\frac{44496}{y^3}\big)
+6016-\frac{18720}{y}+\frac{30627}{y^2}
-\frac{39474}{y^3}+\frac{19683}{y^4}\big],
\\
F_1^{\soeji{8}}&=\frac{y\tilde{p}}{12}e_2
+\frac{(y\tilde{p})^2}{576}
\big[5e_2^2
-e_2\big(12-\frac{24}{y}\big)
+7-\frac{10}{y}+\frac{10}{y^2}\big]
\\
&+\frac{(y\tilde{p})^3}{31104}
\big[39e_2^3
-e_2^2\big(180-\frac{360}{y}\big)
+e_2\big(369-\frac{878}{y}+\frac{878}{y^2}\big)
-276+\frac{712}{y}-\frac{480}{y}+\frac{320}{y^3}\big]
\\
&+\frac{(y\tilde{p})^4}{1492992}
\big[309e_2^4
-e_2^3\big(2088-\frac{4176}{y}\big)
+e_2^2\big(6858-\frac{18724}{y}+\frac{18724}{y^2}\big)
\\
&-e_2\big(12336-\frac{39632}{y}+\frac{44880}{y^2}-\frac{29920}{y^3}\big)
+9513-\frac{33420}{y}+\frac{42510}{y^2}-\frac{18180}{y^3}
+\frac{9090}{y^4}\big],
\end{align*}}
where we recall $\tilde{p}=p\psi(\tau)$.
We see that the genus one \gw partition function
$Z_{1;n}$ takes the form predicted in (\ref{Zgn}), that is,
\begin{equation*}
Z_{1;n}=(Z_{0;1})^n \vp^{2n} {\cal P}_{1;n}(e_2,y^{-1}),
\end{equation*}
where ${\cal P}_{1;n}$ 
is a degree $n$ polynomial in $e_2$ and  $y^{-1}$.

We will give the partition functions
only for the $\E_7$ and $\E_8$ models below.
\paragraph{$\E_7$ model}
{\allowdisplaybreaks
\begin{align}
Z^{(7)}_{1;1}&= \frac{\Zseven}{24}A(E_2+A),
\nn\\
Z^{(7)}_{1;2} &= \frac{\Zseven^2}{4608}
(10E_2^2A^2+36E_2AB-4E_2A^3+27B^2+27A^2B),
\nn\\
Z^{(7)}_{1;3} &= \frac{\Zseven^3}{165888}
(26E_2^3A^3-42E_2^2A^4+180E_2^2A^2B
-81E_2A^3B+387E_2AB^2+84E_2A^5
\nn\\
&
+99A^4B+216B^3+315B^2A^2-32A^6),
\nn\\
Z^{(7)}_{1;4} &= \frac{\Zseven^4}{63700992}
(824A^4E_2^4+8352A^3E_2^3B-2272A^5E_2^3
+30312A^2E_2^2B^2-13176A^4E_2^2B
\nn\\
&
+6528A^6E_2^2-9728A^7E_2+44496AE_2B^3-8640A^3E_2B^2
+32112A^5E_2B+6016A^8
\nn\\
&
+19683B^4+30627A^4B^2-18720A^6B+39474A^2B^3), 
\nn\\
Z^{(7)}_{1;5} &= \frac{\Zseven^5}{9555148800}
(10970A^5E_2^5+145800A^4E_2^4B-42350A^6E_2^4
-1055250A^4E_2^2B^2
\nn\\
&
+748350A^3E_2^3B^2+151200A^7E_2^3-345800A^8E_2^2
+1240650A^6E_2^2B
-389250A^5E_2^3B
\nn\\
&
+3230811A^5E_2B^2
-1961316A^7BE_2+507584A^9E_2+2008395AE_2B^4-340032A^{10}
\nn\\
&
+1930635A^2B^4
+729000B^5+2538540A^4B^3-2195397A^6B^2+1432080A^8B
\nn\\
&
+1817100A^2E_2^2B^3-208440A^3E_2B^3).
\nn
\end{align}}
\paragraph{$\E_8$ model}
{\allowdisplaybreaks
\begin{align}
Z_{1;1}^{(8)}&=
\frac{1}{12\varphi^{12}}
E_2E_4,\nn \\
Z_{1;2}^{(8)}&=
\frac{1}{1152\varphi^{24}}
(9E_4^3+24E_2E_4E_6+10E_2^2E_4^2+5E_6^2),
\nn \\
Z_{1;3}^{(8)}&=
\frac{1}{62208\varphi^{36}}
(472E_4^3E_6+80E_6^3+299E_2E_4^4+439E_2E_4E_6^2
+360E_2^2E_4^2E_6 
+78E_2^3E_4^3),
\nn\\
Z_{1;4}^{(8)}&=
\frac{1}{11943936\varphi^{48}}
(37448E_2^2E_4^2E_6^2+
68768E_2E_4^4E_6+
29920E_2E_4E_6^3+
13809E_4^6              \nn \\
&+57750E_4^3E_6^2+
17416E_2^2E_4^5+
4545E_6^6+
16704E_2^3E_4^3E_6+
2472E_2^4E_4^4),\nn\\
Z_{1;5}^{(8)}&=
\frac{1}{895795200\varphi^{60}}
(4102280E_2E_4^4E_6^2
+808765E_2E_4E_6^4
+1378600E_2^2E_4^2E_6^3
+103760E_6^5 \nn \\
&+2111000E_2^2E_4^5E_6             
+951950E_2^3E_4^3E_6^2
+720057E_2E_4^7
+338950E_2^3E_4^6
+1749528E_4^4E_6
\nn \\
&+32910E_2^5E_4^5
+2340520E_4^3E_6^3
+291600E_2^4E_4^4E_6
).
\nn
\end{align}}

\subsection{Modular anomaly equation}
To get the modular anomaly equation of genus one \cite{HST},
we have only to notice that 
the genus one potential $F_1$ has $E_2(\tau)$-dependence
both through $z_1$ and through $c_n(\tau)$,
where we are considering $(\pa z_1/\pa \tilde{p})$
as a function of $z_1$ and $c_n$ by (\ref{atode}).

The contribution of the former to the derivative
$(\pa F_1/\pa E_2)$ is
\begin{equation*}
\left(\frac{\pa z_1}{\pa E_2}\right)
\left(\frac{\pa F_1}{\pa z_1}\right)=
\frac{1}{12h}(\vT_pz_1)(\vT_pF_1),
\end{equation*} 
where we have used (\ref{F0-kakikae}),
while the latter
\begin{equation*}
-\frac12\sum_{m=1}^{\infty}
\left(\frac{\pa c_m}{\pa E_2}\right)
\frac{\pa }{\pa c_m}
\left(\sum_{n=1}^{\infty}c_nz_1^n
+\log(1+\sum_{n=1}^{\infty}nc_nz_1^n)\right)
=\frac{1}{24h}\vT_p(\vT_p+1)F_0.
\end{equation*}
Then we see that the anomaly equation for genus one takes 
the following form:
\begin{equation}
h\left(\frac{\pa F_1}{\pa E_2}\right)
=\frac{1}{12}(\vT_pF_0)(\vT_pF_1)+\frac{1}{24}\vT_p(\vT_p+1)F_0,
\end{equation} 
which can be rewritten in terms of the \gw partition functions as
\begin{equation*}
h\frac{\pa Z_{1;n}(\tau)}{\pa E_2(\tau)}
=\frac{1}{24}\sum_{h=0}^{1}\sum_{k=1}^{n-1}
k(n\!-\!k)Z_{h;k}(\tau)Z_{1-h;n-k}(\tau)
+\frac{1}{24}n(n\!+\!1)Z_{0;n}(\tau),
\end{equation*}
which takes the from just predicted in (\ref{modular-anomaly}).

\subsection{Elliptic instanton numbers}
Let $N^{\inst}_{1;n,m}\in \text{\bf Z}$ 
be the genus one instanton number of bidegree $(n,m)$,
and
\begin{equation}
Z^{\inst}_{1;n}(\tau)
=\sum_{m=0}^{\infty}
N^{\inst}_{1;n,m} q^{m}
\end{equation}
be its generating function.
According to \cite{BCOV1}, 
we have the following decomposition of the genus one
\gw partition function: 
\begin{equation}
Z_{1;n}(\tau)
=\sum_{k|n}\left(
\sigma_{-1}(k)Z^{\inst}_{1;\frac{n}{k}}(k\tau)
+\frac{1}{12}k^{-1}Z^{\inst}_{0;\frac{n}{k}}(k\tau)\right).
\end{equation}
The inversion of this equation is given by
\begin{equation}
Z^{\inst}_{1;n}(\tau)=
\sum_{k|n}
\left(
a_{-1}(k) Z_{1; \frac{n}{k}}(k\tau)
-\frac{1}{12}a_{-3}(k)Z_{0; \frac{n}{k}}(k\tau)
\right),
\end{equation}
where we have introduced the arithmetic functions
$a_{l}(n):=\sum_{m|n}\mu(m)\mu(n/m)m^l$.

We give the generating functions of the genus one
instanton numbers for each model.
\paragraph{$\E_5$ model}
{\allowdisplaybreaks 
\begin{align}
Z^{\inst}_{1;1}
&=
-8\,q^{4}
-32\,q^{5}
-80\,q^{6}
-192\,q^{7}
-464\,q^{8}
-1024\,q^{9}
-2080\,q^{10}
-\cdots,
\nn\\
Z^{\inst}_{1;2}
&=
 18\,q^{4}
+192\,q^{5}
+1040\,q^{6}
+4352\,q^{7}
+15752\,q^{8}
+51328\,q^{9}
+153448\,q^{10}
+\cdots,
\nn\\
Z^{\inst}_{1;3}
&=
-16\,q^{4}
-384\,q^{5}
-3920\,q^{6}
-26848\,q^{7}
-145440\,q^{8}
-671936\,q^{9}
\nn\\
&-2754816\,q^{10}
-\cdots,
\nn\\
Z^{\inst}_{1;4}
&=
 5\,q^{4}
+320\,q^{5}
+6320\,q^{6}
+71168\,q^{7}
+577264\,q^{8}
+3758848\,q^{9}
+20853184\,q^{10}
+\cdots,
\nn\\
Z^{\inst}_{1;5}
&=
-96\,q^{5}
-4640\,q^{6}
-93056\,q^{7}
-1170496\,q^{8}
-10922336\,q^{9}
-82513280\,q^{10}
-\cdots.
\nn
\end{align}}
%
\paragraph{$\E_6$ model}
{\allowdisplaybreaks 
\begin{align}
Z^{\inst}_{1;1}
&=
-6\, q^3
-54\, q^4
-162\, q^5
-528\, q^6
-1566\, q^7
-3888\, q^8
-9414\, q^9
-21870\, q^{10}
-\cdots,
\nn\\
Z^{\inst}_{1;2}
&=
 9\, q^3
+243\, q^4
+2322\, q^5
+13824\, q^6
+68283\, q^7
+290466\, q^8
+1094580\, q^9
\nn\\
&
+3785940\, q^{10}
+\cdots,
\nn\\
Z^{\inst}_{1;3}
&=
-4\, q^3
-324\, q^4
-7290\, q^5
-85458\, q^6
-700164\, q^7
-4599990\, q^8
-25682910\, q^9
\nn\\
&
-126394182\, q^{10}
-\cdots,
\nn\\
Z^{\inst}_{1;4}
&=
135\, q^4
+8262\, q^5
+194532\, q^6
+2729754\, q^7
+27756027\, q^8
+226001070\, q^9
\nn\\
&
+1557055332\, q^{10}
+\cdots,
\nn\\
Z^{\inst}_{1;5}
&=
-3132\, q^5
-185346\, q^6
-4812210\, q^7
-78689502\, q^8
-948813714\, q^9
\nn\\
&
-9183023298\, q^{10}
-\cdots.
\nn
\end{align}}
\paragraph{$\E_7$ model}
{\allowdisplaybreaks 
\begin{align}
Z^{\inst}_{1;1}
&=
-4\, q^2
-112\, q^3
-564\, q^4
-3056\, q^5
-11108\, q^6
-40528\, q^7
-123112\, q^8
\nn\\
&
-367552\, q^9
-989236\, q^{10}
-\cdots,
\nn\\
Z^{\inst}_{1;2}
&=
 3\, q^2
+336\, q^3
+9018\, q^4
+101088\, q^5
+862098\, q^6
+5657664\, q^7
+32067860\, q^8
\nn\\
&
+158512832\, q^9
+712084479\, q^{10}
+\cdots,
\nn\\
Z^{\inst}_{1;3}
&=
-224\, q^3
-20496\, q^4
-640032\, q^5
-10716104\, q^6
-128761968\, q^7
-1208615256\, q^8
\nn\\
&
-9504050688\, q^9
-64763400720\, q^{10}
-\cdots,
\nn\\
Z^{\inst}_{1;4}
&=
 12042\, q^4
+1116896\, q^5
+41444664\, q^6
+903550592\, q^7
+14095889180\, q^8
\nn\\
&
+172098048640\, q^9
+1743551210128\, q^{10}
+\cdots,
\nn\\
Z^{\inst}_{1;5}
&=
-574896\, q^5
-57707124\, q^6
-2511634800\, q^7
-66979775872\, q^8
\nn\\
&
-1286028782768\, q^9
-19346827285068\, q^{10}
-\cdots.
\nn
\end{align}}
%
\paragraph{$\E_8$ model}
{\allowdisplaybreaks 
\begin{align}
Z^{\inst}_{1;1}
&=
-2\, q
-510\, q^2
-11780\, q^3
-142330\, q^4
-1212930\, q^5
-8207894\, q^6
-46981540\, q^7
\nn\\
&
-236385540\, q^8
-1072489860\, q^9
-4467531670\, q^{10}
-\cdots,
\nn\\
Z^{\inst}_{1;2}
&=
762\, q^2
+205320\, q^3
+11361870\, q^4
+317469648\, q^5
+5863932540\, q^6
\nn\\
&
+81295293600\, q^7
+909465990330\, q^8
+8597134346400\, q^9
\nn\\
&
+70867771453026\, q^{10}
+\cdots,
\nn\\
Z^{\inst}_{1;3}
&=
-246788\, q^3
-76854240\, q^4
-6912918432\, q^5
-323516238180\, q^6
\nn\\
&
-9882453271500\, q^7
-221876231766660\, q^8
-3933705832711600\, q^9
\nn\\
&
-57747806496416088\, q^{10}
-\cdots,
\nn\\
Z^{\inst}_{1;4}
&=
76413073\, q^4
+27863327760\, q^5
+3478600115600\, q^6
+234196316814400\, q^7
\nn\\
&
+10330930335961770\, q^8
+332747064864457152\, q^9
\nn\\
&
+8378290954495817152\, q^{10}
+\cdots,
\nn\\
Z^{\inst}_{1;5}
&=
 -23436186174\, q^5
-9930641443350\, q^6
-1585090167772500\, q^7
\nn\\
&
-140688512133882000\, q^8
-8255877490179586950\, q^9
\nn\\
&
-353737948953627859770\, q^{10}
-\cdots.
\nn
\end{align}}
We see that for the principal series,
$N^{\inst}_{1; n,n}$ coincides with  the genus one, degree $n$
instanton number of the $E_N$ \dP model first obtained in
\cite{LMW}.
Genus one instanton numbers for the $E_8$ model 
have been computed in \cite{KMV}.
\paragraph{$\E_0$ model}
{\allowdisplaybreaks 
\begin{align}
Z^{\inst}_{1;1}
&=
-18\, q^3
-72\, q^4
-252\, q^5
-774\, q^6
-2106\, q^7
-5292\, q^8
-12564\, q^9
\nn\\
&
-28278\, q^{10}
-\cdots,
\nn\\
Z^{\inst}_{1;2}
&=
 108\, q^3
+1152\, q^4
+7812\, q^5
+41022\, q^6
+181656\, q^7
+710856\, q^8
+2526516\, q^9
\nn\\
&
+8310492\, q^{10}
+\cdots,
\nn\\
Z^{\inst}_{1;3}
&=
-336\, q^3
-7368\, q^4
-85284\, q^5
-700896\, q^6
-4602090\, q^7
-25679052\, q^8
\nn\\
&
-126406392\, q^9
-562694940\, q^{10}
-\cdots,
\nn\\
Z^{\inst}_{1;4}
&=
 630\, q^3
+26343\, q^4
+496404\, q^5
+6119388\, q^6
+57190644\, q^7
+437749110\, q^8
\nn\\
&
+2875241088\, q^9
+16711846956\, q^{10}
+\cdots,
\nn\\
Z^{\inst}_{1;5}
&=
-756\, q^3
-59976\, q^4
-1817298\, q^5
-33012216\, q^6
-430550244\, q^7
\nn\\
&
-4429221912\, q^8
-38028172446\, q^9
-282776491026\, q^{10}
-\cdots.
\nn
\end{align}}
We have checked that 
$\{N^{\inst}_{1; 3n,n}\}=\{0,0,-10,231, -4452,\cdots\}$ 
coincides with
the genus one, degree $n$ instanton number of the 
$\text{\bf P}^2$ model first obtained in \cite{LMW}.
\paragraph{$\E_{\one}$ model}
{\allowdisplaybreaks 
\begin{align}
Z^{\inst}_{1;1}
&=
-16\,q^4
-32\,q^5
-112\,q^6
-224\,q^7
-608\,q^8
-1152\,q^9
-2576\,q^{10}
-\cdots,
\nn\\
Z^{\inst}_{1;2}
&=
 84\,q^4
+424\,q^5
+2264\,q^6
+8176\,q^7
+29364\,q^8
+88416\,q^9
\nn\\
&
+260360\,q^{10}
+\cdots,
\nn\\
Z^{\inst}_{1;3}
&=
-224\,q^4
-2208\,q^5
-17392\,q^6
-95872\,q^7
-467376\,q^8
-1947008\,q^9
\nn\\
&
-7471488\,q^{10}
-\cdots,
\nn\\
Z^{\inst}_{1;4}
&=
 350\,q^4
+6272\,q^5
+72512\,q^6
+576704\,q^7
+3778068\,q^8
+20848384\,q^9
\nn\\
&
+102392928\,q^{10}
+\cdots,
\nn\\
Z^{\inst}_{1;5}
&=
-336\,q^4
-10976\,q^5
-188880\,q^6
-2130016\,q^7
-18652816\,q^8
-134027488\,q^9
\nn\\
&
-833043952\,q^{10}
-\cdots.
\nn
\end{align}}
We have checked that 
$
\{N^{\inst}_{1; 2n,n}\}=\{0,0,0,9,136, 1616, 17560,\cdots\}
$ 
coincides with
the genus one, degree $n$ instanton number of the 
$\text{\bf P}^1\!\times\!\text{\bf P}^1$ model listed in \cite{CKYZ}.

\section{Higher Genus Partition Functions}
\label{higher-genus-partition}
In contrast to the genus zero or genus one case,
we cannot evaluate directly the higher genus 
\gw partition functions \cite{BCOV2}.
However, the modular anomaly equation (\ref{modular-anomaly})
invented in \cite{HST} is so powerful that it determines 
the partition function $Z^{(N)}_{g;n}(\tau)$
up to finite constants.  
\subsection{Partition functions as modular forms}
In this subsection, we propose a conjecture 
on the form of the partition functions $Z_{g;n}(\tau)$
of the six models in terms of the modular forms.

First we define $\chi^{(N)}$ for each of the six models by
{\allowdisplaybreaks
\begin{alignat}{3}
\chi^{(0)}(\tau)&=\frac{9q^{\frac16}}{\eta(\tau)^4},
\quad &
\chi^{(\one)}(\tau)&=\frac{8q^{\frac14}}{\eta(\tau)^2\eta(2\tau)^2},
\quad & 
\chi^{(5)}(\tau)
&=\frac{4q^{\frac12}\eta(2\tau)^4}{\eta(\tau)^4\eta(4\tau)^4},
\nn\\ 
\chi^{(6)}(\tau)&=\frac{3q^{\frac12}}{\eta(\tau)^3\eta(3\tau)^3},
\quad & 
\chi^{(7)}(\tau)&=\frac{2q^{\frac12}}{\eta(\tau)^4\eta(2\tau)^4},
\quad & 
\chi^{(8)}(\tau)&=\frac{q^{\frac12}}{\eta(\tau)^{12}}.
\label{kai-no-teigi}
\end{alignat}} 
Then we propose the conjectured forms of the partition functions
of them:
{\allowdisplaybreaks
\begin{align}
Z_{g;n}^{(0)}(\tau)&=(\chi^{(0)}(\tau))^n\,
P_{2g-2+2n}^{(0)}(E_2(\tau),\varpi^{(6)}(\tau),H(\tau)),
\\
Z_{g;n}^{(\one)}(\tau)&=(\chi^{(\one)}(\tau))^n\,
P_{2g-2+2n}^{(\one)}(E_2(\tau),\vat_3(2\tau)^4,\vat_4(2\tau)^4),
\\
Z_{g;n}^{(5)}(\tau)&=(\chi^{(5)}(\tau))^n\,
P_{2g-2+2n}^{(5)}(E_2(\tau),\vat_3(2\tau)^4,\vat_4(2\tau)^4),
\\
Z_{g;n}^{(6)}(\tau)&=(\chi^{(6)}(\tau))^n\,
P_{2g-2+3n}^{(6)}(E_2(\tau),\varpi^{(6)}(\tau),H(\tau)),
\\
Z_{g;n}^{(7)}(\tau)&=(\chi^{(7)}(\tau))^n\,
P_{2g-2+4n}^{(7)}(E_2(\tau),A(\tau),B(\tau)),
\\
Z_{g;n}^{(8)}(\tau)&=(\chi^{(8)}(\tau))^n\,
P_{2g-2+6n}^{(8)}(E_2(\tau),E_4(\tau),E_6(\tau)),
\end{align}}
where each $P^{(N)}$ is a polynomial over $\text{\bf Q}$
in three variables the subscript of which   
shows its weight as a quasi-modular form.

\paragraph{$\E_7$ model}
We list the genus two partition functions of
the $\E_7$ model.
{\allowdisplaybreaks
\begin{align}
Z^{\soeji{7}}_{2;1} &= \frac{\Zseven}{5760}A(6A^2+3B+5E_2^2+10AE_2),
\nn
\\ 
Z_{2;2}^{\soeji{7}} &= \frac{\Zseven^2}{1658880}
(-64A^5+999A^3B+2349AB^2+190E_2^3A^2+30E_2^2A^3+810E_2^2AB
\nn
\\
&
+132A^4E_2+1251A^2BE_2+1215E_2B^2),
\nn\\ 
Z^{\soeji{7}}_{2;3} &= \frac{\Zseven^3}{318504960}
(10561A^7-27183A^5B+222723A^3B^2+273699AB^3+4600E_2^4A^3
\nn\\
&
-5920E_2^3A^4
+36480E_2^3A^2B-288E_2^2A^3B+105840E_2^2AB^2+20544E_2^2A^5
\nn\\
&
+200988E_2B^2A^2
-12100E_2A^6+61704E_2A^4B+103680E_2B^3), 
\nn\\
Z^{\soeji{7}}_{2;4} &= \frac{\Zseven^4}{61152952320}
(-1622467A^9+6147828A^7B-5446746A^5B^2+33187428A^3B^3
\nn\\
&
+26235333AB^4
+108800A^4E_2^5+1236480E_2^4A^3B-280320E_2^4A^5
+7149024E_2^2A^5B
\nn\\
&
+5454720E_2^3B^2A^2
+1067008E_2^3A^6-1639296E_2^3A^4B+1937232E_2^2A^3B^2
\nn\\
&
-2044976E_2^2A^7+10817280E_2^2AB^3+18498600E_2A^4B^2-8131560E_2A^6B
\nn\\
&
+24054408E_2A^2B^3+2625048E_2A^8+8048160E_2B^4), 
\nn\\
Z^{\soeji{7}}_{2;5} &= \frac{\Zseven^5}{17612050268160}
(-142044480E_2^4A^5B+621872640E_2^3A^6B-342771840E_2^3A^4B^2
\nn\\
&
+355628304E_2^2A^9
+55837440E_2^5A^4B+332170560E_2^4A^3B^2+991837440E_2^3A^2B^3
\nn\\
&
+873400320E_2B^5
-1325802096E_2^2A^7B+727775280E_2^2A^3B^3+1478062080E_2^2AB^4
\nn\\
&
-462511536E_2A^{10}
+2359788336E_2^2A^5B^2-2810004192E_2A^6B^2
+292244077A^{11}
\nn\\
&
+3887980560E_2A^2B^4
+4762800000E_2A^4B^3+1930802688E_2A^8B+3804160A^5E_2^6
\nn\\
&
-1325620701A^9B+2471168610A^7B^2-1132668090A^5B^3+6150153825A^3B^4
\nn\\
&
+3302730855AB^5-14400000E_2^5A^6+62142720E_2^4A^7-172019840E_2^3A^8).
\nn
\end{align}}
\paragraph{$\E_8$ model} We give the partition functions of 
genus up to  five. 
{\allowdisplaybreaks
\begin{align}
Z^{\soeji{8}}_{2;1}&=\frac{1}{1440\varphi^{12}}
E_4(E_4+5E_2^2),
\nn\\
Z^{\soeji{8}}_{2;2}&=\frac{1}{207360\varphi^{24}}
(417 E_2 E_4^3  
+ 190 E_4^2  E_2^3  
+ 540 E_2^2  E_4 E_6 
+ 225 E_2 E_6^2
+ 356 E_4^2  E_6),
\nn\\
Z^{\soeji{8}}_{2;3}&=\frac{1}{2488320\varphi^{36}}
 (575 E_2^4  E_4^3  
+ 3040 E_2^3  E_4^2  E_6 
+ 4690 E_2^2  E_4 E_6^2
+ 3548 E_2^2  E_4^4  
+ 1600 E_6^3  E_2 
\nn\\
& 
+ 10176 E_6 E_4^3  E_2
+ 2231 E_4^5
+ 5244 E_4^2  E_6^2 ),
\nn\\
Z^{\soeji{8}}_{2;4}&= \frac{1}{179159040\varphi^{48}}
(77280 E_2^4  E_6 E_4^3  
+ 209200 E_2^2  E_6^3  E_4
+ 547760 E_2^2  E_6 E_4^4  
+ 214811 E_4^6  E_2
\nn\\
&  
+ 203900 E_2^3  E_6^2  E_4^2
+ 103252 E_4^5  E_2^3
+ 827230 E_6^2  E_4^3  E_2 
+ 10200 E_2^5  E_4^4  
+ 57375 E_6^4  E_2
\nn\\
&
+ 420616 E_4^5  E_6 
+ 314360 E_4^2  E_6^3 ),
\nn\\
Z^{\soeji{8}}_{2;5}&=
\frac{1}{12899450880\varphi^{60}}
 (15422230 E_6^4  E_4^2  
+ 43101209 E_6^2  E_4^5  
+ 5522085 E_4^8
+ 1903680 E_6^5  E_2
\nn\\
& 
+ 18947800 E_2^3  E_4^5  E_6
+ 1744920 E_2^5  E_4^4  E_6 
+ 50040570 E_2^2  E_6^2  E_4^4  
+ 6480025 E_2^4  E_4^3  E_6^2
\nn\\
& 
+ 11149400 E_2^3  E_4^2  E_6^3 
+ 8437860 E_2^2  E_4 E_6^4
+ 51231560 E_6^3  E_4^3  E_2 
+ 42541168 E_4^6  E_6 E_2
\nn\\
&
+ 2482715 E_4^6  E_2^4  
+ 9555018 E_4^7  E_2^2  
+ 178320 E_2^6  E_4^5 ),
\nn\\
%
Z^{\soeji{8}}_{3;1}&=\frac{1}{362880\varphi^{12}}
 E_4 (4 E_6 + 21 E_2 E_4 + 35 E_2^3 ),
\nn\\
Z^{\soeji{8}}_{3;2}&=\frac{1}{34836480\varphi^{24}}
 (14984 E_4^2  E_6 E_2 
 + 8925 E_2^2  E_4^3  
 + 2275 E_4^2  E_2^4
 + 7560 E_2^3  E_4 E_6 
 + 4725 E_2^2  E_6^2  
\nn\\
&
 + 3540 E_4^4
 + 4071 E_4 E_6^2 ),
\nn\\
Z^{\soeji{8}}_{3;3}&=\frac{1}{209018880\varphi^{36}}
 (138104 E_4^4  E_6 
 + 224024 E_6 E_4^3  E_2^2
 + 36400 E_2^4  E_4^2  E_6
 + 224456 E_4^2  E_6^2  E_2 
\nn\\
&
 + 49584 E_4 E_6^3  
 + 68460 E_2^3  E_4 E_6^2
 + 55006 E_2^3  E_4^4  
 + 6055 E_2^5  E_4^3  
 + 97431 E_4^5  E_2
\nn\\
&
 + 33600 E_6^3  E_2^2 ),
\nn\\
 Z^{\soeji{8}}_{3;4}&=
\frac{1}{90296156160\varphi^{48}}
 (28134630 E_4^7  
 + 151049093 E_4^4  E_6^2  
 + 25488295 E_4 E_6^4
 + 966630 E_2^6  E_4^4
\nn\\
&
 + 189296376 E_6^2  E_4^3  E_2^2  
 + 8172360 E_2^5  E_6 E_4^3
 + 31388000 E_2^3  E_6^3  E_4 
 + 88718416 E_2^3  E_6 E_4^4
\nn\\
&
 + 24977155 E_2^4  E_6^2  E_4^2  
 + 13366787 E_4^5  E_2^4  
 + 12119625 E_6^4  E_2^2  
 + 137926976 E_4^2  E_6^3  E_2
\nn\\
&
 + 51557313 E_4^6  E_2^2  
 + 192353224 E_4^5  E_6 E_2),
\nn\\
Z^{\soeji{8}}_{3;5}&=
\frac{1}{1083553873920\varphi^{60}}
 (274848600 E_6^5  E_4 
 + 2868277704 E_6^3  E_4^4
 + 1662616800 E_6 E_4^7 
\nn\\
&
 + 635585864 E_2^4  E_4^5  E_6
 + 323470350 E_2^3  E_4 E_6^4  
 + 349176520 E_2^4  E_4^2  E_6^3
 + 41643000 E_2^6  E_4^4  E_6 
\nn\\
&
 + 2109910578 E_2^3  E_6^2  E_4^4
 + 174368705 E_2^5  E_4^3  E_6^2
 + 3866100 E_2^7  E_4^5
 + 101077200 E_6^5  E_2^2
\nn\\
&
 + 424873884 E_4^7  E_2^3  
 + 1739056502 E_6^4  E_4^2  E_2
 + 5180110741 E_6^2  E_4^5  E_2 
 + 70310947 E_4^6  E_2^5
\nn\\
&
 + 3045375184 E_6^3  E_4^3  E_2^2  
 + 2693483096 E_4^6  E_6 E_2^2
 + 696828225 E_4^8  E_2),
\nn\\
%
Z^{\soeji{8}}_{4;1}&=\frac{1}{87091200\varphi^{12}}
 E_4 (39 E_4^2  + 80 E_2 E_6 + 210 E_2^2  E_4 + 175 E_2^4 ),
\nn\\
Z^{\soeji{8}}_{4;2}&=
  \frac{1}{2090188800\varphi^{24}}
(53220 E_2 E_4^4  
 + 112540 E_2^2 E_4^2  E_6 
 + 45185 E_2^3  E_4^3
 + 7385 E_4^2  E_2^5
\nn\\
&
 + 28350 E_2^4  E_4 E_6 
 + 23625 E_2^3  E_6^2
 + 61065 E_2 E_4 E_6^2  
 + 6300 E_6^6  
+ 49402 E_6 E_4^3 ),
\nn\\
Z^{\soeji{8}}_{4;3}&= 
 \frac{1}{75246796800\varphi^{36}}
 ( 3164700 E_2^4  E_4 E_6^2  
 + 8993259 E_4^5  E_2^2
 + 14111840 E_6^2  E_4^3  
 + 806400 E_6^4
\nn\\
&
 + 25171632 E_2 E_6 E_4^4
 + 13855280 E_2^3  E_6 E_4^3  
 + 8963520 E_2 E_6^3  E_4 
  + 20453520 E_2^2  E_6^2  E_4^2
\nn\\
&
 + 4014627 E_4^6  
 + 208985 E_2^6  E_4^3  
 + 2016000 E_6^3  E_2^3
 + 1417920 E_2^5  E_4^2  E_6 
 + 2638125 E_2^4  E_4^4 ),
\nn\\
Z^{\soeji{8}}_{4;4}&=
  \frac{1}{5417769369600\varphi^{48}}
 (3336940980 E_2^3  E_4^3  E_6^2  
 + 7817234620 E_2 E_6^2  E_4^4
 + 3248768730 E_6^3  E_4^3
\nn\\
& 
 + 5085796952 E_2^2  E_4^5  E_6
 + 101280375 E_6^5  
 + 3550525000 E_2^2  E_4^2  E_6^3
 + 1290318725 E_2 E_4 E_6^4  
\nn\\
&
 + 936363912 E_4^6  E_2^3
 + 1481276055 E_4^7  E_2 
 + 2912603799 E_4^6  E_6 
 + 1216807640 E_2^4  E_4^4  E_6
\nn\\
&
 + 152620090 E_2^5  E_4^5  
 + 78676080 E_2^6  E_6 E_4^3
 + 410158000 E_2^4  E_6^3  E_4 
 + 274844990 E_2^5  E_6^2  E_4^2
\nn\\
&
 + 8381520 E_2^7  E_4^4  
 + 202702500 E_6^4  E_2^3 ),
\nn\\
Z^{\soeji{8}}_{4;5}&=
  \frac{1}{52010585948160\varphi^{60}}
 (16869986640 E_6^5  E_4 E_2 
 + 8944068536 E_2^5  E_4^5  E_6
 + 1167070464 E_6^6
\nn\\
&
 + 2035152000 E_6^5  E_2^3
 + 436442160 E_2^7  E_4^4  E_6 
 + 854577430 E_4^6  E_2^6  
+ 114133172104 E_6^2  E_4^4
\nn\\
&
 + 183172864792 E_6^3  E_4^4  E_2
  + 36942885 E_2^8  E_4^5
 + 5146355025 E_2^4  E_4 E_6^4  
  + 11890359900 E_4^9
\nn\\
&
 + 7455500881 E_4^7  E_2^4  
 + 23616142080 E_4^8  E_2^2  
 + 60902666801 E_6^4  E_4^3
 + 2043907670 E_2^6  E_4^3  E_6^2  
\nn\\
&
 + 4688369560 E_2^5  E_4^2  E_6^3
 + 54769592870 E_6^4  E_4^2  E_2^2
 + 66152468720 E_6^3  E_4^3  E_2^3
\nn\\
& 
 + 60955175392 E_4^6  E_6 E_2^3  
 + 109420106696 E_6 E_4^7  E_2
 + 170157797734 E_6^2  E_4^5  E_2^2
\nn\\
&
 + 35736239660 E_2^4  E_6^2  E_4^4 ),
\nn\\
%
Z^{\soeji{8}}_{5;1}&=\frac{1}{11496038400\varphi^{12}}
 E_4 (136 E_4 E_6 + 429 E_4^2  E_2 
+ 440 E_2^2  E_6 + 770 E_2^3  E_4 + 385 E_2^5),
\nn\\
Z^{\soeji{8}}_{5;2}&=
  \frac{1}{3310859059200\varphi^{24}} 
(4510275 E_2^2  E_4^3  
 + 10553400 E_2^2  E_4^4
 + 2494800 E_6^3  E_2 
  + 3358995 E_4^5  
\nn\\
&
 + 14869360 E_2^3  E_4^2  E_6 
 + 12090870 E_2^2  E_4 E_6^2
 + 19568568 E_6 E_4^3  E_2
 + 2245320 E_2^5  E_4 E_6 
\nn\\
&
+ 7083727 E_4^2  E_6^2  
 + 512050 E_4^2  E_2^6
+ 2338875 E_2^2  E_6^2),
\nn\\
Z^{\soeji{8}}_{5;3}&=
\frac{1}{9932577177600\varphi^{36}}
 (935093824 E_6^2  E_4^3  E_2 
 + 233170300 E_2^4  E_6 E_4^3
 + 296640960 E_2^2  E_6^3  E_4 
\nn\\
&
 + 837550728 E_2^2  E_6 E_4^4
 + 453680480 E_2^3  E_6^2  E_4^2  
 + 16385600 E_2^6  E_4^2  E_6
 + 42513240 E_2^5  E_4 E_6^2  
\nn\\
&
 + 201151929 E_4^5  E_2^3
 + 36275085 E_2^5  E_4^4  
 + 53222400 E_6^4  E_2 
 + 266767491 E_4^6  E_2
\nn\\
&
 + 405268284 E_4^5  E_6 
 + 268326944 E_4^2  E_6^3  
 + 33264000 E_6^3  E_2^4  
 + 2155615 E_2^7  E_4^3),
\nn\\
Z^{\soeji{8}}_{5;4}&=
  \frac{1}{2860582227148800\varphi^{48}} 
( 12207942670 E_2^6  E_4^5
+ 523849095 E_2^8  E_4^4  
 + 156150752805 E_4^8
\nn\\
&
 + 113811930320 E_2^5  E_4^4  E_6 
+ 1311485716360 E_4^6  E_6 E_2 
 + 1760563778482 E_2^2  E_6^2  E_4^4  
\nn\\
&
 + 286289201000 E_2^2  E_4 E_6^4  
 + 381058740370 E_2^4  E_4^3  E_6^2
 + 1449394307792 E_6^3  E_4^3  E_2
\nn\\
&
 + 1106487740990 E_6^2  E_4^5
 + 44575839000 E_6^5  E_2 
 + 109025587484 E_4^6  E_2^4
\nn\\
&
 +774483173328 E_2^3  E_4^5  E_6
 + 531170439360 E_2^3  E_4^2  E_6^3
 + 5431290480 E_2^7  E_6 E_4^3
\nn\\
&
 + 37160939200 E_2^5  E_6^3  E_4
 + 337421738130 E_4^7  E_2^2  
+ 21439577390 E_2^6  E_6^2  E_4^2
\nn\\
&
 + 22344052500 E_6^4  E_2^4
 + 344998537324 E_6^4  E_4^2 ),
\nn\\
Z^{\soeji{8}}_{5;5}&=
\frac{1}{102980960177356800\varphi^{60}}
(31511006810584 E_6^5  E_4^2
 + 177751951656248 E_6^3  E_4^5
\nn\\
&
 + 78175349827680 E_6 E_4^8
 + 344664297670 E_4^6  E_2^7  
 + 21179704043952 E_4^8  E_2^3
\nn\\
&
 + 4104656416113 E_4^7  E_2^5
 + 41030103891064 E_4^6  E_6 E_2^4  
 + 1264302270000 E_6^5  E_2^4
\nn\\
&
 + 2891093990400 E_6^6  E_2
 + 31336414684620 E_4^9  E_2 
 + 155081412353885 E_6^4  E_4^3  E_2
\nn\\
&
 + 296146234031236 E_6^2  E_4^6  E_2 
 + 43310617469240 E_6^3  E_4^3  E_2^4
 + 19202491494120 E_2^5  E_6^2  E_4^4  
\nn\\
&
 + 46735606475470 E_6^4  E_4^2  E_2^3  
 + 2649529315125 E_2^5  E_4 E_6^4
 + 154506124080 E_2^8  E_4^4  E_6 
\nn\\
&
 + 803244450470 E_2^7  E_4^3  E_6^2
 + 4107192009800 E_2^6  E_4^5  E_6
+ 2083500320440 E_2^6  E_4^2  E_6^3
\nn\\
&
 + 149437965048686 E_6^2  E_4^5  E_2^3  
 + 21211745049000 E_6^5  E_4 E_2^2
 + 144355295784864 E_6 E_4^7  E_2^2  
\nn\\
&
 + 11970104685 E_2^9  E_4^5
+ 236773842080568 E_6^3  E_4^4  E_2^2).
\nn
\end{align}}

\subsection{\gop invariants}
We can extract from the higher genus \gw partition function
$Z_{g;n}$ an important integer-invariants,
the \gop invariants \cite{GV2}, which we will explain 
very briefly. For more details on this subject, see
\cite{HST,BP,HST2,KKV}.

Let us first consider a BPS state in M theory compactified 
on a \cy threefold $X$  
which is realized by the  M2-brane wrapped around a
holomorphic curve $C$.
In addition to the Abelian gauge charge  
corresponding to the homology class $[C]\in H_2(X)$,
such a state carries also a quantum number of the 5D little group
$SO(4)=SU(2)_{L}\!\times\!SU(2)_{R}$.

To clarify the origin of the Lorentz quantum number,
let ${\cal M}_{\beta}$ be the moduli space of curves
in $X$ with a fixed homology class $\beta\in H_2(X)$,
and 
$\pi\colon\hat{\cal M}_{\beta}\to {\cal M}_{\beta}$ 
the extended moduli space with its Jacobian fibration,
which means that $\hat{\cal M}_{\beta}$ parametrizes all the 
pairs of a curve of the fixed homology class $\beta$ 
and a flat line bundle on it. 
Namely, $\hat{\cal M}_{\beta}$ is the appropriate 
moduli space for a BPS M2 brane with its Abelian charge fixed.

The BPS states for this degree of freedom 
arise from quantization of the moduli space, that is,
the cohomology group $H^*(\hat{\cal M}_{\beta};\text{\bf C})$ 
represented  by the harmonic differential forms.
The $SU(2)_{L}$ and $SU(2)_{R}$ Lorentz quantum numbers come from
the Lefshetz $SU(2)$ actions for the fibre and base direction 
of the Jacobian fibration 
$\pi:\hat{\cal M}_{\beta}\to {\cal M}_{\beta}$
respectively.   

We take the following representation of
the Lorentz spin content of the BPS states with fixed $\beta$:
\[
H^*(\hat{\cal M}_{\beta};\text{\bf C})=\sum_{h=0}^{g}
I_h\otimes U_{h;\beta},\qquad
I_h:=\left[V_{1/2}^{L}\oplus 2V_{0}^{L}\right]^{\otimes h},
\]  
where $g$ is the maximum value of the genus that 
a curve of the fixed homology class $\beta$ can have,
and $V_j$ the irreducible $SU(2)$ module of spin $j$.
Let $U_{h;\beta}=\oplus_{j}N_{h,j;\beta}V_{j}^{R}$ be 
the irreducible decomposition of the $SU(2)_R$ module above. 
Then $N_{h,j;\beta}\in \text{\bf Z}_{\geq 0}$ 
is the multiplicity of the BPS states with 
the $SO(4)$ Lorentz quantum number
$I_{h+1}\otimes V_{j}^{R}$ 
and the Abelian gauge charge $\beta\in H_2(X)$. 

The \gop invariant $N_{h;\beta}^{\gv}$
with fixed $h\in \text{\bf Z}$ and $\beta\in H_2(X)$ is then 
given by  the index with respect to
the $SU(2)_R$ on $U_{h;\beta}$, that is,  
$N_{h;\beta}^{\gv}:
=\sum_{j}\text{e}^{2\pi \ti j}
(2j\!+\!1)N_{h,j;\beta}$.

It has been found in \cite{GV2} that the instanton part of 
the full partition function of IIA topological string on 
$X$ \cite{BCOV2} can be obtained by  
\begin{equation}
\sum_{g=0}^{\infty}x^{2g-2}F_g
=\underset{\beta\in H_2(X)}{{\sum}'}
\sum_{h=0}^{\infty}
\sum_{m=1}^{\infty}
N_{h;\beta}^{\text{\tiny GV}}\,
\frac{1}{m}
\Big[ 2\sin\big(\frac{mx}{2}\big)\Big]^{2h-2}
\text{e}^{2\pi \ti m\langle J, \beta\rangle},
\label{GV-original}
\end{equation}
where $J\in H^2(X;\text{\bf C})$ 
is the complexified \K class of $X$.

{}From now on we turn to the investigation 
of the \gop invariants of one of the six 
local two-parameter models of the $\E_9$ almost \dP surface.
First let $N_{g; n,m}^{\gv}$ be the \gop invariant of
genus $g$ and bidegree $(n,m)$, 
and define its generating function by 
\begin{equation}
Z^{\gv}_{g;n}(\tau)
=\sum_{m=0}^{\infty}N_{g; n,m}^{\gv}q^{m}.
\end{equation}

We can show from (\ref{GV-original}) that
the partition function of the genus $g$ \gw
invariants $Z_{g;n}(\tau)$ admits the following decomposition
into the generating functions 
of \gop invariants of genus $h\leq g$: 
\begin{equation}  
Z_{g;n}(\tau)
=\sum_{h=0}^{g}\beta_{g,h}
\sum_{k|n}k^{2g-3}Z_{h;\frac{n}{k}}^{\gv}(k\tau),
\label{GV}
\end{equation}
where $\beta_{g,h}$ is the rational number 
defined by the following expansion: 
\begin{equation}
\left(\frac{\sin(x/2)}{(x/2)}\right)^{2h-2}
=\sum_{g=h}^{\infty}\beta_{g,h}\, x^{2(g-h)}.
\label{rational-beta}
\end{equation}
Note that $\beta_{g,0}$ coincides with the one given earlier 
in (\ref{leading}).

Now we give the M\"obius inversion formula of (\ref{GV})
following \cite[Prop.2.1]{BP}.
To this end, let us first define 
the rational number $\alpha_{g,h}$ by
\begin{equation}
\left(\frac{\text{arcsin}(x/2)}{(x/2)}\right)^{2h-2}
=\sum_{g=h}^{\infty}\alpha_{g,h}\, x^{2(g-h)}.
\end{equation}
The M\"obius inversion for the \gop invariants 
can then be written as
\begin{equation}
Z^{\gv}_{g;n}(\tau)=
\sum_{h=0}^{g}\alpha_{g,h}
\sum_{k|n}\mu(k)k^{2h-3}Z_{h;\frac{n}{k}}(k\tau).
\label{BP}
\end{equation}
Let us take the $\E_8$ model and substitute 
the leading term (\ref{leading}) of the partition function 
in (\ref{BP}).
Then we see for each $(g,n)\ne (0,1)$, 
\begin{equation}
Z_{g;n}^{\gv (8)}(\tau)
=\sum_{h=0}^{g}\alpha_{g,h}\beta_{h,0}\,n^{2h-3}
\sum_{k|n}\mu(k)+O(q^n)
=O(q^n).
\end{equation}

To be more explicit, we describe below the decompositions
of the \gw partition functions 
$Z_{g;n}(\tau)$ into the generating functions of 
\gop invariants $Z^{\gv}_{g;  n}(\tau)$
(\ref{GV}) and  their M\"obius inversions (\ref{BP}) for lower genera.
\paragraph{Genus zero}
For the genus zero case, (\ref{GV}) and (\ref{BP}) read
\begin{equation*}
Z_{0;n}(\tau)=
\sum_{k|n} k^{-3} Z_{0; \frac{n}{k}}^{\gv}(k\tau),
\quad
Z^{\gv}_{0;  n}(\tau)=
\sum_{k|n} \mu(k) k^{-3} Z_{0; \frac{n}{k}}(k\tau),
\end{equation*}
which shows that $Z^{\inst}_{0; n}(\tau)=Z^{\gv}_{0; n}(\tau)$,
that is, the genus zero \gop invariants are nothing but the
numbers of rational instantons.
\paragraph{Genus one}
For the genus one cases, we have   
\begin{align*}
Z_{1;  n}(\tau)&=
\sum_{k|n}k^{-1}
\left(
Z^{\gv}_{1;  \frac{n}{k}}(k\tau)
+\frac{1}{12}Z^{\gv}_{0;  \frac{n}{k}}(k\tau)
\right),
\\
Z^{\gv}_{1;  n}(\tau)&=
\sum_{k|n}\mu(k)
\left(
               k^{-1} Z_{1;  \frac{n}{k}}(k\tau)
-\frac{1}{12}\,k^{-3} Z_{0;  \frac{n}{k}}(k\tau)
\right).
\end{align*}
The transformation formulae  between $Z^{\gv}_{1; n}$ and
$Z^{\inst}_{1; n}$ are 
\begin{equation}
Z^{\gv}_{1;n}(\tau)
=\sum_{k|n}Z^{\inst}_{1;  \frac{n}{k}}(k\tau),\quad
Z^{\inst}_{1;  n}(\tau)
=\sum_{k|n}\mu(k)\,Z^{\gv}_{1;  \frac{n}{k}}(k\tau).
\end{equation}
\paragraph{Genus two}
For the genus two case,
\begin{align*}
Z_{2; n}(\tau)&=
\sum_{k|n}k\left( Z^{\gv}_{2; \frac{n}{k}}(k\tau)
+\frac{1}{240}Z^{\gv}_{0; \frac{n}{k}}(k\tau)\right),\\
Z^{\gv}_{2; n}(\tau)&=
\sum_{k|n}\mu(k)\left( 
kZ_{2; \frac{n}{k}}(k\tau)
-\frac{1}{240}\,k^{-3} Z_{0; \frac{n}{k}}(k\tau)\right).
\end{align*}
%
On the other hand, we have also the  genus two instanton numbers 
$N^{\inst}_{2;n,m}$ \cite[(7.7)]{BCOV2}, 
the generating function of which
$Z^{\inst}_{2; n}(\tau)
=\sum_{m=0}^{\infty}N^{\inst}_{2;n,m}q^{m}$
is defined through
\begin{equation*}
Z_{2; n}(\tau)=Z^{\inst}_{2; n}(\tau)+\frac{1}{240}
\sum_{k|n}kZ^{\inst}_{0;\frac{n}{k}}(k\tau).
\end{equation*}
We see that the two partition functions 
$Z^{\gv}_{2;n}(\tau)$ and $Z^{\inst}_{2;n}(\tau)$  
are related each other by
\begin{equation}
Z^{\inst}_{2;n}(\tau)
=\sum_{k|n}k Z^{\gv}_{2;\frac{n}{k}}(k\tau),\quad
Z^{\gv}_{2; n}(\tau)
=\sum_{k|n}\mu(k)kZ^{\inst}_{2; \frac{n}{k}}(k\tau).
\end{equation} 
\paragraph{Genus three}
Finally for the genus three case,
{\allowdisplaybreaks 
\begin{align*}
Z_{3; n}(\tau)&=
\sum_{k|n}k^3
\left(
Z^{\gv}_{3; \frac{n}{k}}(k\tau)
-\frac{1}{12}\, Z^{\gv}_{2; \frac{n}{k}}(k\tau)
+\frac{1}{6048}\,Z^{\gv}_{0; \frac{n}{k}}(k\tau)
 \right),
\\
Z^{\gv}_{3; n}(\tau)&=
\sum_{k|n}\mu(k)
\left(
                   k^3    Z_{3; \frac{n}{k}}(k\tau)
+\frac{1}{12}\,    k      Z_{2; \frac{n}{k}}(k\tau)
-\frac{31}{60480}\,k^{-3} Z_{0; \frac{n}{k}}(k\tau)
 \right).
\end{align*}}

The formula  (\ref{BP}) enables us to 
convert the \gw partition function $Z_{g;n}(\tau)$
to the generating function of the \gop invariants 
$Z_{g;n}^{\gv}(\tau)$.
Let $N_{g;n}^{\gv}(B_N)$ be the \gop invariant 
of the local $\E_N$ del Pezzo model of genus $g$ and degree $n$.
Based on the calculation of several $Z_{g;n}^{\gv}(\tau)$
using (\ref{BP}), we propose the following conjecture
for the \gop invariants of the local $\E_9$ del Pezzo models:
\begin{alignat}{2}
&\E_0:\ \ &  N^{\gv}_{g;3n,n}&=N^{\gv}_{g;n}(\text{\bf P}^2),
\\
&\E_{\one}:\ \ &  N^{\gv}_{g;2n,n}&
=N^{\gv}_{g;n}(\text{\bf P}^1\!\times\!\text{\bf P}^1),
\\
&\E_{N}:\ \ &  N^{\gv}_{g;n,n}&=N^{\gv}_{g;n}(B_N),
\quad N=5,6,7,8.
\end{alignat}
It should be noted that
the evaluation of the left hand side
is much easier than that of the right hand side \cite{KKV}.
We will show some examples of  $Z_{g;n}^{\gv}(\tau)$ below
to see  the integrality of their $q$-expansions. 
\paragraph{$\E_7$ model}
We give the genus two \gop generating  functions.
{\allowdisplaybreaks 
\begin{align}
Z^{\gv }_{2;1}
&=
 6\,q^4
+168\,q^5
+860\,q^6
+4976\,q^7
+18660\,q^8
+72160\,q^9
+226952\,q^{10}
\nn\\
&
+712128\,q^{11}
+\cdots,
\nn\\
Z^{\gv}_{2;2}
&=
-580\,q^4
-12224\,q^5
-171192\,q^6
-1520960\,q^7
-11191692\,q^8
-67475456\,q^9
\nn\\
&
-361410816\,q^{10}
-\cdots,
\nn\\
Z^{\gv}_{2;3}
&=
 986\,q^4
+90952\,q^5
+2505136\,q^6
+43815752\,q^7
+539969082\,q^8
\nn\\
&
+5314601592\,q^9
+43546643132\,q^{10}
+\cdots,
\nn\\
Z^{\gv}_{2;4}
&=
-844\,q^4
-219392\,q^5
-14554008\,q^6
-456217600\,q^7
-9386376248\,q^8
\nn\\
&
-142590577280\,q^9
-1733995192624\,q^{10}
-\cdots,
\nn\\
Z^{\gv}_{2;5}
&=
 116880\,q^5
+22288580\,q^6
+1484462912\,q^7
+53446857696\,q^8
\nn\\
&
+1298602990944\,q^9
+23677762683308\,q^{10}
+\cdots.
\nn
\end{align}}
\paragraph{$\E_8$ model} 
We give  only the genus two and three cases. 
%
{\allowdisplaybreaks 
\begin{align}
Z^{\gv}_{2;1}
&=
  3 \,q^2
+ 772 \,q^3
+ 19467 \,q^4  
+ 257796 \,q^5  
+ 2391067 \,q^6  
+ 17484012 \,q^7
+ 107445366 \,q^8  
\nn\\
&
+ 577157904 \,q^9  
+ 2782194327 \,q^{10}
+\cdots,
\nn\\
Z^{\gv}_{2;2}
&=
-4 \,q^2  
- 25604 \,q^3  
- 3075138 \,q^4  
- 135430120 \,q^5  
- 3449998524 \,q^6
- 61300761264 \,q^7
\nn\\
&
- 839145842528 \,q^8  
- 9401698267600 \,q^9
- 89741934231984 \,q^{10}
-\cdots,
\nn\\
Z^{\gv}_{2;3}
&=
30464 \,q^3  
+ 26356767 \,q^4  
+ 4012587684 \,q^5  
+ 267561063651 \,q^6
+ 10669237946340 \,q^7  
\nn\\
&
+ 296540296415919 \,q^8
+ 6281046300189120 \,q^9  
+ 107386914608369634 \,q^{10}
+\cdots,
\nn\\
Z^{\gv}_{2;4}
&=
-26631112 \,q^4  
- 18669096840 \,q^5  
- 3493725635712 \,q^6
- 315335792669280 \,q^7  
\nn\\
&
- 17502072462748056 \,q^8
- 680822976267281568 \,q^9  
- 20119222969453708672 \,q^{10}
\nn\\
&
- 476723960943969692160 \,q^{11}   
-\cdots,
\nn\\
Z^{\gv}_{2;5}
&=
16150498760 \,q^5  
+ 11074858711765 \,q^6  
+ 2457788116576020 \,q^7
\nn\\
&
+ 280285943363605460 \,q^8
+ 20134110289153178480 \,q^9
\nn\\
&
+ 1021994028815246670450 \,q^{10}   
+\cdots,
\nn\\
Z^{\gv}_{3;1}
&=
-4\,q^3
-1038\,q^4
-28200\,q^5
-403530\,q^6
-4027020\,q^7
-31528152\,q^8
\nn\\
&
-206468416\,q^9
-1176822312\,q^{10}
-\cdots,
\nn\\
Z^{\gv}_{3;2}
&=
 1296\,q^3
+494144\,q^4
+38004700\,q^5
+1400424188\,q^6
+32782202520\,q^7
\nn\\
&
+559061195716\,q^8
+7518370093000\,q^9
+83886353406048\,q^{10}
\nn\\
&
+804126968489640\,q^{11}
+\cdots,
\nn\\
Z^{\gv}_{3;3}
&=
-1548\,q^3
-5707354\,q^4
-1607880090\,q^5
-158684891624\,q^6
-8435743979080\,q^7
\nn\\
&
-294159368706504\,q^8
-7512935612951670\,q^9
-150615781749573158\,q^{10}
\nn\\
&
-2483798853495519960\,q^{11}
-\cdots,
\nn\\
Z^{\gv}_{3;4}
&=
 5889840\,q^4
+8744913564\,q^5
+2548788575530\,q^6
+314635716180400\,q^7
\nn\\
&
+22243167756986804\,q^8
+1053665475134158016\,q^9
+36762786441521664780\,q^{10}
\nn\\
&
+1005501515252382449280\,q^{11}
+\cdots,
\nn\\
Z^{\gv}_{3;5}
&=
-7785768630\,q^5
-8996745286730\,q^6
-2835031032258700\,q^7
\nn\\
&
-420624614518458350\,q^8
-37292995978411176810\,q^9
\nn\\
&
-2255647477866896285790\,q^{10}
-101168121676653460498460\,q^{11}
-\cdots.
\nn
\end{align}}

\subsection{Partition functions as Jacobi forms}
The solution (\ref{simplify})  of the modular anomaly
equation (\ref{modular-anomaly}) implies that 
%
the partition function
$Z_{g;n}(\tau|\mu)$ is completely determined only if we could
fix the anomaly-free part of its numerator (\ref{bunshi-bunbo}),
$\jacobi_{g;n}^0(\tau|\mu)$,
which is an $\E_8$ Weyl-invariant
Jacobi form of weight $2g-2+6n$ and index $n$.
We introduce here some notation:
let ${\cal J}_{k,n}^{\varGamma}(\E_8)$ be the space
of the $\E_8$ Weyl-invariant Jacobi forms 
of weight $k$ and index $n$ for a modular group 
$\varGamma\!\subset\!\SLtwo$;
${\cal J}_{*,*}^{\varGamma}(\E_8):=
\bigoplus_{k,n}{\cal J}_{k,n}^{\varGamma}(\E_8)$
the total space of such forms,
which has a structure of 
a graded $M_{*}(\varGamma)$-algebra.
{\em Unfortunately}, we do not have the generators of 
${\cal J}_{*,*}^{\SLtwo}(\E_8)$ \cite{W} to 
fix $\jacobi^0_{g;n}(\tau|\mu)$
up to finite unknown coefficients.

However a powerful  method to generate 
certain elements of ${\cal J}_{*,*}^{\SLtwo}(\E_8)$ 
from the theta function $\TE$
has been used to obtain
$T_{0;n}^0(\tau|\mu)$ for $n=2,3,4$  in \cite{MNVW}.

We will now explain the method.
First, we note that $\TEn(\tau|\mu):=\TE(n\tau|n\mu)$
is an element of ${\cal J}_{4,n}^{\varGamma_0(n)}(\E_8)$.
Secondly, 
the slash action of $\gamma\in \SLtwo$ on
$G(\tau|\mu)\in {\cal J}_{k,m}^{\varGamma_0(n)}(\E_8)$
is defined by
\begin{equation*}
(G|\gamma)(\tau|\mu):=\frac{1}{(c\tau\!+\!d)^{k}}
\exp\left[-\frac{\pi\ti mc}{c\tau\!+\!d}(\mu|\mu)\right]
G\left(\left.\frac{a\tau\!+\!b}{c\tau\!+\!d}\right|
\frac{\mu}{c\tau\!+\!d}\right),
\quad
\gamma=
\begin{pmatrix}
a & b\\
c & d
\end{pmatrix},
\end{equation*}
which satisfies 
$(G|\gamma_1)|\gamma_2=G|(\gamma_1\gamma_2)$.
Note that $G|\gamma=G$ for any $\gamma\in \varGamma_0(n)$
by definition, and 
$G\in {\cal J}_{k,m}^{\SLtwo}(\E_8)$
if and only if $G|\gamma=G$ for any $\gamma\in \SLtwo$.
Thirdly,
consider the coset space
$\varGamma_0(n)\backslash \SLtwo$,
on which $\SLtwo$ acts as permutation from the right,
and the cardinality of which is 
$c(n):=n\prod_{p|n}(1\!+\!p^{-1})$;
in particular we can take 
$\{I,T,TS,\dots, TS^{p-1}\}$ as its representatives
if $n=p$ is prime \cite{Scho}. 
Then for $f(\tau)\in M_k(\varGamma_0(n))$,
we define $\sigma_{a}^{(n)}(f)(\tau|\mu)$ by
\begin{equation}
\sum_{a=0}^{c(n)}t^{c(n)-a}\sigma_{a}^{(n)}(f)=
\prod_{\gamma\in \varGamma_0(n)\backslash \SLtwo}
\left[t+(f\TEn|\gamma)\right],
\end{equation}
that is, $\sigma_{a}^{(n)}(f)$ 
is the $a$th basic symmetric polynomial
in $\{(f\TEn|\gamma)\}$.
{}From the argument above, we can see that 
$\sigma_{a}^{(n)}(f)\in {\cal J}_{a(k+4),an}^{\SLtwo}(\E_8)$.
Note in particular that $\sigma_1^{(n)}(1)$ 
is the $n$th Hecke transform of $\TE$.
More generally, we can see that any permutation invariant
combination gives an element of ${\cal J}_{*,*}^{\SLtwo}(\E_8)$;
for example, if 
$f_i\in M_{k_i}(\varGamma_0(n))$ for $i=1,2$,
then $\sum_{\gamma}(f_1\TEn|\gamma)\cdot(f_2\TEn|\gamma)$ 
is an element of 
${\cal J}_{8+k_1+k_2,2n}^{\SLtwo}(\E_8)$, and so on.

We will now determine doubly winding partition functions
$T_{g;2}^0(\tau|\mu)\in {\cal J}_{2g+10,2}^{\SLtwo}(\E_8)$
for lower $g$s based on the assumption
that they are obtained by the procedure 
that we have just explained above.
We need to consider only 
$\sigma_1^{(2)}\colon M_{*}(\varGamma_0(2))\rightarrow
{\cal J}_{*+4,2}^{\SLtwo}(\E_8)$,
which is a homomorphism of
$M_{*}(\SLtwo)$-modules.
It turns out that $\jacobi_{g;2}^0$ should be found in the   
free $M_{*}(\SLtwo)$-module generated by the three elements
$\sigma_1^{(2)}(AB)$,
$\sigma_1^{(2)}(B^2)$ and
$\sigma_1^{(2)}(B)=(\TE)^2$;
we do not use 
$\sigma_1^{(2)}(1)$ and
$\sigma_1^{(2)}(A)$ as generators
because the $q$-expansion of them specialized to 
the four $\E_{N\ne 7,8}$ models has a pole;
for $\E_5$ model, for example, 
\begin{align*}
\sigma_1^{(2)}(1)(4\tau|\tau\omega_5)&=
2\left(\frac{\TE(4\tau|\tau\omega_5)}
{\vat_2(2\tau)^4\vat_3(2\tau)^2}\right)^2
(9\vat_3(2\tau)^4+\vat_4(2\tau)^4),
\\
\sigma_1^{(2)}(A)(4\tau|\tau\omega_5)&=
-\frac{1}{4}
\left(\frac{\TE(4\tau|\tau\omega_5)}
{\vat_2(2\tau)^4\vat_3(2\tau)^2}\right)^2
(15\vat_3(2\tau)^8+26\vat_4(2\tau)^4\vat_3(2\tau)^4-\vat_4(2\tau)^8).
\end{align*}
Incidentally, $AB$ and $B^2$ can be expanded as
\begin{equation*}
AB=\frac{1}{3}(E_4A+E_6),
\quad
B^2=\frac{1}{9}(4E_6A+3E_4B+2E_4^2).
\end{equation*}

We list the result of our fitting including the fundamental result
of $\jacobi_{0;2}^0$ in \cite{MNVW}:
{\allowdisplaybreaks
\begin{align}
\jacobi_{0;2}^0&=\frac{1}{12}\sigma_1^{(2)}(AB),
\nn\\
\jacobi_{1;2}^0&=\frac{1}{576}\left[E_4(\TE)^2
+6\sigma_1^{(2)}(B^2)\right],
\nn\\
\jacobi_{2;2}^0&=\frac{1}{51840}\left[26E_6(\TE)^2
+63E_4\sigma_1^{(2)}(AB)\right],
\nn\\
\jacobi_{3;2}^0&=\frac{1}{11612160}
\left[445E_4^2(\TE)^2+832E_6\sigma_1^{(2)}(AB)
+1260E_4\sigma_1^{(2)}(B^2)\right],
\nn\\
\jacobi_{4;2}^0&=\frac{1}{1045094400}
\left[6692E_4E_6(\TE)^2
+13599E_4^2\sigma_1^{(2)}(AB)
+7560E_6\sigma_1^{(2)}(B^2)\right],
\nn\\
\jacobi_{5;2}^0&=\frac{1}{1655429529600}
\left[(603615E_4^3+523520E_6^2)(\TE)^2
+2249856E_4E_6\sigma_1^{(2)}(AB)\right.
\nn\\
&\left.+1844370E_4^2\sigma_1^{(2)}(B^2)\right],
\\
\jacobi_{6;2}^0&=\frac{1}{1506440871936000}
\left[123290398E_4^2E_6(\TE)^2\right.
\nn\\
&\left.+(185440941E_4^3+61328640 E_6^2)\sigma_1^{(2)}(AB)
+180540360E_4E_6\sigma_1^{(2)}(B^2)\right],
\nn\\
\jacobi_{7;2}^0&=\frac{1}{867709942235136000}
\left[(2926360905E_4^4+5095068160E_4E_6^2)(\TE)^2\right.
\nn\\
&\left.+16042720896E_4^2E_6\sigma_1^{(2)}(AB)
+(8745249240E_4^3+3358817280E_6^2)
\sigma_1^{(2)}(B^2)\right].
\nn
\end{align}}
It is then easy to recover $Z_{g;2}(\tau|\mu)$ by the solution 
(\ref{winding-2}) of the modular anomaly equation.
As a consistency check of our procedure, 
we can look into the integrality of
the \gop invariants;
indeed for the $\E_7$ model, we have
{\allowdisplaybreaks\begin{align*}
Z^{\gv (7)}_{3;2}(\tau)&=
5\,q^4+560\,q^5+18350\,q^{6}+
240736\,q^{7}+2479193\,q^{8}+\cdots,
\nn\\
Z^{\gv (7)}_{4;2}(\tau)&=
-896\,q^6-21248\,q^7-354032\,q^{8}
-3578624\,q^{9}-30445968\,q^{10}+\cdots,
\nn\\
Z^{\gv (7)}_{5;2}(\tau)&=
7\,q^6+784\,q^7+30124\,q^{8}+
443392\,q^{9}+5276873\,q^{10}+\cdots,
\nn\\
Z^{\gv (7)}_{6;2}(\tau)&=
-1228\,q^8-32064\,q^9-617904\,q^{10}
-6946048\,q^{11}-66942248\,q^{12}-\cdots,
\nn\\
Z^{\gv (7)}_{7;2}(\tau)&=9\,q^8+1008\,q^9+44450\,q^{10}+
720928\,q^{11}+9741094\,q^{12}+\cdots.
\end{align*}}   

We also find the triply winding partition function in the same manner: 
\begin{align*}
T_{0;3}^0&=\frac{1}{864}
\left[20\sigma_1^{(3)}(H^4)+972\eta^{24}\sigma_1^{(3)}(1)
-3E_4(\TE)^3 \right],
\\
T_{1;3}^0&=\frac{1}{2592}
\left[24\sigma_1^{(3)}(H^4(\varpi^{(6)})^2)-E_6(\TE)^3\right],
\end{align*}
where $T_{0;3}^0$ is again the result of \cite{MNVW}.

\section{Seiberg--Witten Curve}
\label{seiberg-witten-curve}
\subsection{Periods of rational elliptic surfaces}
Local mirror of the IIA string on $K_{B_9}$ with
the \K moduli (\ref{canonical}) is the IIB string
on the degenerate Calabi--Yau threefold 
given by the two equations in 
$(\ox,\oy,\tilde{x},\tilde{y},u)$ \cite{MNVW}:
\begin{align}
(\oy)^2&=4(\ox)^3-f(u;\tau,\mu)\ox-g(u;\tau,\mu),
\label{Weierstrass}
\\
\tilde{x}\tilde{y}&=u-u^{*},
\nn
\end{align}
where the first equation (\ref{Weierstrass}) itself
describes the family of rational elliptic surfaces
$S_9$ in Weierstrass form,
the nine moduli $(\tau,\mu)$ of which should be  encoded as
{\allowdisplaybreaks\begin{align}
f(u;\tau,\mu)&=\sum_{i=0}^{4}f_{4-i}(\tau,\mu)u^i,
\quad
f_0(\tau)=\frac{4}{3}\pi^4E_4(\tau),
\label{ef}\\
g(u;\tau,\mu)&=\sum_{i=0}^{6}g_{6-i}(\tau,\mu)u^i,
\quad
g_0(\tau)=\frac{8}{27}\pi^6E_6(\tau).
\label{gii}
\end{align}}
The determination of the precise forms of $f$ and $g$ 
for given moduli $(\tau,\mu)$ will be discussed 
in section~\ref{inverse-problem}.
The base of the elliptic fibration 
$\pi\colon S_9\mapsto \text{\bf P}^1$
is the $u$-plane and we have a rational two-form
$\varOmega=\td \ox/\oy\wedge \td u$ on it inherited  
{}from the holomorphic three-form 
$\varOmega\wedge \td \tilde{x}/\tilde{x}$
on the Calabi--Yau threefold
through the Poincar\'e residue.
Note that the pair $S_9$ given by (\ref{Weierstrass}) and 
$\varOmega$ is nothing but the ingredients
of the Seiberg--Witten curve \cite{SeibWi} 
that describes 4D E-string \cite{Ga1,GMS}. 
%
There exists a $\text{\bf C}^{*}$-action on $S_9$
which preserves the two-form $\varOmega$ \cite{Ga1,GMS}:
\begin{equation}
(\ox,\oy,u)\mapsto (\lambda^2 \ox,\lambda^3 \oy,\lambda u),
\quad \lambda\in \text{\bf C}^{*}.
\label{rescaling}
\end{equation}

Let $E_u:=\pi^{-1}(u)$ be the fiber at $u$;
the leading terms $f_0$ and $g_0$ are fixed 
by the physical requirement 
that $\Einf$, the fiber at infinity,
has the modulus $\tau$.

Let us introduce the coordinates at $u=\infty$ by
$(x,y,t)=(\ox u^2,\oy u^3,1/u)$, in terms of which the defining
equation of $\Einf$ can be written in a canonical form:
\begin{equation}
\Einf:\
y^2=4x^3-f_0(\tau)x-g_0(\tau).
\end{equation}
There exits a pair of one-cycles on $\Einf$  
$(\alpha,\beta)$ such that
\begin{equation}
\int_{\alpha}\frac{\td x}{y}=1,\qquad
\int_{\beta}\frac{\td x}{y}=\tau.
\label{futsuu-no-shuki}
\end{equation}

The rational two-form $\varOmega$ now takes the form
$\varOmega=\td t/t\wedge \td x/y$, from which we see
that the Poinca\'e residue of $\varOmega$ along $\Einf$
is nothing but the canonical one-form on $\Einf$:
\begin{equation}
\text{Res}_{\Einf}(\varOmega)
=\frac{\td x}{y}.
\label{residue}
\end{equation}

We can identify $\Einf$ with the complex torus
$\text{\bf C}/(\text{\bf Z}\tau\!+\!\text{\bf Z})$ through the uniformization
$(x,y)=(\wp(\tau|\nu),\wp'(\tau|\nu))$,
where $\nu$ is the coordinate of the covering space
$\text{\bf C}$ of the torus and 
the Weierstrass $\wp$ function is defined by 
\[
\wp(\tau|\nu)=\frac{1}{\nu^2}
+\underset{\omega\in \text{\bf Z}\tau+\text{\bf Z}}{{\sum}'}
\left[\frac{1}{(\nu-\omega)^2}-\frac{1}{\omega^2}\right].
\]
Note that $\td x/y$ is pulled-back by this isomorphism 
to $\td \nu$, which makes (\ref{futsuu-no-shuki}) rather trivial.
The point $\nu\!\mod{\text{\bf Z}\tau+\text{\bf Z}}$ of the torus
will be frequently used to refer to 
the point $(\wp(\tau|\nu),\wp'(\tau|\nu))$ of $\Einf$
below.

We also introduce the homogeneous coordinates $(x_0,x_1,x_2)$
of $\text{\bf P}^2$ with  $x=x_1/x_0$, $y=x_2/x_0$, 
so that we can realize $\Einf$ as a plane curve 
defined by the ternary cubic $P$:
\begin{equation}
P(x_0,x_1,x_2):=x_0x_2^2-4x_1^3+f_0(\tau)x_0^2x_1+g_0(\tau)x_0^3.
\label{original-cubic}
\end{equation} 

The twelve normalized periods 
$(1,\tau,\sigma,\mu,\partial_{\sigma} F_0)$
of the E-string can be obtained as the periods 
of the Seiberg--Witten curve (\ref{Weierstrass}) 
\cite{Ga1,GMS,LMW,MNW,MNVW,HI}.

To see this,
let $(\alpha(u),\beta(u))$ be the standard symplectic basis of
the fiber $H_1(E_u)$ which extends the one 
$(\alpha,\beta)$ at $u=\infty$.
It is clear from (\ref{futsuu-no-shuki}) and (\ref{residue}) 
that the two periods $1$, $\tau$ can be recovered by the integrals
\cite{HI}:
\begin{equation}
1=-\frac{1}{2\pi \ti}
\oint_{u=\infty}\td u
\oint_{\alpha(u)}
\frac{\td \ox}{\oy},
\qquad
\tau
=-\frac{1}{2\pi \ti}
\oint_{u=\infty}\td u \oint_{\beta(u)}
\frac{\td \ox}{\oy}.
\label{torus-no-shuki}
\end{equation}

Evaluation of the Wilson lines
$\mu=\sum_{i=1}^{8}\mu_i\omega_i$   
needs a more careful study of periods,
since it deals with a seemingly divergent
integral due to the pole of $\varOmega$.

Notice first that $\varOmega$ is a holomorphic two-form on 
$S_9\!-\!\Einf$. Then we are naturally lead to
define the period map $\hat{\varrho}$ \cite{Lo} of $\varOmega$ by 
\begin{equation*}
\hat{\varrho}\colon H_2(S_9\!-\!\Einf)
\lr \text{\bf C},
\qquad
\hat{\chi}(D):=
\frac{1}{2\pi\ti}
\int_{D}\varOmega.
\end{equation*}
The non-trivial part of the homology exact sequence of the pair
$(S_9,S_9\!-\!\Einf)$:
\[
\cdots  \lr            H_i(S_9\!-\!\Einf)
\lr                    H_i(S_9)
\lr                    H_i(S_9,S_9\!-\!\Einf)
 \lr                   H_{i-1}(S_9\!-\!\Einf)
\lr \cdots,
\]
can be written as
\begin{equation}
0\to H_1(\Einf)
\overset{\partial_*}{\lr}
H_2(S_9\!-\!\Einf)
\overset{i_*}{\lr} 
H_2(S_9)
\overset{j_*}{\lr} 
H_0(\Einf)\to 0,
\end{equation}
where we used the Poincar\'e duality 
$H_i(S_9,S_9\!-\!\Einf)\cong H^{4-i}(\Einf)$
and $H_1(S_9\!-\!\Einf)\cong 0$.

There are two points:
first, $\partial_{*}\alpha$ and $\partial_{*}\beta$
are just the elements of $H_2(S_9\!-\!\Einf)$
appeared in (\ref{torus-no-shuki}), from which 
it follows immediately that  
$\hat{\varrho}(\partial_*\alpha)=1$, and 
$\hat{\varrho}(\partial_*\beta)=\tau$, that is,
$\hat{\varrho}\cdot \partial_{*}$
is nothing but the period map of 
$\Einf$ by $\td x/y$;
second, 
$j_{*}\colon  H_2(S_9)\to H_0(\Einf)\cong \text{\bf Z}$
simply counts the intersection number 
of a divisor with $\Einf$.
As the homology class of $\Einf$ is $\fiber$, 
we see that  
$\text{Ker}\,j_{*}=\text{Im}\,i_{*}$ 
can be identified with $L(\E_8^{(1)})$.  

We conclude that  $\hat{\varrho}$ induces the homomorphism
of additive groups:
\begin{equation}
\varrho:L(\E_8^{(1)})\rightarrow 
\text{\bf C}\mod{\text{\bf Z}\tau+\text{\bf Z}}.
\end{equation}
In other words, we can regularize systematically 
the integral of the rational form $\varOmega$ 
over $L(\E_8^{(1)})$ 
at the expense of the additive ambiguity   
$\text{\bf Z}\tau+\text{\bf Z}$. 


It is possible to describe $\chi$
quite  explicitly in terms of the 
moduli of $S_9$.
Let 
$p_i:=(\wp(\tau|\nu_i),\wp'(\tau|\nu_i))$
be the intersection point in $S_9$  of $\Einf$ with 
the $i$th exceptional divisor $\ed_i$.
Then the rational elliptic surface $S_9$ is obtained 
by blow-up of $\text{\bf P}^2$ at these nine points, 
which also satisfy  
\begin{equation}
\sum_{i=1}^{9}\nu_i=0
\mod{\text{\bf Z}\tau+\text{\bf Z}}
\label{wagazero}
\end{equation}
because they are the intersection points of
two cubics in $\text{\bf P}^2$, that is, there exists 
another cubic $Q$ such that 
$\{\nu_i\}=\{P=0\}\cap \{Q=0\}$.
The rational elliptic surface $S_9$ in question is then 
expressed as the hypersurface 
$P+tQ=0$ in $\text{\bf P}^2\!\times\!\text{\bf P}^1$;
this is the same as the realization of the $\E_0$ model,
but it is the complex structure that matters here.

We can show that $\varrho$ is given by \cite{Lo,Sakai} 
\begin{align}
\varrho(\ed_i\!-\!\ed_j)&=\nu_i-\nu_j,  
\label{zensha}
\\
\varrho(l\!-\!\ed_i\!-\!\ed_j\!-\!\ed_k)&=-\nu_i-\nu_j-\nu_k,
\label{kousha}
\end{align}
where $\nu_i$s are defined only up to addition of 
$\text{\bf Z}\tau+\text{\bf Z}$. 

Let us see how  (\ref{zensha}) is obtained.   
Choose a path $\gamma_{i,j}$ which connects
$p_j$ and $p_i$ on $\Einf$ and 
let $T_{i,j}$ be a closed tubular neighbourhood  
of $\gamma_{i,j}$ in $S_9\!-\!\Einf$,
such that each of
$\ed_i\cap T_{i,j}=D_i$ and 
$\ed_j\cap T_{i,j}=D_j$ is its fiber. 
Then $(\ed_i\!-\!D_i)\cup
\partial T_{i,j}
\cup(\ed_j\!-\!D_j)$ becomes an oriented topological manifold 
homologous to $\ed_i-\ed_j$ and disjoint from $\Einf$.
In physical terms, 
$\partial T_{i,j}$ is a ``wormhole''
connecting the two ``universes'' $\ed_i$ and $\ed_j$,
with the singular points 
$p_i$, $p_j$ replaced by a ``black and white hole''.
Now the evaluation of the left hand side of (\ref{zensha}) 
proceeds as follows:
\begin{equation*}
 \frac{1}{2\pi \ti}\int_{\ed_i-\ed_j}\varOmega
=\frac{1}{2\pi \ti}\int_{\partial T_{i,j}}\varOmega
=\int_{\gamma_{i,j}}\text{Res}_{\Einf}\varOmega
=\int_{p_j}^{p_i}\frac{\td x}{y},
\end{equation*}
which yields precisely the right hand side of (\ref{zensha}),
the ambiguity of which comes from 
the choice of the path $\gamma_{i,j}$.

As for (\ref{kousha}), 
we first combine the divisor in question as 
$(l\!-\!\ed_i\!-\!\ed_j)-\ed_k$.
We can then take as $l$ 
the transform of the line on $\text{\bf P}^2$
that passes through the two points $\{p_i,p_j\}$.
Now we see that $(l\!-\!\ed_i\!-\!\ed_j)$ is an effective curve
intersecting with $\Einf$ at 
$-\!p_i\!-\!p_j$, the coordinates of which are 
$(\wp(\tau|\nu),\wp'(\tau|\nu))$, with  
$\nu:=\!-\!\nu_i\!-\!\nu_j$.
At this point, the problem is reduced to (\ref{zensha}),
so that (\ref{kousha}) follows.
We note that as a consequence of (\ref{kousha}), 
\begin{align*}
\varrho(\fiber)&=
\varrho(l\!-\!\ed_1\!-\!\ed_2\!-\!\ed_3)
+\varrho(l\!-\!\ed_4\!-\!\ed_5\!-\!\ed_6)
+\varrho(l\!-\!\ed_7\!-\!\ed_8\!-\!\ed_9)
\\
&=-\sum_{i=1}^{9}\nu_i=0 \mod{\text{\bf Z}\tau+\text{\bf Z}},
\end{align*}
which is consistent with the fact 
that we can take a generic fiber 
$E_u$ as a representative of $\fiber$, thus 
$\varOmega|_{E_u}$ vanishes identically,
since $E_u$ is a holomorphic curve disjoint from $\Einf$. 

We do not have much to say about the Seiberg--Witten periods
\cite{SeibWi}, $\sigma$ and $\partial_{\sigma} F_0$.
These are {\it not} periods of $S_9$ in the sense above,
and roughly given by
\begin{equation*}
\sigma=-\int\td u  \oint_{\alpha(u)}\frac{\td \ox}{\oy},
\quad
\frac{1}{(2\pi \ti)^2}
\left(\frac{\partial F_0}{\partial \sigma}\right)
 =\int\td u \left(
 \oint_{\beta(u)}\frac{\td \ox}{\oy}
-\tau \int_{\alpha(u)}\frac{\td \ox}{\oy}
\right),
\end{equation*}
a detailed account of which we have already given 
for the six two-parameter models in
(\ref{phi0}), (\ref{phi1}) and (\ref{phiN}) for $\sigma$
and in (\ref{tokuni0}), (\ref{tokuni1}), and (\ref{tokuni})
for $\partial_{\sigma}F_0$,
with the correspondence of the bare parameters
$z_1\propto u^{-1}$ in mind.
Explicit evaluation and instanton expansion
of these  Seiberg--Witten integrals in terms of modular forms
has been done for the $\E_8$ model in \cite{MNW}. 

\subsection{Wilson lines}
Based on the results in the last subsection,
we find the Wilson lines \cite{SeibWi,Ga1,GMS,MNVW} to be
\begin{equation}
\mu_i=\frac{1}{2\pi \ti}\,  \varrho([\alpha_i])
=\begin{cases}
\nu_i-\nu_{i+1},\quad   &i=1,\dots,7,\\
-\nu_1-\nu_2-\nu_3,\quad  &i=8,
\end{cases}
\end{equation}
where $[\alpha_i]\in H_2(S_9)$ is defined in (\ref{bunkai}).

The Euclidean coordinates (\ref{euclid})
of the Wilson lines $(m_i)$ 
and the base points of the cubic pencil 
$\nu_i=\Einf\cap \ed_i$ 
are related to each other by
\begin{align}
\begin{pmatrix}
m_1\\
m_2\\
m_3\\
m_4\\
m_5\\
m_7\\
m_8
\end{pmatrix}
&=-\frac{1}{2}
\begin{pmatrix}
1 & 2 & 0 & 0 & 0 & 0 & 0 & 0 \\ 
1 & 0 & 2 & 0 & 0 & 0 & 0 & 0 \\ 
1 & 0 & 0 & 2 & 0 & 0 & 0 & 0 \\ 
1 & 0 & 0 & 0 & 2 & 0 & 0 & 0 \\ 
1 & 0 & 0 & 0 & 0 & 2 & 0 & 0 \\ 
1 & 0 & 0 & 0 & 0 & 0 & 2 & 0 \\ 
1 & 0 & 0 & 0 & 0 & 0 & 0 & 2 \\ 
1 & 2 & 2 & 2 & 2 & 2 & 2 & 2 
\end{pmatrix}
\begin{pmatrix}
\nu_1 \\
\nu_2 \\
\nu_3  \\
\nu_4 \\
\nu_5  \\
\nu_6 \\
\nu_7 \\
\nu_8
\end{pmatrix},
\label{m-nu}
\\
\begin{pmatrix}
\nu_1 \\
\nu_2 \\
\nu_3  \\
\nu_4 \\
\nu_5  \\
\nu_6 \\
\nu_7 \\
\nu_8  
\end{pmatrix}
&=\frac{1}{6}
\begin{pmatrix}
-2 & -2 & -2 & -2 & -2 & -2 & -2 & \sk 2 \\
-5 & \sk 1 & \sk 1 & \sk 1 & \sk 1 & \sk 1 & \sk 1 & -1 \\
\sk 1 & -5 & \sk 1 & \sk 1 & \sk 1 & \sk 1 & \sk 1 & -1 \\
\sk 1 & \sk 1 & -5 & \sk 1 & \sk 1 & \sk 1 & \sk 1 & -1 \\
\sk 1 & \sk 1 & \sk 1 & -5 & \sk 1 & \sk 1 & \sk 1 & -1 \\
\sk 1 & \sk 1 & \sk 1 & \sk 1 & -5 & \sk 1 & \sk 1 & -1 \\
\sk 1 & \sk 1 & \sk 1 & \sk 1 & \sk 1 & -5 & \sk 1 & -1 \\
\sk 1 & \sk 1 & \sk 1 & \sk 1 & \sk 1 & \sk 1 & -5 & -1
\end{pmatrix}
\begin{pmatrix}
m_1\\
m_2\\
m_3\\
m_4\\
m_5\\
m_7\\
m_8
\end{pmatrix}.
\label{nu-m}
\end{align}

\section{Inverse Problem}
\label{inverse-problem}
\subsection{General strategy}
We are interested in the following problem:
given the eight Wilson lines $\mu=\sum_{i=1}^{8}\mu_i\omega_i$, 
or equivalently the nine  points $(\nu_i)_{i=1}^{9}$
on the torus $\text{\bf C}/(\text{\bf Z}\tau\!+\!\text{\bf Z})$,
which satisfy (\ref{wagazero}),
find the corresponding Seiberg--Witten curve
(\ref{Weierstrass}).
In other words, we want to know explicitly the two functions
$f(u;\tau,\mu)$ in (\ref{ef}) and $g(u;\tau,\mu)$ in (\ref{gii})
as functions of the moduli.

We will solve this problem in two steps \cite{GMS}:
in the first step, we obtain the cubic pencil
$P+tQ$ in $\text{\bf P}^2$  for the given base points $(\nu_i)$,
which is achieved by consideration based on 
the elliptic function theory;
and in the second step, we transform the cubic pencil
into the Weierstrass form (\ref{Weierstrass}),
with the help of the classical theory of algebraic invariants.

\paragraph{First step}
We claim that the curve defined by the 
cubic $Q(x_0,x_1,x_2)$ shown below
passes through the nine points 
$\{(1,\wp(\tau|\nu_i),\wp'(\tau|\nu_i))\}$, where 
$(\nu_i)$ are taken to be generic except 
(\ref{wagazero}):
\begin{equation}
Q=
\begin{vmatrix}
x_0^3   & x_0^2x_1  & x_0^2x_2  &  x_0x_1^2  & x_0x_1x_2
& x_0x_2^2 & x_1^2x_2   &  x_1x_2^2   & x_2^3 \\
1 & \wp_1 & \wp_1'  & \wp_1^2  & \wp_1\wp_1' & (\wp_1')^2 
& \wp_1^2\wp_1' & \wp_1(\wp_1')^2 & (\wp_1')^3\\
1 & \wp_2 & \wp_2'  & \wp_2^2 & \wp_2\wp_2'& (\wp_2')^2 
& \wp_2^2\wp_2' & \wp_2(\wp_2')^2 & (\wp_2')^3\\
1 & \wp_3 & \wp_3'  & \wp_3^2 & \wp_3\wp_3'& (\wp_3')^2 
& \wp_3^2\wp_3' & \wp_3(\wp_3')^2 & (\wp_3')^3\\
1 & \wp_4 & \wp_4'  & \wp_4^2 & \wp_4\wp_4'& (\wp_4')^2 
& \wp_4^2\wp_4' & \wp_4(\wp_4')^2 & (\wp_4')^3\\
1 & \wp_5 & \wp_5'  & \wp_5^2 & \wp_5\wp_5'& (\wp_5')^2 
& \wp_5^2\wp_5' & \wp_5(\wp_5')^2 & (\wp_5')^3\\
1 & \wp_6 & \wp_6'  & \wp_6^2 & \wp_6\wp_6'& (\wp_6')^2 
& \wp_6^2\wp_6' & \wp_6(\wp_6')^2 & (\wp_6')^3\\
1 & \wp_7 & \wp_7'  & \wp_7^2 & \wp_7\wp_7'& (\wp_7')^2 
& \wp_7^2\wp_7' & \wp_7(\wp_7')^2 & (\wp_7')^3\\
1 & \wp_8 & \wp_8'  & \wp_8^2 & \wp_8\wp_8'& (\wp_8')^2 
& \wp_8^2\wp_8' & \wp_8(\wp_8')^2 & (\wp_8')^3
\end{vmatrix},
\label{ippan-kai}
\end{equation}   
where we have used the abbreviated notation
$\wp_i=\wp(\tau|\nu_i)$, $\wp_i'=\wp'(\tau|\nu_i)$.

The proof is quite simple if we recall some  fundamental
theorems of the elliptic function theory \cite{HC};
first $h(\tau|\nu):=Q(1,\wp(\tau|\nu),\wp'(\tau|\nu))$ 
is an elliptic function with the only pole 
of ninth order at $\nu\!=\!0$;
then as $h(\tau|\nu)$ has eight simple zeros at 
$\nu\!=\!\nu_1,\dots,\nu_8$ by construction,
the ninth zero should be $\nu\!=\!-\sum_{i=1}^{8}\nu_i\!=\!\nu_9$.

In fact, $h(\tau|\nu)$ admits the following concise expression 
\cite[III, p.98--99]{TM}: 
\begin{equation}
h(\tau|\nu_0)=2^{10}\,
\frac{\sigma(\tau|\sum_{i=0}^{8}\nu_i)
\prod_{i<j}\sigma(\tau|\nu_j\!-\!\nu_i)}%
{\prod_{i=0}^{8}\sigma(\tau|\nu_i)^9},
\label{hnokosiki}
\end{equation}
where $\sigma(\tau|\nu)$ is the Weierstrass sigma function
\[ 
\sigma(\tau|\nu):=\nu
\underset{\omega\in \text{\bf Z}\tau+\text{\bf Z}}{{\prod}'}
\left(1-\frac{\nu}{\omega}\right)
\text{e}^{(\frac{\nu}{\omega})+\frac12(\frac{\nu}{\omega})^2}
=\frac{1}{2\pi}\,\text{e}^{\frac16(\pi\nu)^2E_2(\tau)}
\frac{\vat_1(\tau|\nu)}{\eta(\tau)^3}.
\]
The $E_2(\tau)$ factors in the sigma functions
cancel out in (\ref{hnokosiki}).
Curiously, we encounter the same function
$\vat_1(\tau|\nu)/\eta(\tau)^3$ as in (\ref{ansatz-for-genus-g}).

The cubic curve $Q=0$ intersects with $\Einf$
at the nine points $\{(1,\wp_i,\wp_i')\}$.
Note that $Q$ never coincides with the original cubic $P$
since $Q$ lacks the $x_1^3$ term \cite{GMS}.

Therefore we have found 
for the given moduli parameters $(\tau,\mu)$
the rational elliptic surface $S_9$
in the form of a cubic pencil $P+tQ$, 
where $P$ is given in (\ref{original-cubic})
and $Q$ in (\ref{ippan-kai}).

If $(\nu_i)$ are not generic, e.g., if $\nu_i=\nu_j$, 
then the right hand side of (\ref{ippan-kai})
vanishes identically. However an appropriate limiting procedure
such as $\nu_j\to \nu_i$ should still produce a non-trivial solution.
In fact, in the following subsection we treat several models
with degenerate Wilson line parameters, 
where $Q$ factorizes into lower degree polynomials.

\paragraph{Second step}
What is really in need is the Seiberg--Witten form 
(\ref{Weierstrass}) of $S_9$,
which is given by the cubic pencil at this point.
In principle, we can find a coordinate transformation 
of $(x_0,x_1,x_2)$ by $GL(3;\text{\bf C})$
which takes the cubic pencil $P+tQ$ above into
the Weierstrass form \cite{GMS},
but it seems very difficult to perform this task in general.
We can in fact skip this difficulty to reach 
the Weierstrass form (\ref{Weierstrass}) directly.

To see this, first recall that 
there exists a natural action of 
$GL(3;\text{\bf C})$ on the space of the
ternary cubic forms, a general member $R$ of which we 
write as
\begin{equation}
R:=\sum_{p+q+r=3}\left(\frac{3!}{p!q!r!}\right)
a_{pqr}\,x_0^px_1^qx_2^r.
\label{general-cubic}
\end{equation}
It is well-known that the ring of the projective invariants
are the polynomial ring over $\text{\bf C}$ 
generated by the two basic invariants ${\cal S}$ and
${\cal T}$, that is,
$(\text{\bf C}[a_{pqr}])^{{PGL}(3;\text{\bf C})}
=\text{\bf C}[{\cal S},{\cal T}]$
which can be obtained by the formula \cite[II.7]{Hilb}:
\begin{align}
\text{Hess}\left\{\alpha R+\frac{\beta}{36}\text{Hess}(R)\right\}
&=
\left(432\,\alpha^2\beta{\cal S}+1728\,\beta^3{\cal S}^2
-216\,\alpha\beta^2{\cal T}\right)R
\nn \\
&+\left(\alpha^3+2\,\beta^3{\cal T}-12\,\alpha\beta^2{\cal S}\right)
\text{Hess}(R),
\end{align}
where $\text{Hess}(R):=
|(\partial_{i}\partial_{j}R)|$
is the Hessian of $R$, which is another cubic.
We give in (\ref{es}) and (\ref{tee}) the explicit forms of 
these two invariants, which indeed coincide with those 
in \cite[Prop.4.4.7]{St} and \cite[Exm.4.5.3]{St}.

Any {\em generic} cubic $R$ can then be transformed by  
$GL(3;\text{\bf C})$ to the Weierstrass form: 
\begin{equation}
R\mapsto x_2^2x_0-4x_1^3+fx_1x_0^2+gx_0^3,
\qquad
f=\frac{27}{4}{\cal S},\ 
g=\frac{27}{64}{\cal T}.
\end{equation}  
This technique enables us to find the Seiberg--Witten form 
(\ref{Weierstrass}) of the rational elliptic surface $S_9$
given by a cubic pencil $P+tQ$.

\paragraph{Fixed Wilson lines}
In the remaining of this subsection, 
we take two models with fixed Wilson lines
as a warming-up exercise.
 
Let us first take the $\E_8$ model the Wilson lines of which 
are $\nu_i=0$ for all $i$.
Thus we need a cubic $Q(x_0,x_1,x_2)$ which intersects
with $\Einf$ nine times at $(x_0,x_1,x_2)=(0,1,0)$
corresponding to $\nu=0$.
The triple line $Q=x_0^3$ does the job, simply because
$\nu=0$ is one of the inflection points of $\Einf$.

The resulting cubic pencil 
$P+tQ$ can easily be converted to the Weierstrass form 
in the original variables
$(\ox,\oy)$:
\begin{equation}
(\oy)^2=4(\ox)^3-f_0(\tau)u^4\ox
-(g_0(\tau)u^6+u^5),
\label{SW-e-hachi}
\end{equation}
which is the Seiberg--Witten curve of the $\E_8$ model
found in \cite{GMS}.

Next consider the model with the nine inflection points
$\{\tfrac{1}{3}(a\tau+b)|a,b=0,1,2\}$
as the base points of the cubic pencil.
Note that these nine points sum up to zero.

We can find easily also in this case the cubic $Q$;
it is simply the Hessian of $P$:
\begin{equation}
\text{Hess}(P)=-8\left(f_0(\tau)^2x_0^3+36g_0(\tau)x_0^2x_1
+12f_0(\tau)x_0x_1^2-12x_1x_2^2\right).
\end{equation} 
This follows from the fact that  the four $x$ coordinates 
$\{\wp(\tau|\frac{1}{3}(a\tau\!+\!b))|(a,b)\ne (0,0)\}$ 
are the roots of the equation:
$4x^4-2f_0(\tau)x^2-4g_0(\tau)x-f_0(\tau)^2/12=0$.

The calculation of the basic algebraic invariants
${\cal S}$, ${\cal T}$ of the cubic $P+t\text{Hess}(P)$
yields the Weierstrass form (\ref{Weierstrass}) 
 of the cubic pencil with
\begin{align}
f(u;\tau)=&\frac{4}{3}\pi^4
\left(E_4(\tau)u^4-4E_6(\tau)u^3+6E_4(\tau)^2u^2-4E_4E_6(\tau)u
+4E_6(\tau)^2-3E_4(\tau)^3\right),
\nn\\
g(u;\tau)=&\frac{8}{27}\pi^6
\left(E_6(\tau)u^6-6E_4(\tau)^2u^5+15E_4E_6(\tau)u^4
-20E_6(\tau)^2u^3+15E_4^2E_6(\tau)u^2\right.
\nn\\
&\left.
+6E_4(2E_6^2-3E_4^3)(\tau)u
+E_6(9E_4^3-8E_6^2)(\tau)
\right).
\end{align} 
Recall here that 
$f_0(\tau)=4/3\pi^4E_4(\tau)$,
$g_0(\tau)=8/27\pi^6E_6(\tau)$.

We can see that the partition function 
of the singly-winding sector $Z_{g,1}(\tau)$ vanishes
identically for this model due to the theta function 
formula \cite[II, p.148]{TM}, the physical meaning of which
is yet to be clarified.

\subsection{Several models with a few Wilson lines}
 \paragraph{Two Wilson lines}
The model with two Wilson lines, that is,
$\mu=m_1\text{\bf e}_1+m_2\text{\bf e}_2$ has been investigated 
in \cite{MNVW}, where the Seiberg--Witten curve were found
and the partition function $Z_{0,1}(\tau|m_i)$ 
has been derived directly from it.

We find 
the nine base points of the cubic pencil representing
the rational elliptic surface to be 
$\{\nu_i\}=\{0,0,0,0,0, m_{+},-m_{+},m_{-},-m_{-}\}$, 
where $m_{\pm}=(m_1\pm m_2)/2$.
The cubic $Q$ which passes through these nine points
can easily be identified. In fact, the nine points are 
decomposed into the three triples 
$\{0,0,0\}$,
$\{0,m_{+},-m_{+}\}$,
$\{0,m_{-},-m_{-}\}$,
each of which defines a line in $\text{\bf P}^2$:
\begin{equation}
L_0=x_0,\quad
L_{+}=x_1-\wp(\tau|m_{+})x_0,\quad
L_{-}=x_1-\wp(\tau|m_{-})x_0.
\end{equation}
Thus the cubic $Q$ is found to be 
$Q=x_0(x_1-\wp(\tau|m_{+})x_0)(x_1-\wp(\tau|m_{-})x_0)$.

Now calculation of the basic algebraic invariants 
of the ternary cubic $P+tQ$
tells  us its Weierstrass form (\ref{Weierstrass}):
\begin{align}
f(u;\tau,m_{\pm})&=f_0u^4
-(\wp_{+}+\wp_{-})u^3+\frac{1}{12}u^2,\\
g(u;\tau,m_{\pm})&=g_0u^6+
(\wp_{+}\wp_{-}+\frac{f_0}{12})u^5
-\frac{1}{12}(\wp_{+}+\wp_{-})u^4+\frac{1}{216}u^3,
\end{align}
where $\wp_{\pm}=\wp(\tau|m_{\pm})$.
Indeed, using the formula 
\[
\frac{1}{\pi^2}\wp(\tau|m)=
-\frac{1}{3}(\vat_3(\tau)^4+\vat_2(\tau)^4)
+\vat_3(\tau)^2\vat_2(\tau)^2
\left(\frac{\vat_4(\tau|m)}{\vat_1(\tau|m)}\right)^2,
\]
and a rescaling (\ref{rescaling}), it can be shown that
the above form coincides with
the one in \cite[(A.1)]{MNVW}, which was obtained 
through a reasoning quite different from the one here, 
modulo some misprints.

\paragraph{Three Wilson lines}
Consider the model with the Wilson lines 
$\mu=\sum_{i=1}^{3}m_i\,\text{\bf e}_i$, 
where $m_i$ are chosen to be generic.
A $W(\E_8)$ action simplifies the nine base points of the cubic pencil
to be
$\{\nu_i\}=\{0,0,0,0,0,\zeta_0,\zeta_1,\zeta_2,\zeta_3\}$, where
\begin{equation}
(\zeta_0,\zeta_1,\zeta_2,\zeta_3)
=\frac{1}{2}
(m_1,m_2,m_3)
\begin{pmatrix}
-1 & -1 & \sk 1 & \sk 1 \\
-1 & \sk 1 & -1 & \sk 1 \\
-1 & \sk 1 & \sk 1 & -1
\end{pmatrix}.
\end{equation}
We see immediately that $\{0,0,0\}$ 
determines the line $x_0=0$ and the other six points
$\{0,0,\zeta_0,\zeta_1,\zeta_2,\zeta_3\}$
the conic of the form $C=a_0x_0^2+a_1x_0x_1+a_2x_0x_2+a_3x_1^2$,
the precise coefficients of which are determined by the formula
\begin{align}
&C(x_0,x_1,x_2)=
\begin{vmatrix}
x_0^2 & x_0x_1 &  x_0x_2 & x_1^2\\
1     & \wp_1  &  \wp_1' & \wp_1^2 \\
1     & \wp_2   & \wp_2' & \wp_2^2 \\
1     & \wp_3   & \wp_3' & \wp_3^2 
\end{vmatrix},
 \\
a_0=&
\begin{vmatrix}
\wp_1  & \wp_1' & \wp_1^2 \\
\wp_2  & \wp_2' & \wp_2^2 \\
\wp_3 & \wp_3'  & \wp_3^2
\end{vmatrix},
\
a_1=
-\begin{vmatrix}
1  & \wp_1' & \wp_1^2 \\
1  & \wp_2' & \wp_2^2 \\
1 & \wp_3'  & \wp_3^2
\end{vmatrix},
\
a_2=
\begin{vmatrix}
1 & \wp_1   & \wp_1^2 \\
1 & \wp_2   & \wp_2^2 \\
1 & \wp_3  & \wp_3^2
\end{vmatrix},
\
a_3=-\begin{vmatrix}
1 &\wp_1  & \wp_1'  \\
1 & \wp_2  & \wp_2'  \\
1 & \wp_3 & \wp_3'  
\end{vmatrix},
\nn
\end{align}
where $\wp_i=\wp(\tau|\zeta_i)$, $\wp_i'=\wp'(\tau|\zeta_i)$.

The cubic passing through the nine points are
determined to be 
\[Q=x_0(a_0x_0^2+a_1x_0x_1+a_2x_0x_2+a_3x_1^2)\]
and the Seiberg--Witten form (\ref{Weierstrass})
of it is written in terms of $(a_0,a_1,a_2,a_3)$ as 
\begin{align}
f(u;\tau,m_i)&=f_0u^4+a_1u^3+\frac{1}{12}a_3^2u^2,\\
g(u;\tau,m_i)&=g_0u^6+(a_0+\frac{f_0}{12}a_3)u^5
+\frac{1}{12}(a_1a_3-3a_2^2)u^4+\frac{1}{216}a_3^3u^3.
\end{align}
It is an amusing exercise to see
that under the limit 
$\text{Im}\,\tau\to +\infty$:
\begin{equation*}
\wp(\tau|\zeta) \rightarrow -\frac{\pi^2}{3}
+\frac{\pi^2}{\sin^2(\pi\zeta)},
\quad
\wp'(\tau|\zeta) \rightarrow 
-\frac{2\pi^3\cos(\pi\zeta)}{\sin^3(\pi\zeta)},
\end{equation*}
the curve above reduces to the trigonometric one 
with three Wilson lines \cite[(2.8)]{MNW2}.
%
\paragraph{Three Wilson lines II}
If we consider the model with 
$\mu=
 m_1(\text{\bf e}_1-\text{\bf e}_2)
+m_2(\text{\bf e}_3-\text{\bf e}_4)
+m_3(\text{\bf e}_5-\text{\bf e}_6)$,
the nine base points of the cubic pencil are given by
$\{0,0,0,\pm m_1,\pm m_2,\pm m_3\}$.
As the line intersecting with $\Einf$
with the three points $\{0,\pm m_i\}$
is given by $x_1-\wp_ix_0=0$, where we set 
$\wp_i:=\wp(\tau|m_i)$,
the cubic $Q$ we need is found to be
\begin{equation}
Q=(x_1-\wp_1x_0)(x_1-\wp_2x_0)(x_1-\wp_3x_0).
\end{equation}

The computation of the algebraic invariants of the cubic pencil
$P+tQ$ yields the Weierstrass form (\ref{Weierstrass}) with
\begin{align}
f(u;\tau,m_i)&=f_0u^4+\frac14(4\sigma_2-f_0)u^3+\frac{1}{12}
(\sigma_1^2-3\sigma_2)u^2,
\\
g(u;\tau,m_i)&=g_0u^6-\frac{1}{12}(f_0\sigma_1+\sigma_3)u^5
+\frac{1}{48}(3g_0+f_0\sigma_1-4\sigma_1\sigma_2)u^4
\nn\\
&-\frac{1}{432}(2\sigma_1^3-9\sigma_1\sigma_2+27\sigma_3)u^3,
\end{align}
where
$\sigma_1=\wp_1+\wp_2+\wp_3$,
$\sigma_2=\wp_1\wp_2+\wp_2\wp_3+\wp_3\wp_1$,
and $\sigma_3=\wp_1\wp_2\wp_3$.

\paragraph{Four Wilson lines}
Let us consider the model with the Wilson lines
given in the Euclidean coordinates 
$\mu=\sum_{i=1}^{4}m_i\,\text{\bf e}_i$,
where any of $m_i$ or $m_i\!-\!m_j$ is non-zero.
After a suitable $W(\E_8)$ action,
the nine base points 
of the cubic pencil become
$\{\nu_i\}=\{-2\zeta_0,\zeta_0,\zeta_0,
\zeta_0,\zeta_0,\zeta_1,\zeta_2,\zeta_3,\zeta_4\}$,
where
\begin{equation}
(\zeta_0,\zeta_1,\zeta_2,\zeta_3,\zeta_4)
=\frac{1}{6}
(m_1,m_2,m_3,m_4)
\begin{pmatrix}
 1  & -5 & \sk 1 & \sk 1 & \sk 1 \\
 1  & \sk 1 & -5  & \sk 1 & \sk 1 \\
 1  & \sk 1 & \sk 1  & -5 & \sk 1 \\
 1  & \sk 1 & \sk 1  & \sk 1 & -5 
\end{pmatrix}.
\end{equation}
The first three points $\{-2\zeta_0,\zeta_0,\zeta_0\}$
defines a line $L$ tangent to the elliptic curve $\Einf$,
while the remaining six a conic $C$.
If we set $\wp_i=\wp(\tau|\zeta_i)$, $\wp_i'=\wp'(\tau|\zeta_i)$, 
$\wp_{00}=\wp(\tau|2\zeta_0)$, $\wp_{00}'=-\wp'(\tau|2\zeta_0)$,
they are given  by
\begin{align}
L(x_0,x_1,x_2)&=
\begin{vmatrix}
x_0 & x_1 & x_2 \\
1   & \wp_0 & \wp_0' \\
1   & \wp_{00} & \wp_{00}'
\end{vmatrix},
\\
C(x_0,x_1,x_2)&=
\begin{vmatrix}
x_0^2  &  x_0x_1  & x_0x_2 & x_1^2  & x_1x_2  & x_2^2     \\
1      &  \wp_0     &  \wp_0'  & \wp_0^2  & \wp_0\wp_0' & (\wp_0')^2 \\
1      &  \wp_1     &  \wp_1'  & \wp_1^2  & \wp_1\wp_1' & (\wp_1')^2 \\
1      &  \wp_2     &  \wp_2'  & \wp_2^2  & \wp_2\wp_2' & (\wp_2')^2 \\
1      &  \wp_3     &  \wp_3'  & \wp_3^2  & \wp_3\wp_3' & (\wp_3')^2 \\
1      &  \wp_4     &  \wp_4'  & \wp_4^2  & \wp_4\wp_4' & (\wp_4')^2 
\end{vmatrix}.
\end{align}
The cubic in question is found to be  $Q=L\cdot C$.
However, the Weierstrass form of the cubic pencil $P+tQ$
is so complicated that we do not attempt to describe it here.   

\paragraph{Four Wilson lines II}
Finally, let us  take the model with
$\{\nu_i\}=\{0,\pm \zeta_1,\pm \zeta_2,\pm \zeta_3,\pm \zeta_4\}$.
We thus need to find a cubic $Q$ which passes through
the eight points $(1,\pm \wp_i,\wp_i')$, 
where $\wp_i=\wp(\tau|\zeta_i)$, $\wp_i'=\wp'(\tau|\zeta_i)$,
for $i=1,\dots,4$,
as well as $(0,0,1)$.
 
We find that such a  cubic $Q$ is given by 
\begin{equation*}
Q=(g_0-f_0\sigma_1-4\sigma_3)x_0^2x_1
 +(f_0+4\sigma_2)x_0x_1^2
 -\sigma_1x_0x_2^2
 +x_1x_2^2
 +(4\sigma_4-g_0\sigma_1)x_0^3,
\end{equation*}
where $\sigma_i$ is the $i$th basic symmetric polynomial
of $\wp_i$s.
In fact we can see that 
\[
Q(1,x,y)=4(x-\wp_1)(x-\wp_2)(x-\wp_3)(x-\wp_4),
\]
if $y^2=4x^3-f_0x-g_0$, that is, 
if we restrict $Q$ on $\Einf$.
Nevertheless, $Q$ itself is an irreducible cubic in this case.



\vskip6mm\noindent
{\it Acknowledgements}.

\vskip2mm
I benefitted from attending
the workshop  
``{Periods Associated to Rational Elliptic Surfaces and
Elliptic Lie Algebras}'' 
held at International Christian University in June 2001.
I wish to express my gratitude to  many participants 
both of the workshop and of my informal talks  
at Nagoya University for useful comments.
Among them are
Profs.~%
H.  Kanno,
A.  Kato,
S. Kondo,
H. Ohta,
Y. Shimizu,
A. Tsuchiya, and 
Y. Yamada. 
I would also like to thank Dr.~Y.~Ohtake for 
helpful discussions.
My special thanks are due to
the late Prof.~S.-K. Yang for  valuable  advice.
This work was supported in part by Grant-in-Aid for 
Scientific Research 
on Priority Area 707 ``Supersymmetry and Unified Theory of Elementary 
Particles'', Japan Ministry of Education, Science, Sports, Culture
and Technology.

\renewcommand{\thesection}{}
\section{\!\!\!\!\!\!\!Appendix~A \  }
\renewcommand{\theequation}{A.\arabic{equation}}\setcounter{equation}{0}
In this appendix we present 
for completeness the explicit form of 
the generators ${\cal S}$ and 
${\cal T}$ of the algebraic invariants 
of the ternary cubic form $R$ (\ref{general-cubic}). 
{\allowdisplaybreaks
\begin{align}
 {\cal S}&= 
-a_{300}a_{120}a_{012}^2
-2\,a_{210}a_{012}a_{111}^2
+a_{300}a_{030}a_{102}a_{012}
+a_{300}a_{003}a_{120}a_{021}
\nn \\
&+a_{030}a_{003}a_{210}a_{201}
-a_{210}a_{120}a_{102}a_{012}
-a_{210}a_{201}a_{012}a_{021}
-a_{120}a_{201}a_{102}a_{021}
\nn \\
&-a_{030}a_{201}^2a_{012}
-a_{003}a_{120}^2a_{201}
-a_{300}a_{102}a_{021}^2
+a_{210}^2a_{012}^2
-a_{030}a_{210}a_{102}^2
\nn
\\
&
-a_{003}a_{210}^2a_{021}
+a_{120}^2a_{102}^2
+a_{201}^2a_{021}^2
-2\,a_{120}a_{102}a_{111}^2
+a_{111}^4
\nn \\
&-a_{300}a_{030}a_{003}a_{111}
-2\,a_{201}a_{021}a_{111}^2
+a_{300}a_{012}a_{021}a_{111}
+a_{030}a_{201}a_{102}a_{111}
\nn \\
&+a_{003}a_{210}a_{120}a_{111}
+3\,a_{210}a_{102}a_{021}a_{111}
+3\,a_{120}a_{201}a_{012}a_{111},
\label{es}
\\
{\cal T}&= 
-3\,a_{012}^2a_{300}^2a_{021}^2
-24\,a_{111}^2a_{201}^2a_{021}^2
+24\,a_{120}a_{111}^4a_{102}
+4\,a_{120}^3a_{300}a_{003}^2
\nn \\
&-3\,a_{120}^2a_{210}^2a_{003}^2
-27\,a_{210}^2a_{102}^2a_{021}^2
+4\,a_{201}^3a_{030}^2a_{003}
-24\,a_{012}^2a_{210}^2a_{111}^2
\nn \\
&+24\,a_{012}a_{210}a_{111}^4
-24\,a_{120}^2a_{102}^2a_{111}^2
+4\,a_{300}^2a_{021}^3a_{003}
+a_{300}^2a_{030}^2a_{003}^2
\nn \\
&+24\,a_{111}^4a_{201}a_{021}
-12\,a_{120}^2a_{300}a_{003}a_{102}a_{021}
+4\,a_{012}^3a_{300}^2a_{030}
-27\,a_{012}^2a_{120}^2a_{201}^2
\nn \\
&-3\,a_{201}^2a_{030}^2a_{102}^2
+8\,a_{120}^3a_{102}^3
+8\,a_{012}^3a_{210}^3
+24\,a_{210}^2a_{003}a_{201}a_{021}^2
\nn \\
&+4\,a_{300}a_{030}^2a_{102}^3
+12\,a_{210}^2a_{111}^2a_{021}a_{003}
+12\,a_{210}a_{111}^2a_{102}^2a_{030}
-36\,a_{210}a_{111}^3a_{021}a_{102}
\nn \\
&+8\,a_{201}^3a_{021}^3
-8\,a_{111}^6
+4\,a_{210}^3a_{003}^2a_{030}
-24\,a_{300}a_{030}a_{102}^2a_{111}a_{021}
\nn \\
&+12\,a_{201}^2a_{030}a_{102}a_{111}a_{021}
-12\,a_{120}a_{210}a_{102}^3a_{030}
-12\,a_{120}a_{210}a_{111}^3a_{003}
\nn \\
&+24\,a_{120}a_{300}a_{102}^2a_{021}^2
-12\,a_{300}a_{201}a_{021}^3a_{102}
-12\,a_{102}a_{201}a_{111}^3a_{030}
\nn \\
&
-24\,a_{210}^2a_{003}a_{102}a_{111}a_{030}
-24\,a_{300}a_{210}a_{021}^2a_{111}a_{003}
-20\,a_{300}a_{111}^3a_{030}a_{003}
\nn \\
&-12\,a_{201}^2a_{030}a_{210}a_{021}a_{003}
+36\,a_{210}a_{111}^2a_{201}a_{030}a_{003}
+36\,a_{210}a_{102}a_{111}a_{021}^2a_{201}
\nn\\
&+6\,a_{300}a_{030}a_{102}a_{210}a_{021}a_{003}
-6\,a_{300}a_{030}^2a_{102}a_{201}a_{003}
+6\,a_{210}a_{201}a_{021}a_{102}^2a_{030}
\nn\\
&+12\,a_{300}a_{021}^2a_{111}^2a_{102}
+12\,a_{300}a_{201}a_{030}a_{003}a_{111}a_{021}
+12\,a_{012}a_{300}a_{210}a_{030}a_{111}a_{003}
\nn \\
&
+6\,a_{012}a_{300}a_{210}a_{021}^2a_{102}
+12\,a_{012}a_{120}a_{210}^2a_{111}a_{003}
+6\,a_{012}a_{120}a_{201}^2a_{030}a_{102}
\nn \\
&+18\,a_{012}a_{300}a_{030}a_{102}a_{201}a_{021}
+6\,a_{012}a_{120}a_{300}a_{201}a_{030}a_{003}
-12\,a_{012}a_{120}a_{300}a_{030}a_{102}^2
\nn \\
&-12\,a_{012}a_{120}a_{210}a_{111}^2a_{102}
+36\,a_{012}a_{120}a_{201}^2a_{021}a_{111}
+36\,a_{012}a_{120}^2a_{111}a_{102}a_{201}
\nn \\
&-24\,a_{012}a_{120}^2a_{300}a_{111}a_{003}
+6\,a_{012}a_{120}^2a_{210}a_{201}a_{003}
+18\,a_{012}a_{120}a_{300}a_{210}a_{021}a_{003}
\nn \\
&
-60\,a_{012}a_{120}a_{300}a_{111}a_{021}a_{102}
+6\,a_{012}^2a_{120}a_{300}a_{201}a_{021}
-24\,a_{012}^2a_{300}a_{201}a_{111}a_{030}
\nn \\
&-6\,a_{012}a_{120}a_{210}a_{102}a_{201}a_{021}
-12\,a_{012}a_{210}a_{111}^2a_{201}a_{021}
+36\,a_{012}a_{300}a_{030}a_{102}a_{111}^2
\nn \\
&+36\,a_{012}^2a_{120}a_{210}a_{111}a_{201}
-12\,a_{012}^2a_{300}a_{030}a_{102}a_{210}
+12\,a_{012}^2a_{300}a_{111}a_{021}a_{210}
\nn \\
&-24\,a_{120}a_{201}^2a_{030}a_{111}a_{003}
-12\,a_{120}a_{102}a_{201}^2a_{021}^2
+36\,a_{120}a_{210}a_{102}^2a_{111}a_{021}
\nn \\
&
+6\,a_{120}a_{210}^2a_{003}a_{102}a_{021}
-60\,a_{012}a_{102}a_{201}a_{111}a_{030}a_{210}
+36\,a_{012}a_{210}^2a_{111}a_{021}a_{102}
\nn \\
&+12\,a_{120}a_{300}a_{030}a_{102}a_{111}a_{003}
-60\,a_{120}a_{210}a_{201}a_{021}a_{111}a_{003}
+18\,a_{120}a_{210}a_{102}a_{201}a_{030}a_{003}
\nn \\
&
-12\,a_{120}a_{102}a_{111}^2a_{201}a_{021}
+12\,a_{120}a_{201}a_{111}a_{030}a_{102}^2
-12\,a_{120}^3a_{102}a_{201}a_{003}
\nn \\
&-12\,a_{120}a_{300}a_{003}a_{201}a_{021}^2
+36\,a_{120}a_{300}a_{111}^2a_{021}a_{003}
-6\,a_{120}a_{300}a_{003}^2a_{210}a_{030}
\nn \\
&-12\,a_{120}^2a_{102}^2a_{201}a_{021}
+24\,a_{120}^2a_{201}^2a_{021}a_{003}
+12\,a_{120}^2a_{111}^2a_{201}a_{003}
\nn \\
&+12\,a_{012}a_{201}^2a_{030}a_{111}^2
-12\,a_{012}a_{300}a_{111}^3a_{021}
-12\,a_{012}a_{120}^2a_{210}a_{102}^2
-12\,a_{012}a_{201}^3a_{030}a_{021}
\nn \\
&-12\,a_{012}a_{210}a_{201}^2a_{021}^2
-36\,a_{012}a_{120}a_{111}^3a_{201}
-12\,a_{012}a_{210}^3a_{021}a_{003}
+24\,a_{012}a_{210}^2a_{102}^2a_{030}
\nn \\
&-12\,a_{012}^3a_{300}a_{210}a_{120}
+24\,a_{012}^2a_{120}^2a_{300}a_{102}
-12\,a_{012}^2a_{120}a_{210}^2a_{102}
+12\,a_{012}^2a_{120}a_{300}a_{111}^2
\nn \\
&-12\,a_{012}^2a_{210}^2a_{201}a_{021}
+24\,a_{012}^2a_{201}^2a_{030}a_{210}
+12\,a_{120}^2a_{210}a_{102}a_{111}a_{003}
\nn \\
&-12\,a_{012}a_{210}^2a_{201}a_{030}a_{003}
+12\,a_{012}a_{300}a_{201}a_{021}^2a_{111}
-6\,a_{012}a_{300}^2a_{030}a_{021}a_{003}.
\label{tee}
\end{align}}

\newpage

\end{document}